%% file: draft.tex
\documentclass[fleqn, 11pt]{article}

\usepackage[backgroundcolor=white, textsize=tiny, disable]{todonotes}
\usepackage{amssymb, amsmath,amsthm, natbib, indentfirst, endnotes, graphicx, rotating, thmtools, thm-restate}
\usepackage[singlespacing]{setspace}
\usepackage[multiple]{footmisc}
\usepackage{tikz}
\usepackage{pgfplots}
\usepackage{hyperref}
\hypersetup{colorlinks=true, linkcolor=blue, urlcolor=blue, citecolor=blue, hyperfootnotes=false,
    pdfauthor = Nick Janetos and Jan Tilly,
    pdftitle = Reputation Dynamics in a Market for Illicit Drugs,
    pdfsubject = Reputation Dynamics in a Market for Illicit Drugs,
    pdfkeywords = {online reputation, adverse selection, crime}}

\usepackage{thmtools}
\usepackage{versions}
\usepackage{booktabs}
\usepackage{enumitem}
\usepackage{rotating}
\usepackage{color}
\usepackage{datetime}

\usepackage[labelfont=bf]{caption}

\excludeversion{counterfactual}

\excludeversion{moralhazard}

\excludeversion{robustness}

\excludeversion{dictionary}

\newtheorem*{assumption*}{Assumption}

\declaretheorem{proposition}

\newcommand{\floatfoot}[1]{\vskip 0.25cm \footnotesize #1}

\newenvironment{definition}[1][Definition]{\begin{trivlist}
\item[\hskip \labelsep {\bfseries #1}]}{\end{trivlist}} 

\usepackage[nohead]{geometry}
\makeatletter
\def\namedlabel#1#2{\begingroup
   \def\@currentlabel{#2}
\label{#1}\endgroup
}
\newcommand\blfootnote[1]{%
  \begingroup
  \renewcommand\thefootnote{}\footnote{#1}%
  \addtocounter{footnote}{-1}%
  \endgroup
}
\makeatother

\newcommand{\E}{\mathbb{E}}
\newcommand{\N}{\mbox{N}}
\renewcommand{\P}{\Pr}

\newcommand{\categories}{cannabis, MDMA, heroin, and cocaine}
\newcommand{\Categories}{Cannabis, MDMA, heroin, and cocaine}

\title{Reputation Dynamics in a Market for Illicit Drugs}
\author{Nick Janetos \and Jan Tilly\blfootnote{University of Pennsylvania, 160 McNeil building, 3718 Locust Walk, Philadelphia, PA 19104. Email \mbox{\href{mailto:njanetos@econ.upenn.edu}{njanetos@econ.upenn.edu}} and \mbox{\href{mailto:jtilly@econ.upenn.edu}{jtilly@econ.upenn.edu}}. Thanks to
Camilo Garcia-Jimeno,
Qing Gong,
Ami Ko,
Pinar Yildirim,
and workshop participants at the University of Pennsylvania for helpful discussion and comments.}}
\date{\monthname\ \the \year}

\begin{document}

\maketitle

\thispagestyle{empty}

\begin{abstract}
  \input{section/abstract.tex}
\end{abstract}

\strut\


\strut\


\newpage


\newpage

\onehalfspacing

\input{section/introduction.tex}
\input{section/darknetmarkets.tex}
\input{section/data.tex}
\input{section/model.tex}
\input{section/estimation.tex}
\input{section/results.tex}
\input{section/conclusion.tex}

\clearpage
\phantomsection
\addcontentsline{toc}{section}{References}
\bibliographystyle{ecta}
\bibliography{citations}



\appendix

\section{Appendix: Model}\label{app:model}
\input{section/simplemodel.tex}




\end{document}

%% file: section/abstract.tex

We analyze reputation dynamics in an online market for illicit drugs using a novel dataset of prices and ratings.
The market is a black market, and so contracts cannot be enforced.
We study the role that reputation plays in alleviating adverse selection in this market.
We document the following stylized facts:
(i) There is a positive relationship between the price and the rating of a seller.
This effect is increasing in the number of reviews left for a seller.
A mature highly-rated seller charges a 20\% higher price than a mature low-rated seller.
(ii) Sellers with more reviews charge higher prices regardless of rating.
(iii) Low-rated sellers are more likely to exit the market and make fewer sales.
We show that these stylized facts are explained by a dynamic model of adverse selection, ratings, and exit, in which buyers form rational inferences about the quality of a seller jointly from his rating and number of sales.
Sellers who receive low ratings initially charge the same price as highly-rated sellers since early reviews are less informative about quality.
Bad sellers exit rather than face lower prices in the future.
We provide conditions under which our model admits a unique equilibrium.
We estimate the model, and use the result to compute the returns to reputation in the market.
We find that the market would have collapsed due to adverse selection in the absence of a rating system.

%% file: section/introduction.tex

\section{Introduction\label{sec:introduction}}

In recent years, online marketplaces for illicit drugs have emerged due to innovations in digital currencies and encryption.
In these illegal marketplaces, buyers and sellers transact anonymously and the merchandise is shipped by mail.
These marketplaces face significant problems of adverse selection, because based on product descriptions alone, buyers cannot tell good products apart from bad products.
To study the role that reputation plays to alleviate the adverse selection problem in these markets, we collect a novel dataset of prices, reviews, and ratings from one such marketplace.
In this market, buyers can influence the reputation of sellers by leaving reviews, which affect seller ratings.
We document how ratings, prices, and sellers' decisions to exit the marketplace interact.

We then develop a dynamic model of reputation, in which buyers form rational inferences about the quality of a seller jointly from his rating and the number of sales.\footnote{In the theoretical literature, there are different approaches to modeling reputation.
See~\cite{Mailath06} for a survey.
We take an adverse selection approach.
Sellers differ in their types, which we interpret as qualities.
Buyers do not directly observe the quality of sellers.
They observe noisy signals, which are informative about the quality of the seller.
As in~\cite{Ekmekci11}, these signals are modeled as a rating system, in which buyers leave a review after each sale.
The reputation of a seller is the market's belief about his type, conditional on the information revealed by the rating system.} Sellers who initially receive low ratings charge the same price as sellers who initially receive high ratings, since early reviews are less informative about quality.
Bad sellers exit rather than face lower prices in the future.
We provide conditions under which our model admits a unique equilibrium.
We estimate the model, and use the result to compute the returns to reputation in the market. \begin{counterfactual} We find that unsurprisingly, these markets would not exist without reputation mechanisms.
\end{counterfactual}

Black markets, or markets for illegal goods, are of considerable economic importance, and understudied due to difficulties involved in measuring activity in these markets (see e.g.~\cite{Levitt00}).
Online markets for illicit drugs generate annual revenues in the hundreds of millions of US dollars.\footnote{The first such online black market was Silk Road, founded in 2011.~\cite{Soska15} analyze data from Silk Road and impute that the site generated daily revenues of about $\$300,000$ in early 2013, implying annual revenues in excess of $\$100$ million.
Later in 2013, the Silk Road's founder was arrested and the dark net market shut down.
A variety of other market platforms then imitated Silk Road, and some of these market platforms continue to operate.
They comprise a relatively small part of the global drug trade, where global revenue is estimated to be in the hundreds of billions of US dollars.}
Black markets differ from legal markets in a number of important ways.
The aspect we focus on in this paper is the role reputation plays in the operation of these markets.
A buyer may purchase a good and discover it is of low quality.
His only recourse is to then leave a bad review for the seller through a rating system, akin to those used by legal marketplaces such as Amazon and eBay.
In legal markets, a buyer may also receive a refund through the market platform itself (both Amazon and eBay have money-back guarantees), or, in extreme circumstances, may take legal action.
Neither of these channels are available to buyers in a black market.
If lower-ranked sellers receive a lower price, then the rating system works either to enforce cooperation (in the case of moral hazard) or to drive bad sellers out of the market (in the case of adverse selection).
In this paper, we focus on adverse selection.\begin{moralhazard}\footnote{In Appendix~\ref{sec:moralhazard}, we consider a model with both moral hazard and adverse selection.
We estimate this model against the data, and find that reputation dynamics in the data are better explained by adverse selection.}\end{moralhazard}

We collect a novel dataset of prices and ratings for illicit drugs from Agora, a popular online marketplace for a vast range of illegal merchandise.
We develop a web crawler suitable for the dark net and use it to archive listing pages, vendor pages, and reviews.
The dataset covers the time period from January 2014 until the market's closure in August 2015.
Using this dataset, we document the effect a seller's rating has on the prices he charges.

In line with the prior literature that studies legal markets, there is a positive relationship between the price and the rating of a seller (for surveys, see~\cite{Dellarocas02} and~\cite{Cabral12}).
We show that this effect is increasing in the number of reviews left for a seller.
A mature highly-rated seller charges a 20\% higher price than a mature low-rated seller.
For young sellers, the rating has no relationship to the price.
Sellers with more reviews charge a higher price than sellers with a low number of reviews regardless of rating.
Our findings are consistent with adverse selection and buyers who draw rational inferences from ratings: When a seller only has a few reviews, buyers do not draw much inference about the quality of a seller from the rating, but when a seller has many reviews, the rating is more informative, and has a stronger effect on the price.
Bad sellers exit the marketplace rather than face lower prices in the future.

We introduce a dynamic model of adverse selection and ratings that is consistent with the data.
In our model, sellers enter the market, and are assigned a type, unobserved by buyers.
As sales are made, buyers sample the goods, and can leave feedback with the market.
We are agnostic about the exact process through which buyers form opinions and leave feedback.
Following prior theoretical literature on rating systems (see e.g.~\cite{Ekmekci11, Bohren12, Hu15}), we assume only that the market offers some type-dependent, stochastically evolving, publicly available information about each buyer.
In our case this information consists of the rating and the total number of sales made by a seller.

Buyers are short-lived and do not observe the full history of the seller's public characteristics, only the public rating at the time at which the buyer enters the game.
From this information, buyers form inferences about the quality of sellers consistent with the sellers' equilibrium exit strategies, and sellers are paid a price for their product which depends on the inference buyers draw in equilibrium about seller quality.
Our model differs from the existing theoretical literature on rating systems by considering a problem of adverse selection; the sole strategic decision made by a seller is whether or not to exit the game.
For some parameter values there may be multiple equilibria.
We establish restrictions on the discount rate so that the model admits a unique equilibrium.

In our model, the market's rating system acts to separate low and high quality sellers, alleviating adverse selection in the market.
Low quality sellers who are identified as such prefer to exit the market.
This induces a tension between exit and the rating system---if, for example, it were the case that low quality sellers strictly preferred to exit at a low rating, then rational buyers should infer that low rated sellers are in fact high quality, meaning that \emph{low} rated sellers charge a \emph{high} price in equilibrium.
At the same time, exit of low-rated sellers introduces survivorship bias into reduced form estimates of the relationship between price and rating.
These two channels---inference on the part of consumers, and survivorship bias---both bias reduced form estimates of the relationship between price and rating downward.
We show by example that these two effects may be so large that it may happen that some sellers receive a \emph{higher} price if rated poorly, even as the returns to reputation are positive.

We use a nested fixed point maximum likelihood estimator to estimate our model for the subset of cannabis sellers.
Our estimates allow us to identify the proportion of high and low quality sellers in the market, as well as how the composition of the market changes as sellers age.
We estimate the returns to reputation, and show that sellers face significant dynamic penalties from a drop in rating.
In particular, a young high quality seller whose rating drops from 5.0 to 4.99 expects to lose around 2\% of total expected future profits, while a low quality seller expects to lose 11\%.
We consider various counterfactual exercises.
We show that the market would not have functioned without a rating system, meaning that sellers would have exited as soon as possible.

\subsubsection*{Prior literature}

The work in this paper contrasts with previous research in two ways: First, we consider a novel market platform, in which contracts cannot be enforced.
Second, we do not solely focus on the impact a seller's rating has on his price, rather, we focus on the joint role of ratings, exit, prices, and inferences buyers draw about the quality of sellers.

Perhaps the closest paper to ours is~\cite{Saeedi14}, who estimates a structural model of adverse selection and ratings on eBay.
Our approach differs in the following key ways: First, we consider a richer set of publicly observable seller characteristics, including the rating, a feature made available by the relatively more disperse ratings in our dataset compared to eBay.
Second, we model price formation differently.
\citeauthor{Saeedi14} assumes that buyers draw no inferences from the price.
In contrast, we assume that sellers are price takers and paid their expected worth conditional only on publicly available characteristics.

Prior work on the market for illicit drugs has focused mainly on the determinants of prices, using the \textsc{stride} dataset which records prices of seized illicit drugs in the United States.
Using these data,~\cite{Galenianos12} and~\cite{Galenianos15} focus on learning under moral hazard, but do not focus on rating systems.
\cite{Dobkin00} look at the effect of government seizures of methamphetamine precursors on the price of methamphetamines.

Our findings are consistent with the literature on \emph{legal} online markets.
A small but growing literature attempts to estimate the `returns to reputation', i.e.\ the additional value that a highly-rated seller receives in a market over a low-rated seller (see~\cite{Ederington02},
\cite{Melnik02},
\cite{Resnik02},
\cite{Eaton05},
\cite{Eaton05b},
\cite{Houser06},
\cite{Cabral10},
\cite{Yoganarasimhan13},
\cite{Saeedi14},
and~\cite{Jolivet16}).
In almost all of the existing literature, a statistically significant and positive, but small and fragile effect of reputation on price is found.

The rest of this paper is structured as follows.
In Section~\ref{sec:darknetmarkets}, we discuss the institutional details of these markets and provide some historical background.
In Section~\ref{sec:data}, we explain how the data were collected, and establish some stylized facts about the role reputation plays in this market.
In Section~\ref{sec:model}, we present a dynamic model of reputation consistent with these stylized facts. In Section~\ref{sec:estimation}, we estimate the model. In Section~\ref{sec:results}, we use the estimated model to compute the return to reputation in this market. \begin{counterfactual} We provide an estimate of how well the markets would function in the absence of a rating system, and estimate the marginal worth of a so-called `Sybil' attack.\end{counterfactual} We conclude in Section~\ref{sec:conclusion}.

\nocite{Bajari03, Mcdonald02, Dellarocas01, Douceur02, Resnick01, Ghose07, Zhang06, Lee00, Ba02, Resnik06, Bolton04, Ying16}

%% file: section/darknetmarkets.tex

\section[Background]{Background: Dark Net Market Places\label{sec:darknetmarkets}}

The first online black market was Silk Road, founded in 2011 and shut down by law enforcement 2013.
Among its successors is the Agora marketplace, which is the data source for this paper.
The Agora marketplace launched in late 2013, and went on to weather two of the major shocks to dark net markets.
First, in November 2014, the FBI seized the servers of many major markets as part of `Operation Onymous'.
The Agora marketplace was not affected, and subsequently became one of the major platforms for trading on the dark net.
In March 2015, one of Agora's competitors, the Evolution marketplace, closed suddenly, taking the assets held in escrow along with it, after which the Agora marketplace became the largest market platform in the dark net.
Traffic spiked, and several months later, in August 2015, the administrators of the site announced that due to security concerns, they were temporarily taking their operations offline.
Funds were returned to sellers and buyers, and the market servers were shut down.

\begin{figure}[hbt!]
  \frame{\includegraphics[height = 2.4in]{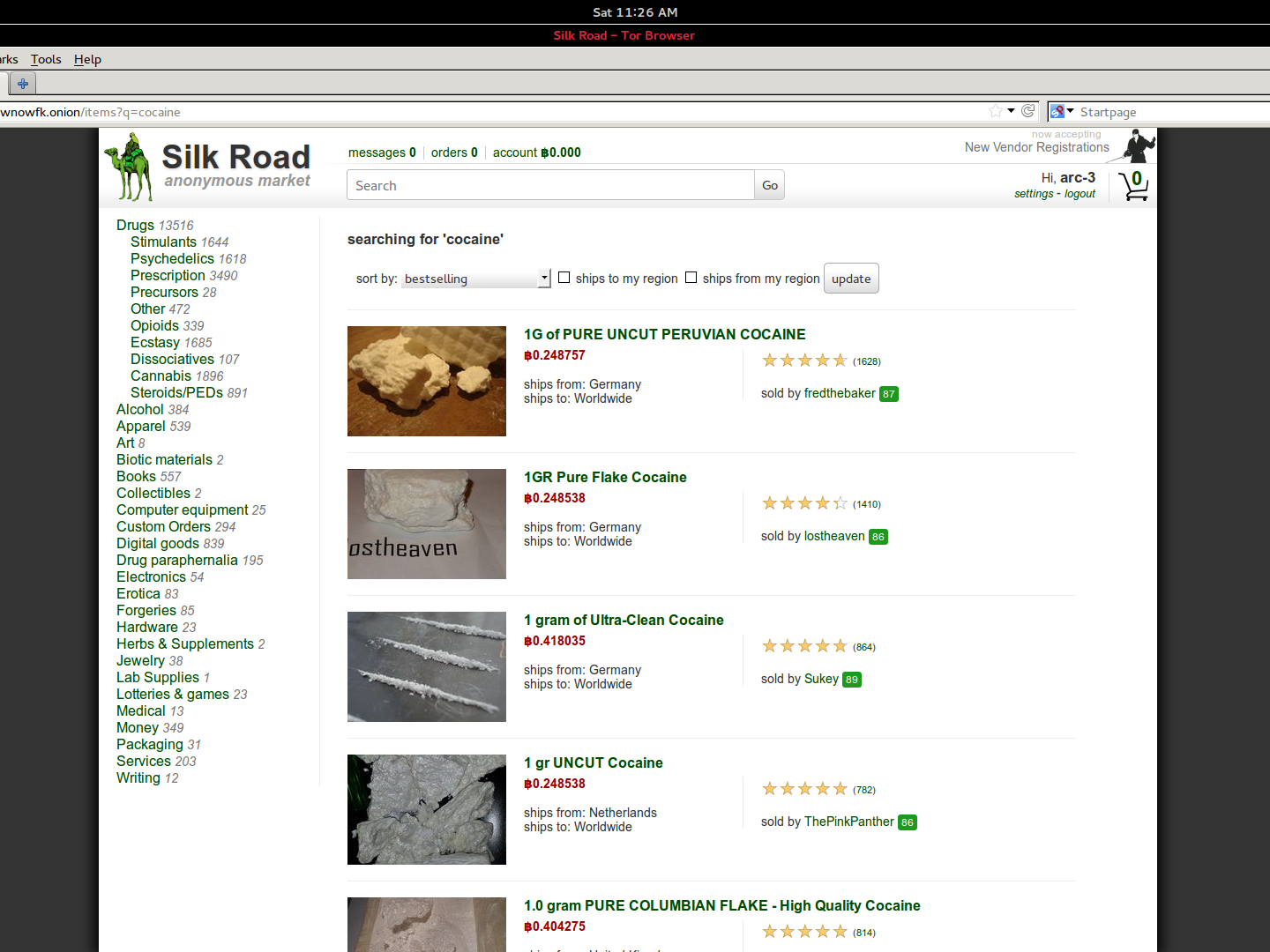}}
  \frame{\includegraphics[height = 2.4in]{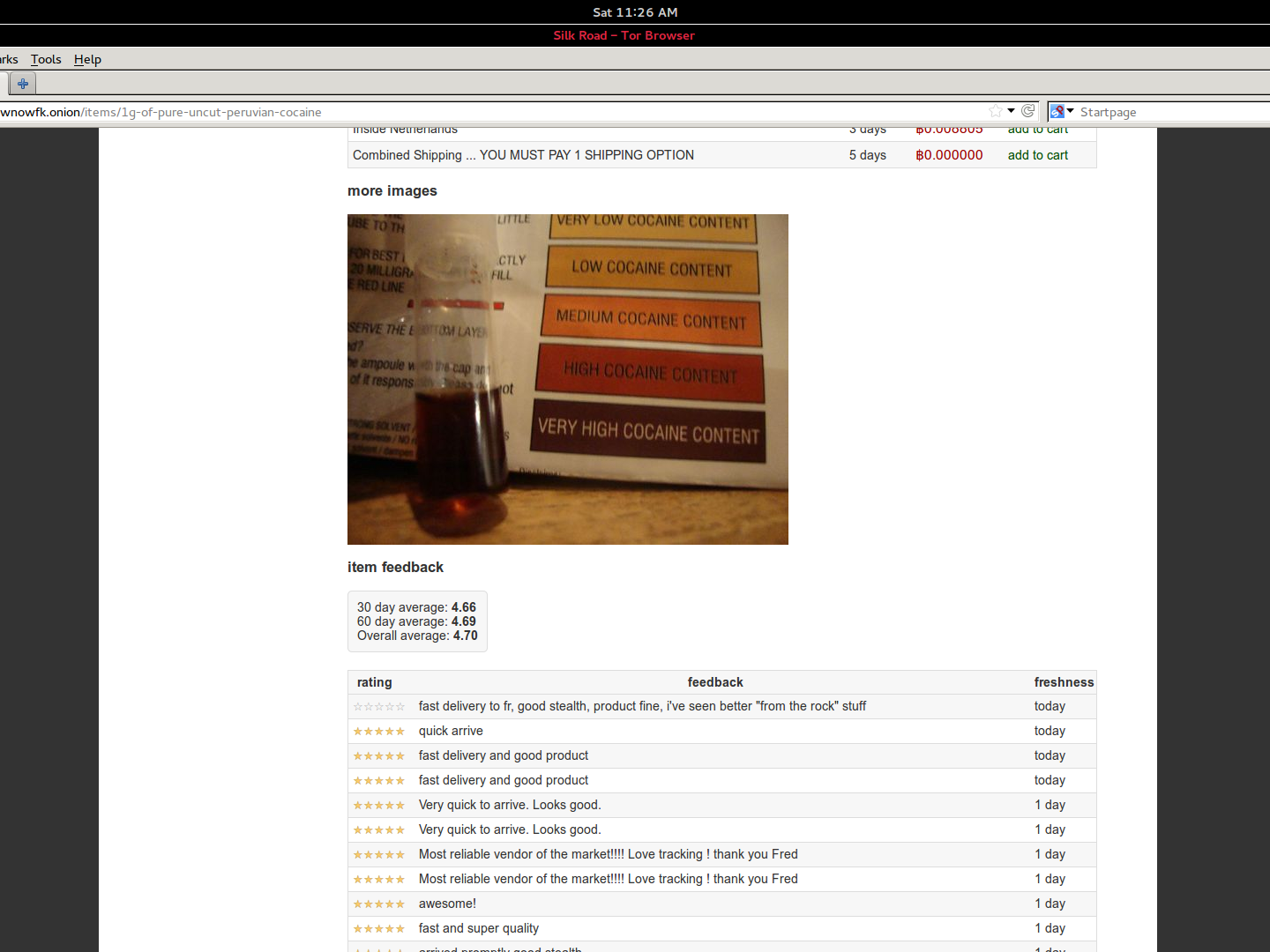}}
  \caption{Screen shots from Silk Road, mid-2015.\label{fig:silkroad}}
  \floatfoot{\textbf{Note:} On the left, we show a screen shot from searching for `cocaine' on Silk Road. On the right, we show a screen shot of the reviews page of a seller, as well as a photo provided by the seller illustrating the high cocaine content of his product.}
\end{figure}

While it was operational, the Agora marketplace was (like most dark net marketplaces, see Figure~\ref{fig:silkroad}) run in a way recognizable to any user of Amazon.
Sellers could, after paying a fee (around \$500), open an account and create listings.
Sellers were free to choose the price for each listing.
Buyers could then see a list of goods for sale using a search tool or by clicking on a category.
Each entry for a listing contained its price, the number of sales made by a seller, the net rating of the seller, and information on the shipping source and destination of the goods.
Potential buyers could then click to see more details about a listing, such as a more detailed description written by the seller, as well as individual reviews left by buyers.

Once a buyer had made a decision, they could click to add the listing to a shopping cart.
All items were denominated and purchased using bitcoins.
To pay, a buyer would transfer their bitcoins directly to the marketplace itself.
Many of the marketplaces offered a bitcoin `tumbler', a bitcoin-laundering algorithm which transfered the bitcoin between many different addresses to anonymize its source location.\footnote{Bitcoins are a non-fungible currency, and in theory it is possible to track its history through every transaction made.} The Agora marketplace did not offer a bitcoin tumbler, rather, participants were encouraged to anonymize their own currency.

Once a transaction had been requested, the marketplace would check to make sure that a buyer had the requisite number of funds held in escrow by the marketplace.
A postal address was sent to the seller, who would then ship the item to the buyer using regular mail.
Items were usually disguised to not draw suspicion from the postal service or law enforcement.
Once the item arrived, buyers were responsible for returning to the marketplace and marking the transaction as having completed, at which point, the marketplace would release the funds held in escrow to the seller.
If the buyer did not mark the transaction as complete, the funds would be released after several weeks, unless the buyer decided to contest the sale, in which case the Agora marketplace offered mediation services between buyer and seller.
We do not observe instances of this occurring, and to our knowledge this was a rare outcome.
Some sellers preferred to bypass the escrow system, advertising themselves as sellers who required buyers to finalize the transactions early, releasing the funds before the goods had actually been received.

%% file: section/data.tex

\section{Data\label{sec:data}}

\subsection[Data Collection]{Data Collection and Summary Statistics\label{sec:datacollection}}

We scrape data from the Agora marketplace at weekly intervals from January 2014 until the market's closure in August 2015.
We develop a web crawler to archive vendor pages, listing pages, and reviews.\footnote{A web crawler is a computer program which visits a link, records all the links it finds there, then visits each of those links, recursively repeating the process in order to visit every page of a website.
Previous research archiving the dark web uses infrequent archives of the entire website to develop a `snapshot' of the site at different times.
Since sites on the dark web are often inaccessible due to network problems, we develop a bespoke web crawler which continually visits the site, cross-referencing the information it finds with previous visits to produce a high-frequency history of the site.} Vendor pages contain the market-reported aggregate rating and the number of sales made.
These vendor pages change frequently as vendors make additional sales.
Listing pages contain detailed production information, including a title and description of the product, its price, a picture, and whether the vendor requires buyers to pay up-front or after arrival of the shipment.
Note that the listing page may change over time, because vendors can edit all the information on the listing page.
Listing pages also contain the entire history of reviews.
For each review we observe and record
\begin{enumerate}
  \item the date at which a review was left,
  \item the price of the good at the time the review was left,
  \item the rating which was left for the review, which is a number between 0 and 5,
  \item the market-reported aggregate rating for the vendor for whom the review was left,
  \item the sort of good on sale, in this paper we focus on \categories, and
  \item various vendor and product characteristics, such as the total sales made by that vendor and whether he required buyers to pay up-front or after the shipment arrived.
\end{enumerate}
For the four product categories that we focus on, \categories, we collected information on $1,482$ vendors, $14,292$ listings, and $572,266$ reviews.

We do not observe transactions directly, but we find that the total number of reviews is a good approximation for the total number of sales made by each particular vendor.
Therefore, we treat each review as a transaction and we will occasionally use the terms review and transaction interchangeably.
We do not observe the time of sale, only the time at which the review was left.
Therefore, the price at the time of sale is imputed by using the price one week before the review was left.

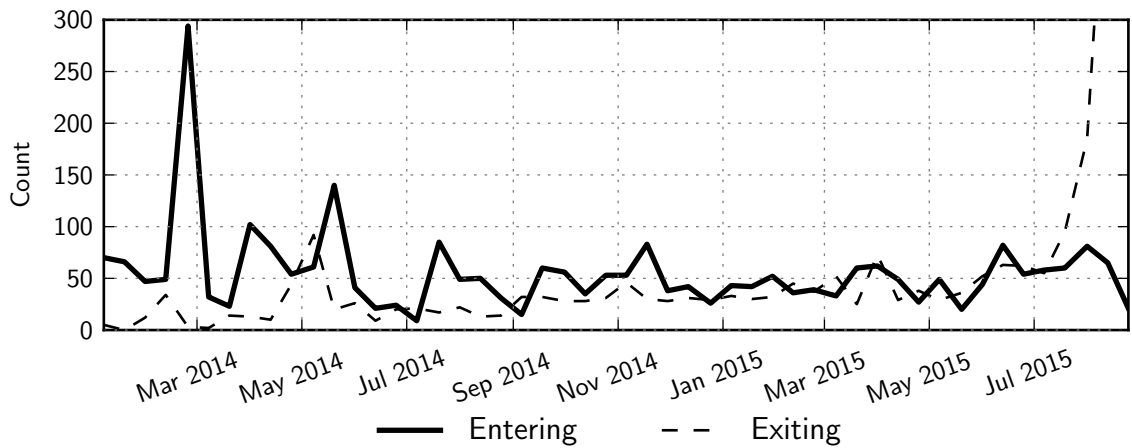
\begin{figure}
  \centerline{
    \input{figures/entryexit.pgf}
  }
  \caption{Distribution of entry and exit by vendors over the course of the market lifespan.\label{fig:entryexit}}
  \floatfoot{\textbf{Note:} The figure shows the number of vendors who enter the market (solid line) and exit the market (dashed line) over time. Entry is defined as the date the first review was observed in the sample, and exit as one week before the last review observed in the sample. The numbers of entries and exits are relatively stable throughout the lifetime of the market. Due to our definition of exit, the number of exits spikes when the market closed.}
\end{figure}

The entry and exit date of vendors is not directly observed.
Entry is not directly observed, because some vendors entered the market before we begin scraping the platform.
Exit is not directly observed, because exit is indistinguishable from inactivity.
We therefore use reviews---our proxy for transactions---to determine entry and exit.
A seller is defined to have entered one week before the first review is left, and exited one week before the last review observed in the sample.
See Figure~\ref{fig:entryexit} for the distribution of entry and exit dates through the lifespan of the market.
Entry and exit rates are relatively homogeneous throughout the lifetime of the market, with the exception that many sellers exited when the market closed.

The Agora market does not have an explicit mechanism for registering the sort of good on sale.
Goods are categorized (e.g., \tt{MDMA---Pills}\rm), but the sort of item on sale and the quantity on sale is not explicitly registered with the site.
To address this, we focus on homogeneous goods and extract quantity information from the listing text by exploiting conventions developed on the marketplace for listing quantities.
For example, a typical listing might be titled
\begin{quote}
  \tt{XTC pills---best quality---fresh from Amsterdam---120mg x5},\rm
\end{quote}
from which the scraper extracts the relevant information that this is a listing for ecstasy (`XTC') pills, each containing 120 milligrams of MDMA, sold in batches of five.
The prices are then normalized to be in dollars per gram, and consistency checks are run on price data to ensure that prices found are broadly consistent with other data.

We choose to focus on the top four categories of goods sold: \Categories.
We report summary statistics for these four categories in Table~\ref{tab:summary-stats}.
In total we observe 1,482 vendors who sell in any of these four product categories.
Among them, 1,080 vendors exclusively sell in one product category, 95 in two, 298 in three, and only four vendors sell in all four product categories.
On average, each seller has received 386 reviews across the four product categories.
The average lifespan of a vendor ranges from approximately 300 to 350 days, where the lifespan per product category refers to the number of days between the first and last posting of a product in a particular category.
A little over one third of all vendors was still active when the market closed in August 2015.
This is roughly equal across product categories.
Cannabis is by far the cheapest product per gram (approximately 12 dollars per gram), Heroin the most expensive (approximately 162 dollars per gram).
For all products, the ratings that vendors receive are very high.
The average rating across product categories is $4.93$ out of $5.00$.
Per product category, the average ratings range from $4.80$ for Heroin to $4.97$ for MDMA.\@

\begin{table}[th!]
    \centerline{
        \input{figures/summary_stats}
    }
\caption{Summary Statistics\label{tab:summary-stats}}
\floatfoot{\textbf{Note:} The table reports summary statistics for the four product categories that we consider: \Categories. The number of vendors across product categories does not sum to the total number of vendors, because some vendors are active in multiple product categories. Share of vendors active when market closed refers to the share of vendors that we observe making sales after August 1, 2015 (the market was shut down in late August). Prices are reported in USD per gram, where we obtain the dollar amounts by converting bitcoin to U.S.\ dollars at the time of the transaction.}
\end{table}

\begin{figure}[h!]
  \centerline{
    \input{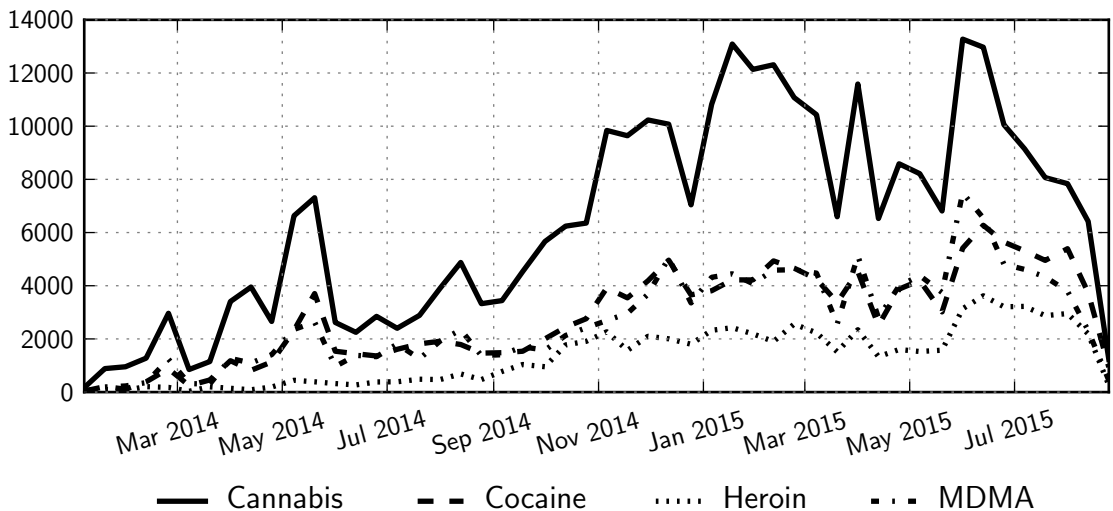}
  }
  \caption{Volume of reviews left, over time, by category.\label{fig:reviewvolume}}
  \floatfoot{\textbf{Note:} The figures show the total number of reviews per week that were left for products that fall into the categories \categories~over the entire lifespan of the marketplace. We use reviews as a proxy for transactions. Figure~\ref{fig:reviewvolume} shows that the market's traffic steadily increased over its lifespan, reaching its peak after the Agora marketplace's main competitors were shut down during `Operation Onymous'.}
\end{figure}

Figure~\ref{fig:reviewvolume} plots the volume of reviews left weekly for each of these four categories.
The figure shows that the market's traffic steadily increased over its lifespan, reaching its peak after the Agora marketplace's main competitors were shut down during `Operation Onymous'.
In the last months of its life, the Agora marketplace began to experience frequent periods during which it was inaccessible due to security precautions taken by its owners.
The marketplace was eventually was shut down, also due to security concerns.

\begin{figure}[h!]
  \centerline{
    \input{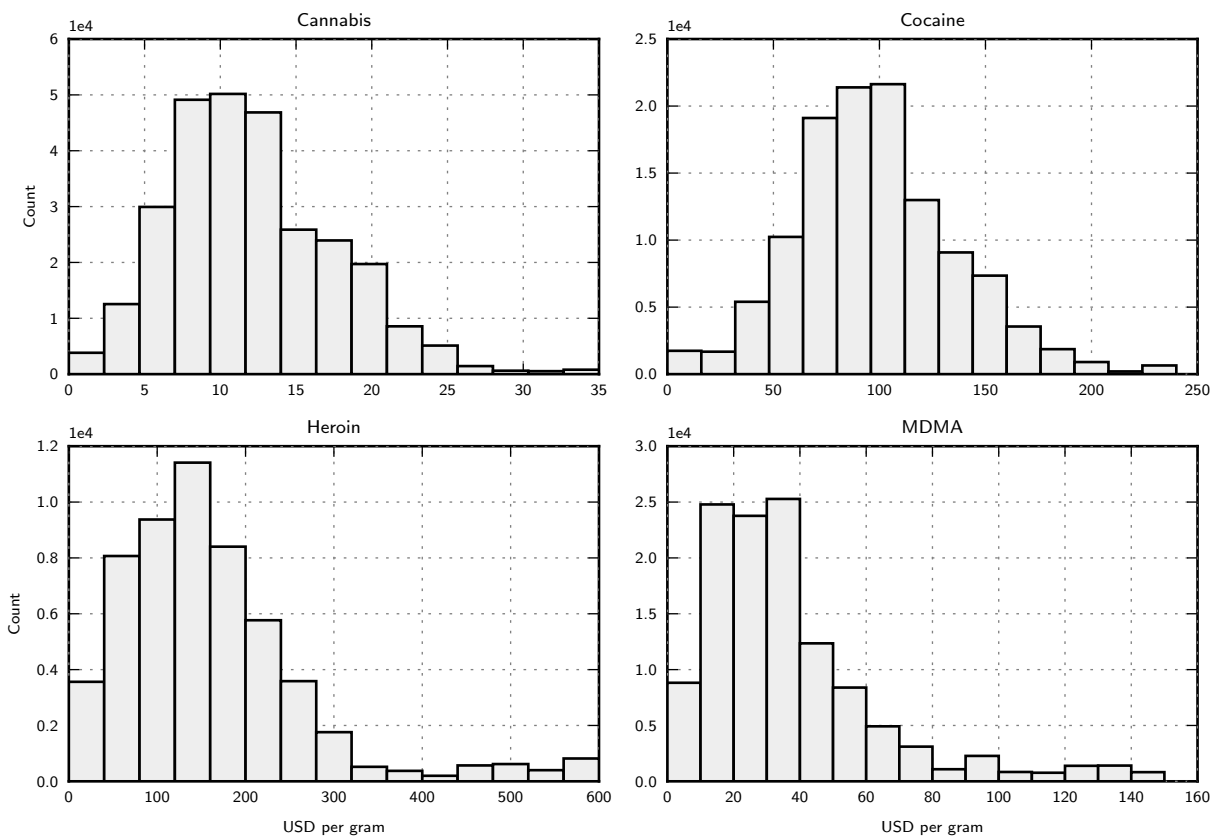}
  }
  \caption{Distribution of the prices of illicit drugs in the Agora market.\label{fig:price}}
  \floatfoot{\textbf{Note:} The figures show the distribution of prices for \categories~over the entire lifespan of the marketplace. The unit of observation is a review, which we use as proxy for transactions. The y-axes are denoted in $10,000$. The x-axes are denoted in USD per gram, where we obtain the dollar amounts by converting bitcoin to U.S.\ dollars at the time of the transaction. The price distributions of \categories~are hump-shaped. The price distribution for Heroin has a long right tail.}
\end{figure}

Figure~\ref{fig:price} shows the derived price at the imputed transaction date, in dollars per gram, for each of the four product categories.
Since prices are denominated in bitcoins, the value of which fluctuated wildly for the duration of the sample, we use the market-reported exchange rate between bitcoins and US dollars to convert all prices to US dollars.
The price distributions or all product categories are humpshaped with a fair degree of dispersion.
Overall, the price distributions look reasonable giving credence to our classification algorithm to obtain products, quantities, and weights as well as the currency conversion.

\begin{figure}
  \centerline{
    \input{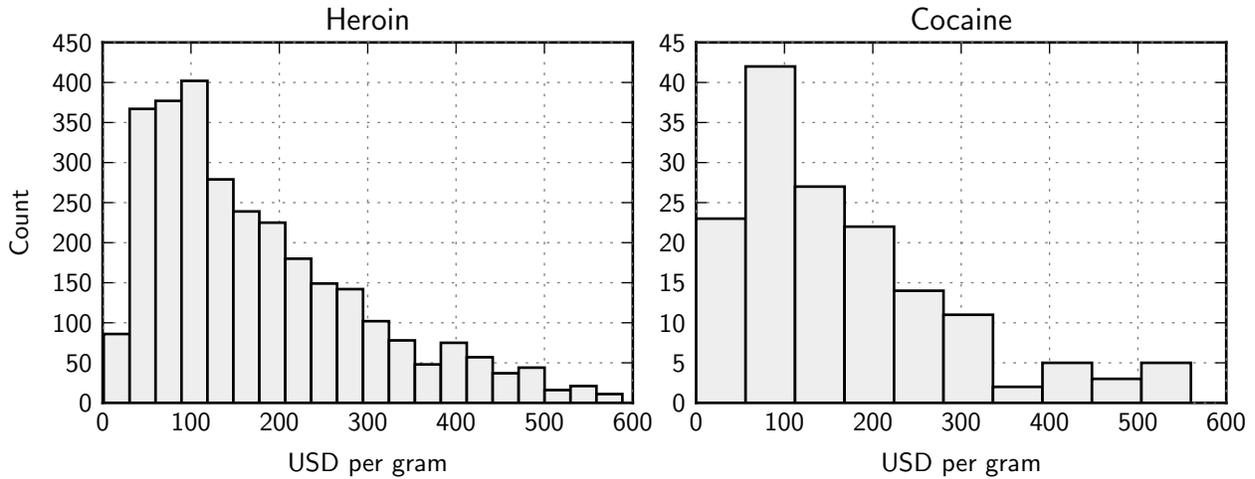}
  }
  \caption{Distribution of the price of heroin and cocaine in the \textsc{stride} dataset\label{fig:pricestride}}
  \floatfoot{\textbf{Note:} The figures show the distribution of the price of heroin and cocaine in the \textsc{stride} dataset. The \textsc{stride} dataset is a dataset with information on prices of drugs that were seized between 2007 and 2013. We obtained the full. The distribution of prices in the \textsc{stride} dataset displays a spread and mean consistent with the prices we find online.}
\end{figure}

As a point of reference, Figure~\ref{fig:pricestride} shows the distribution of prices for Heroin and Cocaine in the \textsc{stride} dataset \citep{Stride}.
The \textsc{stride} dataset is provided by the United States Drug Enforcement Administration with information on prices of drugs that were seized between 2007 and 2013.
We obtained the full \textsc{stride} dataset through a Freedom of Information Act Request.
The price distributions of Heroin and Cocaine---the other product categories, Cannabis and MDMA, are not part of the \textsc{stride} dataset---look similar to the price distributions that we report from the online marketplace.
In particular, they are similarly disperse.
On average, prices in the online marketplace are higher.
We can only speculate why prices in the online marketplace are higher.
It is perceivable that vendors in online markets charge a convenience premium.
Also, the \textsc{stride} data is a selected sample, since it only contains data from seizures, not from transactions.


\begin{figure}[h!]
  \centerline{
    \input{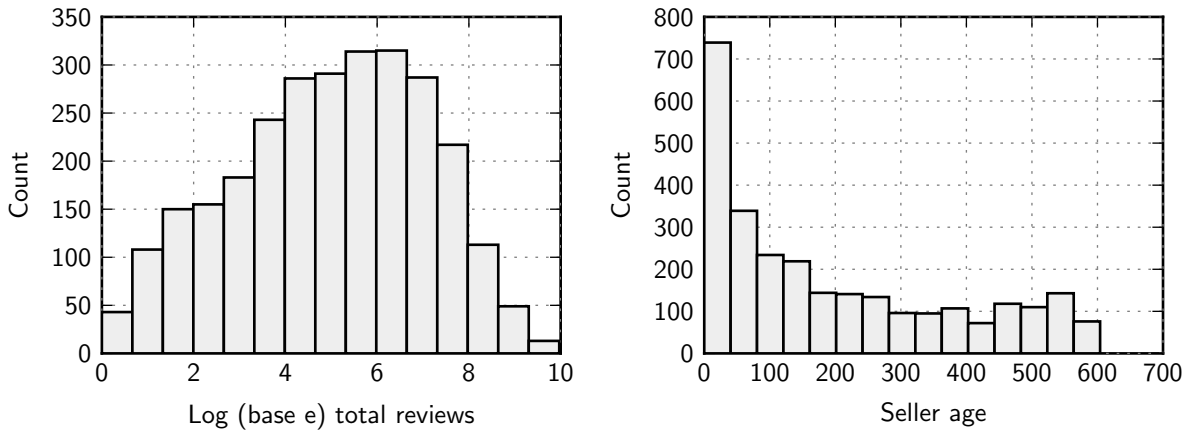}
  }
  \caption{Distribution of seller size and age.\label{fig:salesvolume}}
  \floatfoot{\textbf{Note:} The figure on the left shows the distribution of the log of total sales per seller (proxied by the number of reviews left), where we again restrict our analysis to products in one of the four main categories (\categories). The figure on the right shows the distribution of the final age (in days) of vendors in the sample for vendors who have sold products in one of the four main categories that we focus on.}
\end{figure}

Vendors differ in experience and success in the market. See Figure~\ref{fig:salesvolume} for distributions of sales volume and final age of vendors by the end of sample.
Many vendors entered, made a few sales, and then left the market.
Other vendors stayed for the full duration of the market of approximately 600 days.

\subsection{Stylized Facts\label{sec:stylizedfacts}}
In this section, we establish the following stylized facts from the data about the relationship between a seller's price, rating, and total reviews left.
\begin{enumerate}
  \item There is a positive, significant, but small relationship between the price a seller charges, and a seller's rating.
  Moving from the 25th percentile to the 75th percentile in ratings is associated with a 0.05\% increase in price.
  \item The more reviews left for a seller, the greater the relationship between a seller's price and his rating.
  For the largest sellers in the sample, moving from the 25th percentile to the 75th percentile in ratings is associated with a 19.7\% increase in price.
  \item There is a positive relationship between number of reviews left for a seller and the seller's price.
  Sellers just entering the market charge, on average, 10\% less than mature sellers with more than 5,000 reviews.
  \item The more reviews a seller receives, the smaller the impact of an additional review on his price.
\end{enumerate}

In addition, we look at patterns of exit as a function of rating and total reviews left.
\begin{enumerate}[resume]
  \item A seller's lifespan can be broken into three stages, an entry stage, an exit stage, and a mature stage.
  \item The lower a seller's rating, the more likely he is to exit.
\end{enumerate}

\subsubsection*{Price, rating, and total reviews}

We begin by analyzing the relationship between prices, ratings, and total number of reviews.
We replicate the work of prior literature that studies legal markets and look for reduced form evidence of the impact of the rating on prices.
We estimate the model
\begin{equation}
  \log(\text{Price}_{it}) = \beta \times \text{Rating}_{it} + \gamma \times \text{Other product characteristics}_{it} + \eta_i + \varepsilon_{it},
  \label{eq:simpleregression}
\end{equation}
where $i$ refers to a particular seller, and $t$ to a particular transaction.
The coefficient $\beta$ captures the effect of rating on price.
$\gamma$ is the coefficient vector associated with additional controls, which include factors such as dummies for the locations vendors ship from and to, the type of product on sale, and age of the vendor and time trends.
$\eta_i$ are vendor fixed effects and $\varepsilon_{it}$ is a disturbance term.

\begin{table}
  \centerline{\small
  \input{figures/regression.tex}
  }
  \caption{Regressions of log price residuals on rating, quantity, and sales.\label{tab:regression}}
  \floatfoot{\textbf{Note:} The table shows the coefficient estimates from regressions of the log price on rating, quantity, and the number of sales made.
  The data consist of vendors who sell cannabis, but similar results hold for the other categories of products.
  The unit of observation is a review left, our proxy for the number of transactions.
  Column~(1) reports the estimates of coefficients in equation~\eqref{eq:simpleregression} for the entire sample.
  Column~(2) estimates the coefficient on ratings separately for large (number of transactions $>$50th percentile) and small sellers.
  Columns~(3)~and~(4) repeat the exercise, but with seller fixed effects.}
\end{table}

The first column in Table~\ref{tab:regression} displays the estimates of the coefficients $\beta$ and $\gamma$.
The data consist of vendors who sell cannabis, but similar results hold for the other categories of products.
Consistent with previous literature, a positive, statistically significant, but small effect is found of the rating on the price.
We do not interpret the coefficient estimates in Table~\ref{tab:regression} as causal.
Rather we interpret them as conditional means.
To interpret the point estimate of $0.10$, a seller at the 25th percentile in ratings (around 4.90/5) charges an average 1\% lower price than a seller in the 75th percentile in ratings (or 5/5).

The small size of the relationship between rating and price is puzzling, because a higher rating, all else equal, should be a signal of a more valuable product to consumers.
One resolution to the puzzle is unobserved seller heterogeneity (see e.g.~\cite{Resnik02}).
However, we see all relevant seller characteristics.
Since sellers are anonymous in our data, their seller page contains all information available to buyers.
This is contrast to legal marketplaces, where additional information about sellers might be available elsewhere.\footnote{\cite{Newberry16} investigate this possible explanation for Alibaba's Tmall, China's largest business-to-consumer marketplace.
They find that the impact of ratings decreases with the offline presence of a seller.}

We favor a different explanation for the small size of the relationship between rating and price.
Intuitively, the rating of a seller is only informative about the true quality of the seller, once the seller has accumulated a large number of reviews.
A seller with a high rating and only a small number of reviews may have just got lucky.
A seller with a high rating and a large number of reviews must in fact be delivering a high quality product.

To test this, the second column in Table~\ref{tab:regression} reports the estimate of $\beta$ when we allow the coefficient on ratings to depend on whether a seller falls in the top or bottom 50\% of total sales made.
For sellers in the top 50\% of sales made, the relationship between price and rating strengthens to $1.80$.
To interpret this, a seller moving from the 25th to the 75th percentile in ratings charges an average 19.2\% more.
This fact suggests that the impact of a bad rating on the price increases as more reviews are left.

\begin{figure}
  \centerline{
    \input{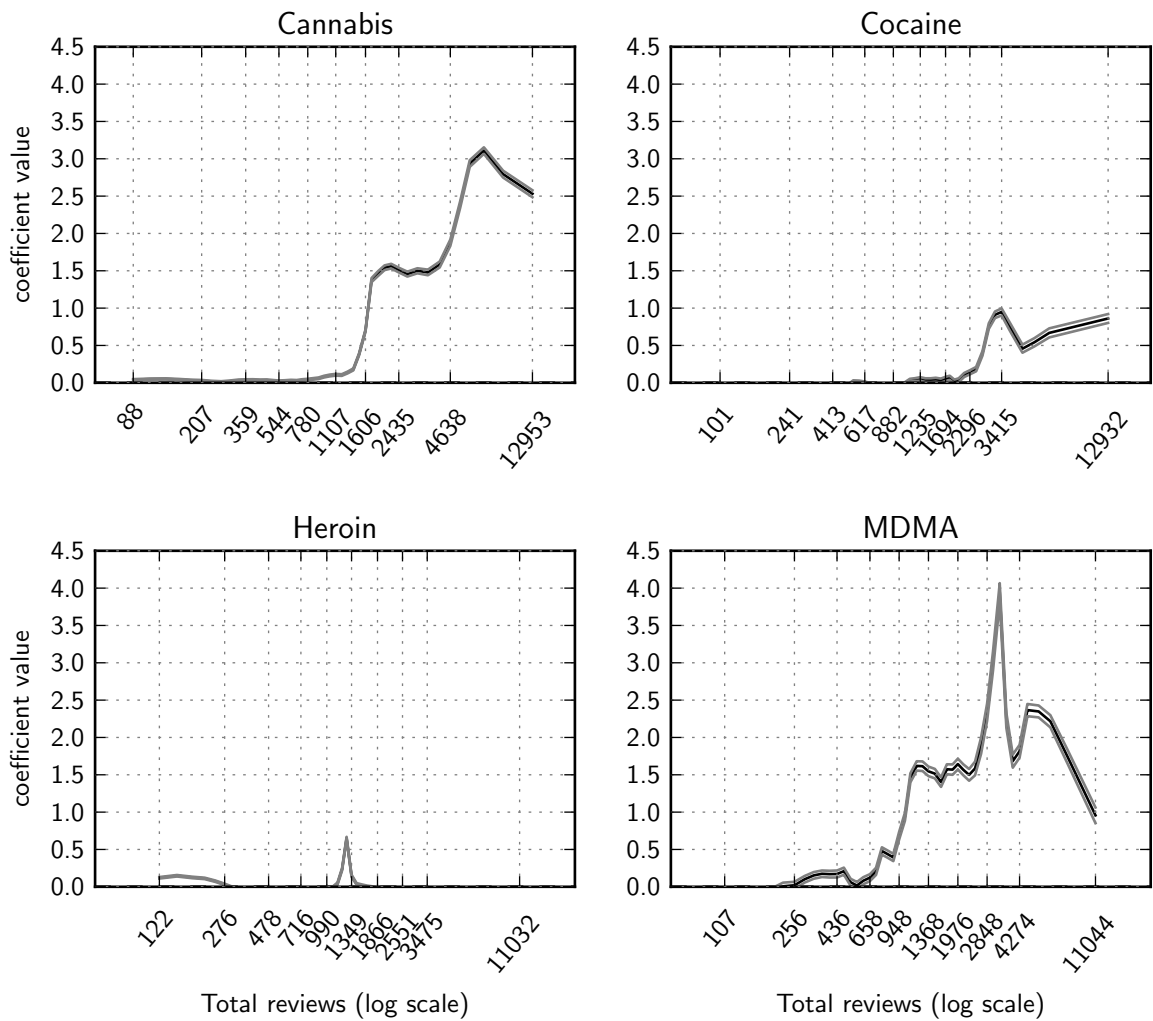}
  }
  \caption{Effect of rating on log price broken down by number of reviews.\label{fig:pricevsrating}}
  \floatfoot{\textbf{Note:} We show the regression coefficient $\beta$ (and their 95\% confidence intervals in gray) from the linear regression model shown in equation~\eqref{eq:simpleregression} broken down by different quantiles of total reviews left.
  Total reviews are denoted with a log scale.
  The effect of rating on price is zero for sellers with few sales.
  The effect is large and positive for sellers with many sales.
  The product category `Heroine' is an outlier.}
\end{figure}

We further highlight this point in Figure~\ref{fig:pricevsrating}.
We report the coefficient estimate of ratings on price, $\beta$, broken down by different quantiles of total reviews left.
The effect of rating on price is zero when only a small number of reviews have been left, but becomes much larger as the number of reviews increases.
For example, after 10,000 reviews, moving from the 25th to the 75th percentile in ratings in the market for cannabis is associated with a 35\% higher price, a large and statistically significant number.

The above supports the hypothesis that buyers treat reviews of sellers as informative signals of the quality of the seller, and draw inferences about seller quality accordingly.
When a seller does not have many reviews, the rating is uninformative about his quality and has no effect on price.
When a seller has many reviews, the rating is very informative and has a large effect on price.
Not only does the effect of rating on price increase as sellers receive more reviews, but a sellers with more reviews also charge higher prices.
For example, a seller with 10,000 total reviews charges on average 35\% more than a seller with only one review.

The results of this section suggest that the absence of a strong effect of rating on price is due to a composition effect.
For young sellers with few reviews, the rating has little bearing on the price.
For older sellers with many reviews, the rating has a substantial effect on prices.

\subsubsection*{Exit, rating dispersion, and total reviews}

 The composition effect is amplified by the fact that poorly rated sellers are more likely to leave the market.
 We investigate the co-evolution of ratings dispersion, ratings, total reviews, and exit.

\begin{figure}
  \centerline{
    \input{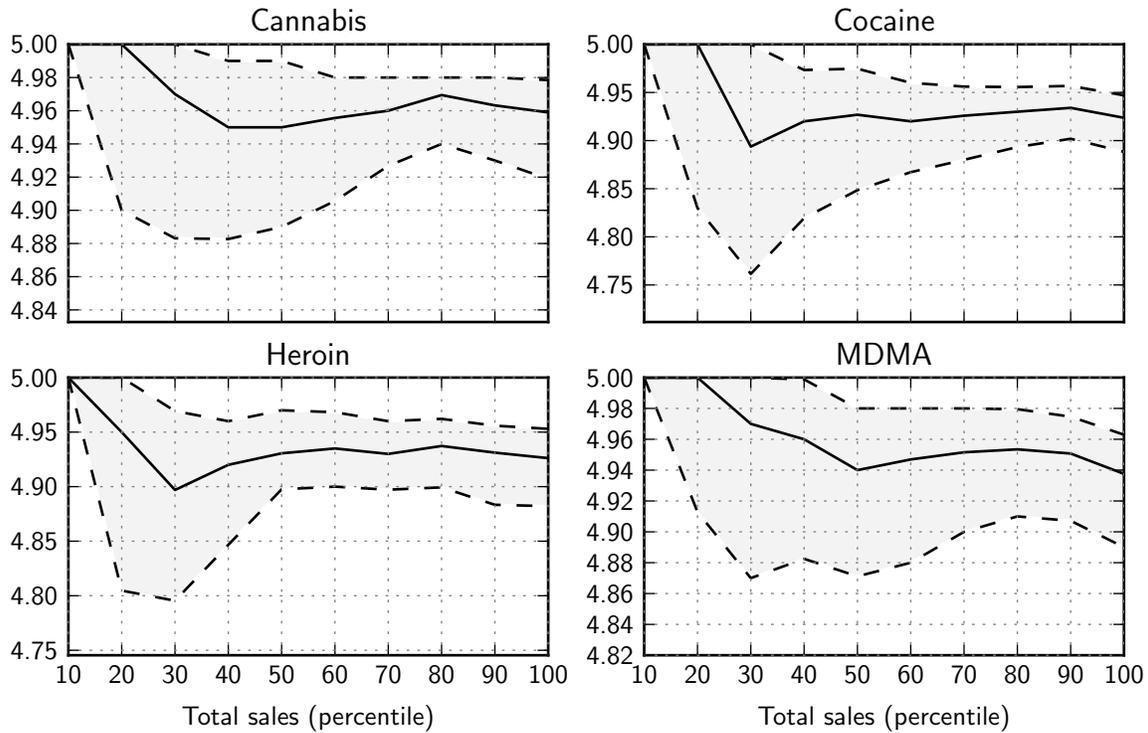}
  }
  \caption{Dispersion of ratings, as a function of sales made.\label{fig:ratingdensity}}
  \floatfoot{\textbf{Note:} The figures show a seller's rating as a function of the total number of sales made.
  Solid line represents the median rating, dashed lines indicate the 30th and 70th percentile of ratings and give an indication of the variance of ratings.
  In the 0th--10th percentile for sales made (the entry stage), most sellers have the top rating of 5.
  In the 10th-30th percentile for sales made, sellers ratings spread apart as more reviews arrive.
  In the 30th-50th percentile (the exit stage), low-rated sellers exit, rather than sell at lower prices, resulting in a narrowing of the rating spread.
  Finally, in the 50th-100th percentile (the mature stage), the spread and median rating stabilize.}
\end{figure}

 In Figure~\ref{fig:ratingdensity}, we plot the distribution of seller's average weekly rating.
 Due to granularity in the ratings left (most sellers are in the 4.90--5.00 range, meaning a total of 10 possible ratings, because average ratings are reported with two decimal points), we smooth ratings by using weekly averages.
 The average rating of a seller tends to follow a `U'-shape, initially dipping before leveling up again.
 As it dips, the variance of ratings in the population increases, and as it rises back up, the variance decreases.
 What causes these U-shaped dynamics? We argue that the answer is two-fold: The design of the rating system itself accounts for the initial dip, while survivorship bias accounts for the subsequent rise.

To see why the design of the rating system causes the initial dip, note that most reviews result in the maximum rating possible, 5/5, the next-most number of reviews result in the worst rating possible, 0/5.
As a result, only a small fraction of sellers observed to be entering the market have anything less than the maximum possible rating of 5.
Therefore, the initial drop in average rating, and corresponding rise in rating dispersion, as seen in Figure~\ref{fig:ratingdensity} in percentiles 10--30, is a consequence of the fact that the vast majority of sellers enter with a perfect rating of 5/5, and so all sellers see their rating drop as they begin to make more sales.


After the rating has fallen and the rating dispersion increases, the average rating rises and the rating dispersion falls.
We argue this is a consequence of survivorship bias.
The sellers most likely to continue selling in the market are the high-rated sellers.
Hence, past some point, the average rating is increasing and the rating dispersion decreasing, as low-rated sellers exit.

\begin{figure}
  \centerline{
    \input{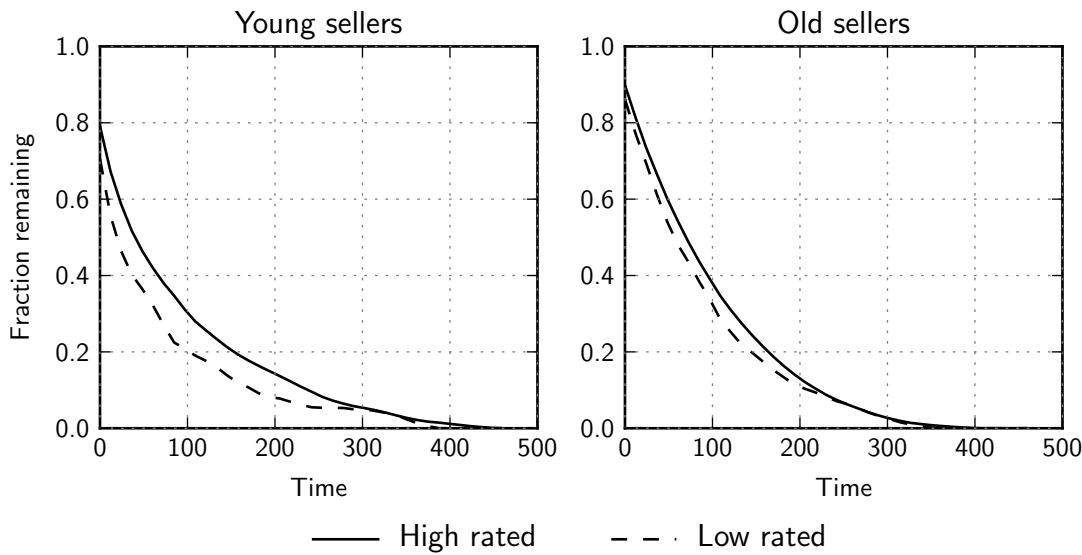}
  }
  \caption{Survival probabilities by age and rating.\label{fig:transitions}}
  \floatfoot{\textbf{Note:} Left: The fraction of young sellers with a high rating (solid) versus sellers with a low rating (dashed) remaining in the market after some time.
  Right: The fraction of old sellers with a high rating remaining versus sellers with a low rating remaining.
  High rating is defined to be higher than 4.95, low rating is defined to be less than 4.95, old is defined as greater than 50 days in the market, young is defined as less than 50 days in the market.}
 \end{figure}

\subsubsection*{Interpretation}

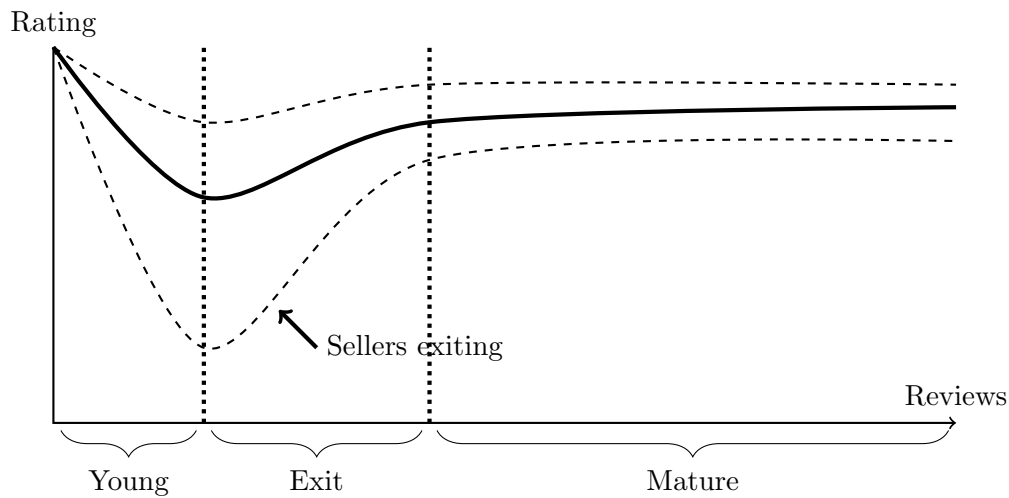
\begin{figure}
  \centerline{
  \begin{tikzpicture}
    \draw[thick, ->] (0, 5) -- (0, 0) -- (12, 0);
    \node[above] at (0, 5) {Rating};
    \node[above] at (12, 0.1) {Reviews};
    \draw[ultra thick] plot[smooth] coordinates {(0, 5) (2, 3) (5, 4) (12, 4.2)};
    \draw[thick, dashed] plot[smooth] coordinates {(0, 5) (2, 1) (5, 3.5) (12, 3.75)};
    \draw[thick, dashed] plot[smooth] coordinates {(0, 5) (2, 4) (5, 4.5) (12, 4.5)};
    \draw[ultra thick, dotted] (2, 0) -- (2, 5);
    \draw[ultra thick, dotted] (5, 0) -- (5, 5);
    \draw[decorate, decoration={brace,amplitude=10pt,mirror}] (0.1, -0.1) -- (1.9, -0.1);
    \draw[decorate, decoration={brace,amplitude=10pt,mirror}] (2.1, -0.1) -- (4.9, -0.1);
    \draw[decorate, decoration={brace,amplitude=10pt,mirror}] (5.1, -0.1) -- (11.9, -0.1);
    \node[below] at (1, -0.5) {Young};
    \node[below] at (3.5, -0.5) {Exit};
    \node[below] at (8.5, -0.5) {Mature};
    \draw[ultra thick, <-] (3, 1.5) -- (3.5, 1);
    \node[right] at (3.5, 1) {Sellers exiting};
  \end{tikzpicture}
  }
  \caption{Graphical illustration of the life-cycle of sellers in the data.\label{fig:lifecycle}}
  \floatfoot{\textbf{Note:} The Figure illustrates how the life-cycle of sellers interacts with ratings and the dispersion thereof.
  We argue that the evolution of ratings and their dispersion is mostly driven by a composition effect.
  Initially, all sellers (good and bad ones alike) start out at the best possible rating.
  As time goes by and sales are made, good sellers receive good reviews and bad sellers receive bad reviews.
  As bad sellers accumulate bad reviews, their rating deteriorates and they exit the market.
  Therefore, as sellers mature, the average rating increases and the dispersion of ratings decreases.}
\end{figure}

In the remainder of the paper, we argue that these stylized facts are consistent with a dynamic model of ratings, exit, and adverse selection.
In the model, when sellers enter the market, they draw a permanent quality type.
Based on this type, they then decide in subsequent periods whether or not to exit the market.
We show how in some equilibria, the seller's life-cycle may be broken into four stages.
See Figure~\ref{fig:lifecycle} for a graphical illustration, and compare to Figure~\ref{fig:ratingdensity} for the corresponding patterns in the data.

Initially, entering sellers receive the highest possible rating for most reviews.
In the entry stage, their rating is an almost entirely uninformative signal of quality.
As more reviews accumulate, buyers move into the young stage.
Here, the dispersion of ratings increases rapidly, and the average rating falls for all sellers, resulting from the fact that all sellers began at the highest possible rating.
Faced with the prospect of lower prices at low ratings, sellers move into the exit stage, in which low-rated sellers exit the market rather than lower their price.
As low quality sellers exit, the rating dispersion narrows, and the average rating increases.
In the mature stage, the mature sellers continue to operate, almost all of them relatively highly rated.
Sellers whose rating falls exit the market rather than face the prospect of lower prices.
Thus, there is a relatively narrow spread in the rating among mature sellers.

The lack of a relationship between rating and price can be explained by a combination of two factors: First, low-rated sellers exit, leading to an omitted variable bias.
If they were to stay in the market and continue to sell a product, they would sell it at a much lower price, but they exit instead.
This channel is consistent with prior literature on reputation which finds strong effects of negative reviews on exit from the markets~\citep{Cabral10}.
Second, there is a composition bias introduced by failing to control for the number of reviews left for a vendor, and the joint relationship between reviews and ratings is non-linear.

%% file: figures/entryexit.pgf
\begingroup%
\makeatletter%
\begin{pgfpicture}%
\pgfpathrectangle{\pgfpointorigin}{\pgfqpoint{6.087009in}{2.551835in}}%
\pgfusepath{use as bounding box, clip}%
\begin{pgfscope}%
\pgfsetbuttcap%
\pgfsetmiterjoin%
\definecolor{currentfill}{rgb}{1.000000,1.000000,1.000000}%
\pgfsetfillcolor{currentfill}%
\pgfsetlinewidth{0.000000pt}%
\definecolor{currentstroke}{rgb}{1.000000,1.000000,1.000000}%
\pgfsetstrokecolor{currentstroke}%
\pgfsetdash{}{0pt}%
\pgfpathmoveto{\pgfqpoint{0.000000in}{0.000000in}}%
\pgfpathlineto{\pgfqpoint{6.087009in}{0.000000in}}%
\pgfpathlineto{\pgfqpoint{6.087009in}{2.551835in}}%
\pgfpathlineto{\pgfqpoint{0.000000in}{2.551835in}}%
\pgfpathclose%
\pgfusepath{fill}%
\end{pgfscope}%
\begin{pgfscope}%
\pgfsetbuttcap%
\pgfsetmiterjoin%
\definecolor{currentfill}{rgb}{1.000000,1.000000,1.000000}%
\pgfsetfillcolor{currentfill}%
\pgfsetlinewidth{0.000000pt}%
\definecolor{currentstroke}{rgb}{0.000000,0.000000,0.000000}%
\pgfsetstrokecolor{currentstroke}%
\pgfsetstrokeopacity{0.000000}%
\pgfsetdash{}{0pt}%
\pgfpathmoveto{\pgfqpoint{0.624509in}{0.759629in}}%
\pgfpathlineto{\pgfqpoint{5.987009in}{0.759629in}}%
\pgfpathlineto{\pgfqpoint{5.987009in}{2.384629in}}%
\pgfpathlineto{\pgfqpoint{0.624509in}{2.384629in}}%
\pgfpathclose%
\pgfusepath{fill}%
\end{pgfscope}%
\begin{pgfscope}%
\pgfpathrectangle{\pgfqpoint{0.624509in}{0.759629in}}{\pgfqpoint{5.362500in}{1.625000in}} %
\pgfusepath{clip}%
\pgfsetbuttcap%
\pgfsetroundjoin%
\pgfsetlinewidth{1.003750pt}%
\definecolor{currentstroke}{rgb}{0.000000,0.000000,0.000000}%
\pgfsetstrokecolor{currentstroke}%
\pgfsetdash{{6.000000pt}{6.000000pt}}{0.000000pt}%
\pgfpathmoveto{\pgfqpoint{0.624509in}{0.786712in}}%
\pgfpathlineto{\pgfqpoint{0.732471in}{0.759629in}}%
\pgfpathlineto{\pgfqpoint{0.840433in}{0.824629in}}%
\pgfpathlineto{\pgfqpoint{0.948396in}{0.943795in}}%
\pgfpathlineto{\pgfqpoint{1.065355in}{0.775879in}}%
\pgfpathlineto{\pgfqpoint{1.173317in}{0.770462in}}%
\pgfpathlineto{\pgfqpoint{1.281654in}{0.835462in}}%
\pgfpathlineto{\pgfqpoint{1.389616in}{0.830045in}}%
\pgfpathlineto{\pgfqpoint{1.497578in}{0.813795in}}%
\pgfpathlineto{\pgfqpoint{1.605541in}{0.997962in}}%
\pgfpathlineto{\pgfqpoint{1.722500in}{1.257962in}}%
\pgfpathlineto{\pgfqpoint{1.830462in}{0.867962in}}%
\pgfpathlineto{\pgfqpoint{1.938424in}{0.900462in}}%
\pgfpathlineto{\pgfqpoint{2.046387in}{0.808379in}}%
\pgfpathlineto{\pgfqpoint{2.154349in}{0.867962in}}%
\pgfpathlineto{\pgfqpoint{2.262311in}{0.873379in}}%
\pgfpathlineto{\pgfqpoint{2.379270in}{0.851712in}}%
\pgfpathlineto{\pgfqpoint{2.487232in}{0.878795in}}%
\pgfpathlineto{\pgfqpoint{2.595195in}{0.830045in}}%
\pgfpathlineto{\pgfqpoint{2.703157in}{0.835462in}}%
\pgfpathlineto{\pgfqpoint{2.811119in}{0.932962in}}%
\pgfpathlineto{\pgfqpoint{2.919081in}{0.932962in}}%
\pgfpathlineto{\pgfqpoint{3.036041in}{0.911295in}}%
\pgfpathlineto{\pgfqpoint{3.144003in}{0.911295in}}%
\pgfpathlineto{\pgfqpoint{3.251965in}{0.927545in}}%
\pgfpathlineto{\pgfqpoint{3.359552in}{1.008795in}}%
\pgfpathlineto{\pgfqpoint{3.467515in}{0.922129in}}%
\pgfpathlineto{\pgfqpoint{3.575477in}{0.911295in}}%
\pgfpathlineto{\pgfqpoint{3.683439in}{0.927545in}}%
\pgfpathlineto{\pgfqpoint{3.800398in}{0.916712in}}%
\pgfpathlineto{\pgfqpoint{3.908361in}{0.938379in}}%
\pgfpathlineto{\pgfqpoint{4.016323in}{0.922129in}}%
\pgfpathlineto{\pgfqpoint{4.124285in}{0.932962in}}%
\pgfpathlineto{\pgfqpoint{4.232247in}{1.003379in}}%
\pgfpathlineto{\pgfqpoint{4.340210in}{0.960045in}}%
\pgfpathlineto{\pgfqpoint{4.457169in}{1.041295in}}%
\pgfpathlineto{\pgfqpoint{4.565506in}{0.895045in}}%
\pgfpathlineto{\pgfqpoint{4.673468in}{1.155045in}}%
\pgfpathlineto{\pgfqpoint{4.781430in}{0.916712in}}%
\pgfpathlineto{\pgfqpoint{4.889393in}{0.965462in}}%
\pgfpathlineto{\pgfqpoint{4.997355in}{0.916712in}}%
\pgfpathlineto{\pgfqpoint{5.114314in}{0.954629in}}%
\pgfpathlineto{\pgfqpoint{5.222276in}{1.041295in}}%
\pgfpathlineto{\pgfqpoint{5.330238in}{1.100879in}}%
\pgfpathlineto{\pgfqpoint{5.438201in}{1.095462in}}%
\pgfpathlineto{\pgfqpoint{5.546163in}{1.057545in}}%
\pgfpathlineto{\pgfqpoint{5.654125in}{1.274212in}}%
\pgfpathlineto{\pgfqpoint{5.771084in}{1.761712in}}%
\pgfpathlineto{\pgfqpoint{5.810630in}{2.394629in}}%
\pgfusepath{stroke}%
\end{pgfscope}%
\begin{pgfscope}%
\pgfpathrectangle{\pgfqpoint{0.624509in}{0.759629in}}{\pgfqpoint{5.362500in}{1.625000in}} %
\pgfusepath{clip}%
\pgfsetrectcap%
\pgfsetroundjoin%
\pgfsetlinewidth{2.007500pt}%
\definecolor{currentstroke}{rgb}{0.000000,0.000000,0.000000}%
\pgfsetstrokecolor{currentstroke}%
\pgfsetdash{}{0pt}%
\pgfpathmoveto{\pgfqpoint{0.624509in}{1.138795in}}%
\pgfpathlineto{\pgfqpoint{0.732471in}{1.117129in}}%
\pgfpathlineto{\pgfqpoint{0.840433in}{1.014212in}}%
\pgfpathlineto{\pgfqpoint{0.948396in}{1.025045in}}%
\pgfpathlineto{\pgfqpoint{1.065355in}{2.352129in}}%
\pgfpathlineto{\pgfqpoint{1.173317in}{0.932962in}}%
\pgfpathlineto{\pgfqpoint{1.281654in}{0.884212in}}%
\pgfpathlineto{\pgfqpoint{1.389616in}{1.312129in}}%
\pgfpathlineto{\pgfqpoint{1.497578in}{1.198379in}}%
\pgfpathlineto{\pgfqpoint{1.605541in}{1.052129in}}%
\pgfpathlineto{\pgfqpoint{1.722500in}{1.090045in}}%
\pgfpathlineto{\pgfqpoint{1.830462in}{1.517962in}}%
\pgfpathlineto{\pgfqpoint{1.938424in}{0.981712in}}%
\pgfpathlineto{\pgfqpoint{2.046387in}{0.873379in}}%
\pgfpathlineto{\pgfqpoint{2.154349in}{0.889629in}}%
\pgfpathlineto{\pgfqpoint{2.262311in}{0.808379in}}%
\pgfpathlineto{\pgfqpoint{2.379270in}{1.220045in}}%
\pgfpathlineto{\pgfqpoint{2.487232in}{1.025045in}}%
\pgfpathlineto{\pgfqpoint{2.595195in}{1.030462in}}%
\pgfpathlineto{\pgfqpoint{2.703157in}{0.927545in}}%
\pgfpathlineto{\pgfqpoint{2.811119in}{0.840879in}}%
\pgfpathlineto{\pgfqpoint{2.919081in}{1.084629in}}%
\pgfpathlineto{\pgfqpoint{3.036041in}{1.062962in}}%
\pgfpathlineto{\pgfqpoint{3.144003in}{0.949212in}}%
\pgfpathlineto{\pgfqpoint{3.251965in}{1.046712in}}%
\pgfpathlineto{\pgfqpoint{3.359552in}{1.046712in}}%
\pgfpathlineto{\pgfqpoint{3.467515in}{1.209212in}}%
\pgfpathlineto{\pgfqpoint{3.575477in}{0.965462in}}%
\pgfpathlineto{\pgfqpoint{3.683439in}{0.987129in}}%
\pgfpathlineto{\pgfqpoint{3.800398in}{0.900462in}}%
\pgfpathlineto{\pgfqpoint{3.908361in}{0.992545in}}%
\pgfpathlineto{\pgfqpoint{4.016323in}{0.987129in}}%
\pgfpathlineto{\pgfqpoint{4.124285in}{1.041295in}}%
\pgfpathlineto{\pgfqpoint{4.232247in}{0.954629in}}%
\pgfpathlineto{\pgfqpoint{4.340210in}{0.970879in}}%
\pgfpathlineto{\pgfqpoint{4.457169in}{0.938379in}}%
\pgfpathlineto{\pgfqpoint{4.565506in}{1.084629in}}%
\pgfpathlineto{\pgfqpoint{4.673468in}{1.095462in}}%
\pgfpathlineto{\pgfqpoint{4.781430in}{1.025045in}}%
\pgfpathlineto{\pgfqpoint{4.889393in}{0.905879in}}%
\pgfpathlineto{\pgfqpoint{4.997355in}{1.025045in}}%
\pgfpathlineto{\pgfqpoint{5.114314in}{0.867962in}}%
\pgfpathlineto{\pgfqpoint{5.222276in}{0.997962in}}%
\pgfpathlineto{\pgfqpoint{5.330238in}{1.203795in}}%
\pgfpathlineto{\pgfqpoint{5.438201in}{1.052129in}}%
\pgfpathlineto{\pgfqpoint{5.546163in}{1.073795in}}%
\pgfpathlineto{\pgfqpoint{5.654125in}{1.084629in}}%
\pgfpathlineto{\pgfqpoint{5.771084in}{1.198379in}}%
\pgfpathlineto{\pgfqpoint{5.879046in}{1.111712in}}%
\pgfpathlineto{\pgfqpoint{5.987009in}{0.873379in}}%
\pgfusepath{stroke}%
\end{pgfscope}%
\begin{pgfscope}%
\pgfsetrectcap%
\pgfsetmiterjoin%
\pgfsetlinewidth{1.003750pt}%
\definecolor{currentstroke}{rgb}{0.000000,0.000000,0.000000}%
\pgfsetstrokecolor{currentstroke}%
\pgfsetdash{}{0pt}%
\pgfpathmoveto{\pgfqpoint{5.987009in}{0.759629in}}%
\pgfpathlineto{\pgfqpoint{5.987009in}{2.384629in}}%
\pgfusepath{stroke}%
\end{pgfscope}%
\begin{pgfscope}%
\pgfsetrectcap%
\pgfsetmiterjoin%
\pgfsetlinewidth{1.003750pt}%
\definecolor{currentstroke}{rgb}{0.000000,0.000000,0.000000}%
\pgfsetstrokecolor{currentstroke}%
\pgfsetdash{}{0pt}%
\pgfpathmoveto{\pgfqpoint{0.624509in}{0.759629in}}%
\pgfpathlineto{\pgfqpoint{5.987009in}{0.759629in}}%
\pgfusepath{stroke}%
\end{pgfscope}%
\begin{pgfscope}%
\pgfsetrectcap%
\pgfsetmiterjoin%
\pgfsetlinewidth{1.003750pt}%
\definecolor{currentstroke}{rgb}{0.000000,0.000000,0.000000}%
\pgfsetstrokecolor{currentstroke}%
\pgfsetdash{}{0pt}%
\pgfpathmoveto{\pgfqpoint{0.624509in}{2.384629in}}%
\pgfpathlineto{\pgfqpoint{5.987009in}{2.384629in}}%
\pgfusepath{stroke}%
\end{pgfscope}%
\begin{pgfscope}%
\pgfsetrectcap%
\pgfsetmiterjoin%
\pgfsetlinewidth{1.003750pt}%
\definecolor{currentstroke}{rgb}{0.000000,0.000000,0.000000}%
\pgfsetstrokecolor{currentstroke}%
\pgfsetdash{}{0pt}%
\pgfpathmoveto{\pgfqpoint{0.624509in}{0.759629in}}%
\pgfpathlineto{\pgfqpoint{0.624509in}{2.384629in}}%
\pgfusepath{stroke}%
\end{pgfscope}%
\begin{pgfscope}%
\pgfpathrectangle{\pgfqpoint{0.624509in}{0.759629in}}{\pgfqpoint{5.362500in}{1.625000in}} %
\pgfusepath{clip}%
\pgfsetbuttcap%
\pgfsetroundjoin%
\pgfsetlinewidth{0.501875pt}%
\definecolor{currentstroke}{rgb}{0.501961,0.501961,0.501961}%
\pgfsetstrokecolor{currentstroke}%
\pgfsetdash{{1.000000pt}{3.000000pt}}{0.000000pt}%
\pgfpathmoveto{\pgfqpoint{1.112213in}{0.759629in}}%
\pgfpathlineto{\pgfqpoint{1.112213in}{2.384629in}}%
\pgfusepath{stroke}%
\end{pgfscope}%
\begin{pgfscope}%
\pgfsetbuttcap%
\pgfsetroundjoin%
\definecolor{currentfill}{rgb}{0.000000,0.000000,0.000000}%
\pgfsetfillcolor{currentfill}%
\pgfsetlinewidth{0.501875pt}%
\definecolor{currentstroke}{rgb}{0.000000,0.000000,0.000000}%
\pgfsetstrokecolor{currentstroke}%
\pgfsetdash{}{0pt}%
\pgfsys@defobject{currentmarker}{\pgfqpoint{0.000000in}{0.000000in}}{\pgfqpoint{0.000000in}{0.055556in}}{%
\pgfpathmoveto{\pgfqpoint{0.000000in}{0.000000in}}%
\pgfpathlineto{\pgfqpoint{0.000000in}{0.055556in}}%
\pgfusepath{stroke,fill}%
}%
\begin{pgfscope}%
\pgfsys@transformshift{1.112213in}{0.759629in}%
\pgfsys@useobject{currentmarker}{}%
\end{pgfscope}%
\end{pgfscope}%
\begin{pgfscope}%
\pgfsetbuttcap%
\pgfsetroundjoin%
\definecolor{currentfill}{rgb}{0.000000,0.000000,0.000000}%
\pgfsetfillcolor{currentfill}%
\pgfsetlinewidth{0.501875pt}%
\definecolor{currentstroke}{rgb}{0.000000,0.000000,0.000000}%
\pgfsetstrokecolor{currentstroke}%
\pgfsetdash{}{0pt}%
\pgfsys@defobject{currentmarker}{\pgfqpoint{0.000000in}{-0.055556in}}{\pgfqpoint{0.000000in}{0.000000in}}{%
\pgfpathmoveto{\pgfqpoint{0.000000in}{0.000000in}}%
\pgfpathlineto{\pgfqpoint{0.000000in}{-0.055556in}}%
\pgfusepath{stroke,fill}%
}%
\begin{pgfscope}%
\pgfsys@transformshift{1.112213in}{2.384629in}%
\pgfsys@useobject{currentmarker}{}%
\end{pgfscope}%
\end{pgfscope}%
\begin{pgfscope}%
\pgftext[x=0.815382in,y=0.379299in,left,base,rotate=20.000000]{\sffamily\fontsize{10.000000}{12.000000}\selectfont Mar 2014}%
\end{pgfscope}%
\begin{pgfscope}%
\pgfpathrectangle{\pgfqpoint{0.624509in}{0.759629in}}{\pgfqpoint{5.362500in}{1.625000in}} %
\pgfusepath{clip}%
\pgfsetbuttcap%
\pgfsetroundjoin%
\pgfsetlinewidth{0.501875pt}%
\definecolor{currentstroke}{rgb}{0.501961,0.501961,0.501961}%
\pgfsetstrokecolor{currentstroke}%
\pgfsetdash{{1.000000pt}{3.000000pt}}{0.000000pt}%
\pgfpathmoveto{\pgfqpoint{1.661021in}{0.759629in}}%
\pgfpathlineto{\pgfqpoint{1.661021in}{2.384629in}}%
\pgfusepath{stroke}%
\end{pgfscope}%
\begin{pgfscope}%
\pgfsetbuttcap%
\pgfsetroundjoin%
\definecolor{currentfill}{rgb}{0.000000,0.000000,0.000000}%
\pgfsetfillcolor{currentfill}%
\pgfsetlinewidth{0.501875pt}%
\definecolor{currentstroke}{rgb}{0.000000,0.000000,0.000000}%
\pgfsetstrokecolor{currentstroke}%
\pgfsetdash{}{0pt}%
\pgfsys@defobject{currentmarker}{\pgfqpoint{0.000000in}{0.000000in}}{\pgfqpoint{0.000000in}{0.055556in}}{%
\pgfpathmoveto{\pgfqpoint{0.000000in}{0.000000in}}%
\pgfpathlineto{\pgfqpoint{0.000000in}{0.055556in}}%
\pgfusepath{stroke,fill}%
}%
\begin{pgfscope}%
\pgfsys@transformshift{1.661021in}{0.759629in}%
\pgfsys@useobject{currentmarker}{}%
\end{pgfscope}%
\end{pgfscope}%
\begin{pgfscope}%
\pgfsetbuttcap%
\pgfsetroundjoin%
\definecolor{currentfill}{rgb}{0.000000,0.000000,0.000000}%
\pgfsetfillcolor{currentfill}%
\pgfsetlinewidth{0.501875pt}%
\definecolor{currentstroke}{rgb}{0.000000,0.000000,0.000000}%
\pgfsetstrokecolor{currentstroke}%
\pgfsetdash{}{0pt}%
\pgfsys@defobject{currentmarker}{\pgfqpoint{0.000000in}{-0.055556in}}{\pgfqpoint{0.000000in}{0.000000in}}{%
\pgfpathmoveto{\pgfqpoint{0.000000in}{0.000000in}}%
\pgfpathlineto{\pgfqpoint{0.000000in}{-0.055556in}}%
\pgfusepath{stroke,fill}%
}%
\begin{pgfscope}%
\pgfsys@transformshift{1.661021in}{2.384629in}%
\pgfsys@useobject{currentmarker}{}%
\end{pgfscope}%
\end{pgfscope}%
\begin{pgfscope}%
\pgftext[x=1.352401in,y=0.370717in,left,base,rotate=20.000000]{\sffamily\fontsize{10.000000}{12.000000}\selectfont May 2014}%
\end{pgfscope}%
\begin{pgfscope}%
\pgfpathrectangle{\pgfqpoint{0.624509in}{0.759629in}}{\pgfqpoint{5.362500in}{1.625000in}} %
\pgfusepath{clip}%
\pgfsetbuttcap%
\pgfsetroundjoin%
\pgfsetlinewidth{0.501875pt}%
\definecolor{currentstroke}{rgb}{0.501961,0.501961,0.501961}%
\pgfsetstrokecolor{currentstroke}%
\pgfsetdash{{1.000000pt}{3.000000pt}}{0.000000pt}%
\pgfpathmoveto{\pgfqpoint{2.209829in}{0.759629in}}%
\pgfpathlineto{\pgfqpoint{2.209829in}{2.384629in}}%
\pgfusepath{stroke}%
\end{pgfscope}%
\begin{pgfscope}%
\pgfsetbuttcap%
\pgfsetroundjoin%
\definecolor{currentfill}{rgb}{0.000000,0.000000,0.000000}%
\pgfsetfillcolor{currentfill}%
\pgfsetlinewidth{0.501875pt}%
\definecolor{currentstroke}{rgb}{0.000000,0.000000,0.000000}%
\pgfsetstrokecolor{currentstroke}%
\pgfsetdash{}{0pt}%
\pgfsys@defobject{currentmarker}{\pgfqpoint{0.000000in}{0.000000in}}{\pgfqpoint{0.000000in}{0.055556in}}{%
\pgfpathmoveto{\pgfqpoint{0.000000in}{0.000000in}}%
\pgfpathlineto{\pgfqpoint{0.000000in}{0.055556in}}%
\pgfusepath{stroke,fill}%
}%
\begin{pgfscope}%
\pgfsys@transformshift{2.209829in}{0.759629in}%
\pgfsys@useobject{currentmarker}{}%
\end{pgfscope}%
\end{pgfscope}%
\begin{pgfscope}%
\pgfsetbuttcap%
\pgfsetroundjoin%
\definecolor{currentfill}{rgb}{0.000000,0.000000,0.000000}%
\pgfsetfillcolor{currentfill}%
\pgfsetlinewidth{0.501875pt}%
\definecolor{currentstroke}{rgb}{0.000000,0.000000,0.000000}%
\pgfsetstrokecolor{currentstroke}%
\pgfsetdash{}{0pt}%
\pgfsys@defobject{currentmarker}{\pgfqpoint{0.000000in}{-0.055556in}}{\pgfqpoint{0.000000in}{0.000000in}}{%
\pgfpathmoveto{\pgfqpoint{0.000000in}{0.000000in}}%
\pgfpathlineto{\pgfqpoint{0.000000in}{-0.055556in}}%
\pgfusepath{stroke,fill}%
}%
\begin{pgfscope}%
\pgfsys@transformshift{2.209829in}{2.384629in}%
\pgfsys@useobject{currentmarker}{}%
\end{pgfscope}%
\end{pgfscope}%
\begin{pgfscope}%
\pgftext[x=1.957384in,y=0.411609in,left,base,rotate=20.000000]{\sffamily\fontsize{10.000000}{12.000000}\selectfont Jul 2014}%
\end{pgfscope}%
\begin{pgfscope}%
\pgfpathrectangle{\pgfqpoint{0.624509in}{0.759629in}}{\pgfqpoint{5.362500in}{1.625000in}} %
\pgfusepath{clip}%
\pgfsetbuttcap%
\pgfsetroundjoin%
\pgfsetlinewidth{0.501875pt}%
\definecolor{currentstroke}{rgb}{0.501961,0.501961,0.501961}%
\pgfsetstrokecolor{currentstroke}%
\pgfsetdash{{1.000000pt}{3.000000pt}}{0.000000pt}%
\pgfpathmoveto{\pgfqpoint{2.767634in}{0.759629in}}%
\pgfpathlineto{\pgfqpoint{2.767634in}{2.384629in}}%
\pgfusepath{stroke}%
\end{pgfscope}%
\begin{pgfscope}%
\pgfsetbuttcap%
\pgfsetroundjoin%
\definecolor{currentfill}{rgb}{0.000000,0.000000,0.000000}%
\pgfsetfillcolor{currentfill}%
\pgfsetlinewidth{0.501875pt}%
\definecolor{currentstroke}{rgb}{0.000000,0.000000,0.000000}%
\pgfsetstrokecolor{currentstroke}%
\pgfsetdash{}{0pt}%
\pgfsys@defobject{currentmarker}{\pgfqpoint{0.000000in}{0.000000in}}{\pgfqpoint{0.000000in}{0.055556in}}{%
\pgfpathmoveto{\pgfqpoint{0.000000in}{0.000000in}}%
\pgfpathlineto{\pgfqpoint{0.000000in}{0.055556in}}%
\pgfusepath{stroke,fill}%
}%
\begin{pgfscope}%
\pgfsys@transformshift{2.767634in}{0.759629in}%
\pgfsys@useobject{currentmarker}{}%
\end{pgfscope}%
\end{pgfscope}%
\begin{pgfscope}%
\pgfsetbuttcap%
\pgfsetroundjoin%
\definecolor{currentfill}{rgb}{0.000000,0.000000,0.000000}%
\pgfsetfillcolor{currentfill}%
\pgfsetlinewidth{0.501875pt}%
\definecolor{currentstroke}{rgb}{0.000000,0.000000,0.000000}%
\pgfsetstrokecolor{currentstroke}%
\pgfsetdash{}{0pt}%
\pgfsys@defobject{currentmarker}{\pgfqpoint{0.000000in}{-0.055556in}}{\pgfqpoint{0.000000in}{0.000000in}}{%
\pgfpathmoveto{\pgfqpoint{0.000000in}{0.000000in}}%
\pgfpathlineto{\pgfqpoint{0.000000in}{-0.055556in}}%
\pgfusepath{stroke,fill}%
}%
\begin{pgfscope}%
\pgfsys@transformshift{2.767634in}{2.384629in}%
\pgfsys@useobject{currentmarker}{}%
\end{pgfscope}%
\end{pgfscope}%
\begin{pgfscope}%
\pgftext[x=2.470931in,y=0.379392in,left,base,rotate=20.000000]{\sffamily\fontsize{10.000000}{12.000000}\selectfont Sep 2014}%
\end{pgfscope}%
\begin{pgfscope}%
\pgfpathrectangle{\pgfqpoint{0.624509in}{0.759629in}}{\pgfqpoint{5.362500in}{1.625000in}} %
\pgfusepath{clip}%
\pgfsetbuttcap%
\pgfsetroundjoin%
\pgfsetlinewidth{0.501875pt}%
\definecolor{currentstroke}{rgb}{0.501961,0.501961,0.501961}%
\pgfsetstrokecolor{currentstroke}%
\pgfsetdash{{1.000000pt}{3.000000pt}}{0.000000pt}%
\pgfpathmoveto{\pgfqpoint{3.316443in}{0.759629in}}%
\pgfpathlineto{\pgfqpoint{3.316443in}{2.384629in}}%
\pgfusepath{stroke}%
\end{pgfscope}%
\begin{pgfscope}%
\pgfsetbuttcap%
\pgfsetroundjoin%
\definecolor{currentfill}{rgb}{0.000000,0.000000,0.000000}%
\pgfsetfillcolor{currentfill}%
\pgfsetlinewidth{0.501875pt}%
\definecolor{currentstroke}{rgb}{0.000000,0.000000,0.000000}%
\pgfsetstrokecolor{currentstroke}%
\pgfsetdash{}{0pt}%
\pgfsys@defobject{currentmarker}{\pgfqpoint{0.000000in}{0.000000in}}{\pgfqpoint{0.000000in}{0.055556in}}{%
\pgfpathmoveto{\pgfqpoint{0.000000in}{0.000000in}}%
\pgfpathlineto{\pgfqpoint{0.000000in}{0.055556in}}%
\pgfusepath{stroke,fill}%
}%
\begin{pgfscope}%
\pgfsys@transformshift{3.316443in}{0.759629in}%
\pgfsys@useobject{currentmarker}{}%
\end{pgfscope}%
\end{pgfscope}%
\begin{pgfscope}%
\pgfsetbuttcap%
\pgfsetroundjoin%
\definecolor{currentfill}{rgb}{0.000000,0.000000,0.000000}%
\pgfsetfillcolor{currentfill}%
\pgfsetlinewidth{0.501875pt}%
\definecolor{currentstroke}{rgb}{0.000000,0.000000,0.000000}%
\pgfsetstrokecolor{currentstroke}%
\pgfsetdash{}{0pt}%
\pgfsys@defobject{currentmarker}{\pgfqpoint{0.000000in}{-0.055556in}}{\pgfqpoint{0.000000in}{0.000000in}}{%
\pgfpathmoveto{\pgfqpoint{0.000000in}{0.000000in}}%
\pgfpathlineto{\pgfqpoint{0.000000in}{-0.055556in}}%
\pgfusepath{stroke,fill}%
}%
\begin{pgfscope}%
\pgfsys@transformshift{3.316443in}{2.384629in}%
\pgfsys@useobject{currentmarker}{}%
\end{pgfscope}%
\end{pgfscope}%
\begin{pgfscope}%
\pgftext[x=3.015374in,y=0.376214in,left,base,rotate=20.000000]{\sffamily\fontsize{10.000000}{12.000000}\selectfont Nov 2014}%
\end{pgfscope}%
\begin{pgfscope}%
\pgfpathrectangle{\pgfqpoint{0.624509in}{0.759629in}}{\pgfqpoint{5.362500in}{1.625000in}} %
\pgfusepath{clip}%
\pgfsetbuttcap%
\pgfsetroundjoin%
\pgfsetlinewidth{0.501875pt}%
\definecolor{currentstroke}{rgb}{0.501961,0.501961,0.501961}%
\pgfsetstrokecolor{currentstroke}%
\pgfsetdash{{1.000000pt}{3.000000pt}}{0.000000pt}%
\pgfpathmoveto{\pgfqpoint{3.865251in}{0.759629in}}%
\pgfpathlineto{\pgfqpoint{3.865251in}{2.384629in}}%
\pgfusepath{stroke}%
\end{pgfscope}%
\begin{pgfscope}%
\pgfsetbuttcap%
\pgfsetroundjoin%
\definecolor{currentfill}{rgb}{0.000000,0.000000,0.000000}%
\pgfsetfillcolor{currentfill}%
\pgfsetlinewidth{0.501875pt}%
\definecolor{currentstroke}{rgb}{0.000000,0.000000,0.000000}%
\pgfsetstrokecolor{currentstroke}%
\pgfsetdash{}{0pt}%
\pgfsys@defobject{currentmarker}{\pgfqpoint{0.000000in}{0.000000in}}{\pgfqpoint{0.000000in}{0.055556in}}{%
\pgfpathmoveto{\pgfqpoint{0.000000in}{0.000000in}}%
\pgfpathlineto{\pgfqpoint{0.000000in}{0.055556in}}%
\pgfusepath{stroke,fill}%
}%
\begin{pgfscope}%
\pgfsys@transformshift{3.865251in}{0.759629in}%
\pgfsys@useobject{currentmarker}{}%
\end{pgfscope}%
\end{pgfscope}%
\begin{pgfscope}%
\pgfsetbuttcap%
\pgfsetroundjoin%
\definecolor{currentfill}{rgb}{0.000000,0.000000,0.000000}%
\pgfsetfillcolor{currentfill}%
\pgfsetlinewidth{0.501875pt}%
\definecolor{currentstroke}{rgb}{0.000000,0.000000,0.000000}%
\pgfsetstrokecolor{currentstroke}%
\pgfsetdash{}{0pt}%
\pgfsys@defobject{currentmarker}{\pgfqpoint{0.000000in}{-0.055556in}}{\pgfqpoint{0.000000in}{0.000000in}}{%
\pgfpathmoveto{\pgfqpoint{0.000000in}{0.000000in}}%
\pgfpathlineto{\pgfqpoint{0.000000in}{-0.055556in}}%
\pgfusepath{stroke,fill}%
}%
\begin{pgfscope}%
\pgfsys@transformshift{3.865251in}{2.384629in}%
\pgfsys@useobject{currentmarker}{}%
\end{pgfscope}%
\end{pgfscope}%
\begin{pgfscope}%
\pgftext[x=3.590947in,y=0.395698in,left,base,rotate=20.000000]{\sffamily\fontsize{10.000000}{12.000000}\selectfont Jan 2015}%
\end{pgfscope}%
\begin{pgfscope}%
\pgfpathrectangle{\pgfqpoint{0.624509in}{0.759629in}}{\pgfqpoint{5.362500in}{1.625000in}} %
\pgfusepath{clip}%
\pgfsetbuttcap%
\pgfsetroundjoin%
\pgfsetlinewidth{0.501875pt}%
\definecolor{currentstroke}{rgb}{0.501961,0.501961,0.501961}%
\pgfsetstrokecolor{currentstroke}%
\pgfsetdash{{1.000000pt}{3.000000pt}}{0.000000pt}%
\pgfpathmoveto{\pgfqpoint{4.396065in}{0.759629in}}%
\pgfpathlineto{\pgfqpoint{4.396065in}{2.384629in}}%
\pgfusepath{stroke}%
\end{pgfscope}%
\begin{pgfscope}%
\pgfsetbuttcap%
\pgfsetroundjoin%
\definecolor{currentfill}{rgb}{0.000000,0.000000,0.000000}%
\pgfsetfillcolor{currentfill}%
\pgfsetlinewidth{0.501875pt}%
\definecolor{currentstroke}{rgb}{0.000000,0.000000,0.000000}%
\pgfsetstrokecolor{currentstroke}%
\pgfsetdash{}{0pt}%
\pgfsys@defobject{currentmarker}{\pgfqpoint{0.000000in}{0.000000in}}{\pgfqpoint{0.000000in}{0.055556in}}{%
\pgfpathmoveto{\pgfqpoint{0.000000in}{0.000000in}}%
\pgfpathlineto{\pgfqpoint{0.000000in}{0.055556in}}%
\pgfusepath{stroke,fill}%
}%
\begin{pgfscope}%
\pgfsys@transformshift{4.396065in}{0.759629in}%
\pgfsys@useobject{currentmarker}{}%
\end{pgfscope}%
\end{pgfscope}%
\begin{pgfscope}%
\pgfsetbuttcap%
\pgfsetroundjoin%
\definecolor{currentfill}{rgb}{0.000000,0.000000,0.000000}%
\pgfsetfillcolor{currentfill}%
\pgfsetlinewidth{0.501875pt}%
\definecolor{currentstroke}{rgb}{0.000000,0.000000,0.000000}%
\pgfsetstrokecolor{currentstroke}%
\pgfsetdash{}{0pt}%
\pgfsys@defobject{currentmarker}{\pgfqpoint{0.000000in}{-0.055556in}}{\pgfqpoint{0.000000in}{0.000000in}}{%
\pgfpathmoveto{\pgfqpoint{0.000000in}{0.000000in}}%
\pgfpathlineto{\pgfqpoint{0.000000in}{-0.055556in}}%
\pgfusepath{stroke,fill}%
}%
\begin{pgfscope}%
\pgfsys@transformshift{4.396065in}{2.384629in}%
\pgfsys@useobject{currentmarker}{}%
\end{pgfscope}%
\end{pgfscope}%
\begin{pgfscope}%
\pgftext[x=4.099234in,y=0.379299in,left,base,rotate=20.000000]{\sffamily\fontsize{10.000000}{12.000000}\selectfont Mar 2015}%
\end{pgfscope}%
\begin{pgfscope}%
\pgfpathrectangle{\pgfqpoint{0.624509in}{0.759629in}}{\pgfqpoint{5.362500in}{1.625000in}} %
\pgfusepath{clip}%
\pgfsetbuttcap%
\pgfsetroundjoin%
\pgfsetlinewidth{0.501875pt}%
\definecolor{currentstroke}{rgb}{0.501961,0.501961,0.501961}%
\pgfsetstrokecolor{currentstroke}%
\pgfsetdash{{1.000000pt}{3.000000pt}}{0.000000pt}%
\pgfpathmoveto{\pgfqpoint{4.944873in}{0.759629in}}%
\pgfpathlineto{\pgfqpoint{4.944873in}{2.384629in}}%
\pgfusepath{stroke}%
\end{pgfscope}%
\begin{pgfscope}%
\pgfsetbuttcap%
\pgfsetroundjoin%
\definecolor{currentfill}{rgb}{0.000000,0.000000,0.000000}%
\pgfsetfillcolor{currentfill}%
\pgfsetlinewidth{0.501875pt}%
\definecolor{currentstroke}{rgb}{0.000000,0.000000,0.000000}%
\pgfsetstrokecolor{currentstroke}%
\pgfsetdash{}{0pt}%
\pgfsys@defobject{currentmarker}{\pgfqpoint{0.000000in}{0.000000in}}{\pgfqpoint{0.000000in}{0.055556in}}{%
\pgfpathmoveto{\pgfqpoint{0.000000in}{0.000000in}}%
\pgfpathlineto{\pgfqpoint{0.000000in}{0.055556in}}%
\pgfusepath{stroke,fill}%
}%
\begin{pgfscope}%
\pgfsys@transformshift{4.944873in}{0.759629in}%
\pgfsys@useobject{currentmarker}{}%
\end{pgfscope}%
\end{pgfscope}%
\begin{pgfscope}%
\pgfsetbuttcap%
\pgfsetroundjoin%
\definecolor{currentfill}{rgb}{0.000000,0.000000,0.000000}%
\pgfsetfillcolor{currentfill}%
\pgfsetlinewidth{0.501875pt}%
\definecolor{currentstroke}{rgb}{0.000000,0.000000,0.000000}%
\pgfsetstrokecolor{currentstroke}%
\pgfsetdash{}{0pt}%
\pgfsys@defobject{currentmarker}{\pgfqpoint{0.000000in}{-0.055556in}}{\pgfqpoint{0.000000in}{0.000000in}}{%
\pgfpathmoveto{\pgfqpoint{0.000000in}{0.000000in}}%
\pgfpathlineto{\pgfqpoint{0.000000in}{-0.055556in}}%
\pgfusepath{stroke,fill}%
}%
\begin{pgfscope}%
\pgfsys@transformshift{4.944873in}{2.384629in}%
\pgfsys@useobject{currentmarker}{}%
\end{pgfscope}%
\end{pgfscope}%
\begin{pgfscope}%
\pgftext[x=4.636253in,y=0.370717in,left,base,rotate=20.000000]{\sffamily\fontsize{10.000000}{12.000000}\selectfont May 2015}%
\end{pgfscope}%
\begin{pgfscope}%
\pgfpathrectangle{\pgfqpoint{0.624509in}{0.759629in}}{\pgfqpoint{5.362500in}{1.625000in}} %
\pgfusepath{clip}%
\pgfsetbuttcap%
\pgfsetroundjoin%
\pgfsetlinewidth{0.501875pt}%
\definecolor{currentstroke}{rgb}{0.501961,0.501961,0.501961}%
\pgfsetstrokecolor{currentstroke}%
\pgfsetdash{{1.000000pt}{3.000000pt}}{0.000000pt}%
\pgfpathmoveto{\pgfqpoint{5.493681in}{0.759629in}}%
\pgfpathlineto{\pgfqpoint{5.493681in}{2.384629in}}%
\pgfusepath{stroke}%
\end{pgfscope}%
\begin{pgfscope}%
\pgfsetbuttcap%
\pgfsetroundjoin%
\definecolor{currentfill}{rgb}{0.000000,0.000000,0.000000}%
\pgfsetfillcolor{currentfill}%
\pgfsetlinewidth{0.501875pt}%
\definecolor{currentstroke}{rgb}{0.000000,0.000000,0.000000}%
\pgfsetstrokecolor{currentstroke}%
\pgfsetdash{}{0pt}%
\pgfsys@defobject{currentmarker}{\pgfqpoint{0.000000in}{0.000000in}}{\pgfqpoint{0.000000in}{0.055556in}}{%
\pgfpathmoveto{\pgfqpoint{0.000000in}{0.000000in}}%
\pgfpathlineto{\pgfqpoint{0.000000in}{0.055556in}}%
\pgfusepath{stroke,fill}%
}%
\begin{pgfscope}%
\pgfsys@transformshift{5.493681in}{0.759629in}%
\pgfsys@useobject{currentmarker}{}%
\end{pgfscope}%
\end{pgfscope}%
\begin{pgfscope}%
\pgfsetbuttcap%
\pgfsetroundjoin%
\definecolor{currentfill}{rgb}{0.000000,0.000000,0.000000}%
\pgfsetfillcolor{currentfill}%
\pgfsetlinewidth{0.501875pt}%
\definecolor{currentstroke}{rgb}{0.000000,0.000000,0.000000}%
\pgfsetstrokecolor{currentstroke}%
\pgfsetdash{}{0pt}%
\pgfsys@defobject{currentmarker}{\pgfqpoint{0.000000in}{-0.055556in}}{\pgfqpoint{0.000000in}{0.000000in}}{%
\pgfpathmoveto{\pgfqpoint{0.000000in}{0.000000in}}%
\pgfpathlineto{\pgfqpoint{0.000000in}{-0.055556in}}%
\pgfusepath{stroke,fill}%
}%
\begin{pgfscope}%
\pgfsys@transformshift{5.493681in}{2.384629in}%
\pgfsys@useobject{currentmarker}{}%
\end{pgfscope}%
\end{pgfscope}%
\begin{pgfscope}%
\pgftext[x=5.241236in,y=0.411609in,left,base,rotate=20.000000]{\sffamily\fontsize{10.000000}{12.000000}\selectfont Jul 2015}%
\end{pgfscope}%
\begin{pgfscope}%
\pgfpathrectangle{\pgfqpoint{0.624509in}{0.759629in}}{\pgfqpoint{5.362500in}{1.625000in}} %
\pgfusepath{clip}%
\pgfsetbuttcap%
\pgfsetroundjoin%
\pgfsetlinewidth{0.501875pt}%
\definecolor{currentstroke}{rgb}{0.501961,0.501961,0.501961}%
\pgfsetstrokecolor{currentstroke}%
\pgfsetdash{{1.000000pt}{3.000000pt}}{0.000000pt}%
\pgfpathmoveto{\pgfqpoint{0.624509in}{0.759629in}}%
\pgfpathlineto{\pgfqpoint{5.987009in}{0.759629in}}%
\pgfusepath{stroke}%
\end{pgfscope}%
\begin{pgfscope}%
\pgfsetbuttcap%
\pgfsetroundjoin%
\definecolor{currentfill}{rgb}{0.000000,0.000000,0.000000}%
\pgfsetfillcolor{currentfill}%
\pgfsetlinewidth{0.501875pt}%
\definecolor{currentstroke}{rgb}{0.000000,0.000000,0.000000}%
\pgfsetstrokecolor{currentstroke}%
\pgfsetdash{}{0pt}%
\pgfsys@defobject{currentmarker}{\pgfqpoint{0.000000in}{0.000000in}}{\pgfqpoint{0.055556in}{0.000000in}}{%
\pgfpathmoveto{\pgfqpoint{0.000000in}{0.000000in}}%
\pgfpathlineto{\pgfqpoint{0.055556in}{0.000000in}}%
\pgfusepath{stroke,fill}%
}%
\begin{pgfscope}%
\pgfsys@transformshift{0.624509in}{0.759629in}%
\pgfsys@useobject{currentmarker}{}%
\end{pgfscope}%
\end{pgfscope}%
\begin{pgfscope}%
\pgfsetbuttcap%
\pgfsetroundjoin%
\definecolor{currentfill}{rgb}{0.000000,0.000000,0.000000}%
\pgfsetfillcolor{currentfill}%
\pgfsetlinewidth{0.501875pt}%
\definecolor{currentstroke}{rgb}{0.000000,0.000000,0.000000}%
\pgfsetstrokecolor{currentstroke}%
\pgfsetdash{}{0pt}%
\pgfsys@defobject{currentmarker}{\pgfqpoint{-0.055556in}{0.000000in}}{\pgfqpoint{0.000000in}{0.000000in}}{%
\pgfpathmoveto{\pgfqpoint{0.000000in}{0.000000in}}%
\pgfpathlineto{\pgfqpoint{-0.055556in}{0.000000in}}%
\pgfusepath{stroke,fill}%
}%
\begin{pgfscope}%
\pgfsys@transformshift{5.987009in}{0.759629in}%
\pgfsys@useobject{currentmarker}{}%
\end{pgfscope}%
\end{pgfscope}%
\begin{pgfscope}%
\pgftext[x=0.568953in,y=0.759629in,right,]{\sffamily\fontsize{10.000000}{12.000000}\selectfont 0}%
\end{pgfscope}%
\begin{pgfscope}%
\pgfpathrectangle{\pgfqpoint{0.624509in}{0.759629in}}{\pgfqpoint{5.362500in}{1.625000in}} %
\pgfusepath{clip}%
\pgfsetbuttcap%
\pgfsetroundjoin%
\pgfsetlinewidth{0.501875pt}%
\definecolor{currentstroke}{rgb}{0.501961,0.501961,0.501961}%
\pgfsetstrokecolor{currentstroke}%
\pgfsetdash{{1.000000pt}{3.000000pt}}{0.000000pt}%
\pgfpathmoveto{\pgfqpoint{0.624509in}{1.030462in}}%
\pgfpathlineto{\pgfqpoint{5.987009in}{1.030462in}}%
\pgfusepath{stroke}%
\end{pgfscope}%
\begin{pgfscope}%
\pgfsetbuttcap%
\pgfsetroundjoin%
\definecolor{currentfill}{rgb}{0.000000,0.000000,0.000000}%
\pgfsetfillcolor{currentfill}%
\pgfsetlinewidth{0.501875pt}%
\definecolor{currentstroke}{rgb}{0.000000,0.000000,0.000000}%
\pgfsetstrokecolor{currentstroke}%
\pgfsetdash{}{0pt}%
\pgfsys@defobject{currentmarker}{\pgfqpoint{0.000000in}{0.000000in}}{\pgfqpoint{0.055556in}{0.000000in}}{%
\pgfpathmoveto{\pgfqpoint{0.000000in}{0.000000in}}%
\pgfpathlineto{\pgfqpoint{0.055556in}{0.000000in}}%
\pgfusepath{stroke,fill}%
}%
\begin{pgfscope}%
\pgfsys@transformshift{0.624509in}{1.030462in}%
\pgfsys@useobject{currentmarker}{}%
\end{pgfscope}%
\end{pgfscope}%
\begin{pgfscope}%
\pgfsetbuttcap%
\pgfsetroundjoin%
\definecolor{currentfill}{rgb}{0.000000,0.000000,0.000000}%
\pgfsetfillcolor{currentfill}%
\pgfsetlinewidth{0.501875pt}%
\definecolor{currentstroke}{rgb}{0.000000,0.000000,0.000000}%
\pgfsetstrokecolor{currentstroke}%
\pgfsetdash{}{0pt}%
\pgfsys@defobject{currentmarker}{\pgfqpoint{-0.055556in}{0.000000in}}{\pgfqpoint{0.000000in}{0.000000in}}{%
\pgfpathmoveto{\pgfqpoint{0.000000in}{0.000000in}}%
\pgfpathlineto{\pgfqpoint{-0.055556in}{0.000000in}}%
\pgfusepath{stroke,fill}%
}%
\begin{pgfscope}%
\pgfsys@transformshift{5.987009in}{1.030462in}%
\pgfsys@useobject{currentmarker}{}%
\end{pgfscope}%
\end{pgfscope}%
\begin{pgfscope}%
\pgftext[x=0.568953in,y=1.030462in,right,]{\sffamily\fontsize{10.000000}{12.000000}\selectfont 50}%
\end{pgfscope}%
\begin{pgfscope}%
\pgfpathrectangle{\pgfqpoint{0.624509in}{0.759629in}}{\pgfqpoint{5.362500in}{1.625000in}} %
\pgfusepath{clip}%
\pgfsetbuttcap%
\pgfsetroundjoin%
\pgfsetlinewidth{0.501875pt}%
\definecolor{currentstroke}{rgb}{0.501961,0.501961,0.501961}%
\pgfsetstrokecolor{currentstroke}%
\pgfsetdash{{1.000000pt}{3.000000pt}}{0.000000pt}%
\pgfpathmoveto{\pgfqpoint{0.624509in}{1.301295in}}%
\pgfpathlineto{\pgfqpoint{5.987009in}{1.301295in}}%
\pgfusepath{stroke}%
\end{pgfscope}%
\begin{pgfscope}%
\pgfsetbuttcap%
\pgfsetroundjoin%
\definecolor{currentfill}{rgb}{0.000000,0.000000,0.000000}%
\pgfsetfillcolor{currentfill}%
\pgfsetlinewidth{0.501875pt}%
\definecolor{currentstroke}{rgb}{0.000000,0.000000,0.000000}%
\pgfsetstrokecolor{currentstroke}%
\pgfsetdash{}{0pt}%
\pgfsys@defobject{currentmarker}{\pgfqpoint{0.000000in}{0.000000in}}{\pgfqpoint{0.055556in}{0.000000in}}{%
\pgfpathmoveto{\pgfqpoint{0.000000in}{0.000000in}}%
\pgfpathlineto{\pgfqpoint{0.055556in}{0.000000in}}%
\pgfusepath{stroke,fill}%
}%
\begin{pgfscope}%
\pgfsys@transformshift{0.624509in}{1.301295in}%
\pgfsys@useobject{currentmarker}{}%
\end{pgfscope}%
\end{pgfscope}%
\begin{pgfscope}%
\pgfsetbuttcap%
\pgfsetroundjoin%
\definecolor{currentfill}{rgb}{0.000000,0.000000,0.000000}%
\pgfsetfillcolor{currentfill}%
\pgfsetlinewidth{0.501875pt}%
\definecolor{currentstroke}{rgb}{0.000000,0.000000,0.000000}%
\pgfsetstrokecolor{currentstroke}%
\pgfsetdash{}{0pt}%
\pgfsys@defobject{currentmarker}{\pgfqpoint{-0.055556in}{0.000000in}}{\pgfqpoint{0.000000in}{0.000000in}}{%
\pgfpathmoveto{\pgfqpoint{0.000000in}{0.000000in}}%
\pgfpathlineto{\pgfqpoint{-0.055556in}{0.000000in}}%
\pgfusepath{stroke,fill}%
}%
\begin{pgfscope}%
\pgfsys@transformshift{5.987009in}{1.301295in}%
\pgfsys@useobject{currentmarker}{}%
\end{pgfscope}%
\end{pgfscope}%
\begin{pgfscope}%
\pgftext[x=0.568953in,y=1.301295in,right,]{\sffamily\fontsize{10.000000}{12.000000}\selectfont 100}%
\end{pgfscope}%
\begin{pgfscope}%
\pgfpathrectangle{\pgfqpoint{0.624509in}{0.759629in}}{\pgfqpoint{5.362500in}{1.625000in}} %
\pgfusepath{clip}%
\pgfsetbuttcap%
\pgfsetroundjoin%
\pgfsetlinewidth{0.501875pt}%
\definecolor{currentstroke}{rgb}{0.501961,0.501961,0.501961}%
\pgfsetstrokecolor{currentstroke}%
\pgfsetdash{{1.000000pt}{3.000000pt}}{0.000000pt}%
\pgfpathmoveto{\pgfqpoint{0.624509in}{1.572129in}}%
\pgfpathlineto{\pgfqpoint{5.987009in}{1.572129in}}%
\pgfusepath{stroke}%
\end{pgfscope}%
\begin{pgfscope}%
\pgfsetbuttcap%
\pgfsetroundjoin%
\definecolor{currentfill}{rgb}{0.000000,0.000000,0.000000}%
\pgfsetfillcolor{currentfill}%
\pgfsetlinewidth{0.501875pt}%
\definecolor{currentstroke}{rgb}{0.000000,0.000000,0.000000}%
\pgfsetstrokecolor{currentstroke}%
\pgfsetdash{}{0pt}%
\pgfsys@defobject{currentmarker}{\pgfqpoint{0.000000in}{0.000000in}}{\pgfqpoint{0.055556in}{0.000000in}}{%
\pgfpathmoveto{\pgfqpoint{0.000000in}{0.000000in}}%
\pgfpathlineto{\pgfqpoint{0.055556in}{0.000000in}}%
\pgfusepath{stroke,fill}%
}%
\begin{pgfscope}%
\pgfsys@transformshift{0.624509in}{1.572129in}%
\pgfsys@useobject{currentmarker}{}%
\end{pgfscope}%
\end{pgfscope}%
\begin{pgfscope}%
\pgfsetbuttcap%
\pgfsetroundjoin%
\definecolor{currentfill}{rgb}{0.000000,0.000000,0.000000}%
\pgfsetfillcolor{currentfill}%
\pgfsetlinewidth{0.501875pt}%
\definecolor{currentstroke}{rgb}{0.000000,0.000000,0.000000}%
\pgfsetstrokecolor{currentstroke}%
\pgfsetdash{}{0pt}%
\pgfsys@defobject{currentmarker}{\pgfqpoint{-0.055556in}{0.000000in}}{\pgfqpoint{0.000000in}{0.000000in}}{%
\pgfpathmoveto{\pgfqpoint{0.000000in}{0.000000in}}%
\pgfpathlineto{\pgfqpoint{-0.055556in}{0.000000in}}%
\pgfusepath{stroke,fill}%
}%
\begin{pgfscope}%
\pgfsys@transformshift{5.987009in}{1.572129in}%
\pgfsys@useobject{currentmarker}{}%
\end{pgfscope}%
\end{pgfscope}%
\begin{pgfscope}%
\pgftext[x=0.568953in,y=1.572129in,right,]{\sffamily\fontsize{10.000000}{12.000000}\selectfont 150}%
\end{pgfscope}%
\begin{pgfscope}%
\pgfpathrectangle{\pgfqpoint{0.624509in}{0.759629in}}{\pgfqpoint{5.362500in}{1.625000in}} %
\pgfusepath{clip}%
\pgfsetbuttcap%
\pgfsetroundjoin%
\pgfsetlinewidth{0.501875pt}%
\definecolor{currentstroke}{rgb}{0.501961,0.501961,0.501961}%
\pgfsetstrokecolor{currentstroke}%
\pgfsetdash{{1.000000pt}{3.000000pt}}{0.000000pt}%
\pgfpathmoveto{\pgfqpoint{0.624509in}{1.842962in}}%
\pgfpathlineto{\pgfqpoint{5.987009in}{1.842962in}}%
\pgfusepath{stroke}%
\end{pgfscope}%
\begin{pgfscope}%
\pgfsetbuttcap%
\pgfsetroundjoin%
\definecolor{currentfill}{rgb}{0.000000,0.000000,0.000000}%
\pgfsetfillcolor{currentfill}%
\pgfsetlinewidth{0.501875pt}%
\definecolor{currentstroke}{rgb}{0.000000,0.000000,0.000000}%
\pgfsetstrokecolor{currentstroke}%
\pgfsetdash{}{0pt}%
\pgfsys@defobject{currentmarker}{\pgfqpoint{0.000000in}{0.000000in}}{\pgfqpoint{0.055556in}{0.000000in}}{%
\pgfpathmoveto{\pgfqpoint{0.000000in}{0.000000in}}%
\pgfpathlineto{\pgfqpoint{0.055556in}{0.000000in}}%
\pgfusepath{stroke,fill}%
}%
\begin{pgfscope}%
\pgfsys@transformshift{0.624509in}{1.842962in}%
\pgfsys@useobject{currentmarker}{}%
\end{pgfscope}%
\end{pgfscope}%
\begin{pgfscope}%
\pgfsetbuttcap%
\pgfsetroundjoin%
\definecolor{currentfill}{rgb}{0.000000,0.000000,0.000000}%
\pgfsetfillcolor{currentfill}%
\pgfsetlinewidth{0.501875pt}%
\definecolor{currentstroke}{rgb}{0.000000,0.000000,0.000000}%
\pgfsetstrokecolor{currentstroke}%
\pgfsetdash{}{0pt}%
\pgfsys@defobject{currentmarker}{\pgfqpoint{-0.055556in}{0.000000in}}{\pgfqpoint{0.000000in}{0.000000in}}{%
\pgfpathmoveto{\pgfqpoint{0.000000in}{0.000000in}}%
\pgfpathlineto{\pgfqpoint{-0.055556in}{0.000000in}}%
\pgfusepath{stroke,fill}%
}%
\begin{pgfscope}%
\pgfsys@transformshift{5.987009in}{1.842962in}%
\pgfsys@useobject{currentmarker}{}%
\end{pgfscope}%
\end{pgfscope}%
\begin{pgfscope}%
\pgftext[x=0.568953in,y=1.842962in,right,]{\sffamily\fontsize{10.000000}{12.000000}\selectfont 200}%
\end{pgfscope}%
\begin{pgfscope}%
\pgfpathrectangle{\pgfqpoint{0.624509in}{0.759629in}}{\pgfqpoint{5.362500in}{1.625000in}} %
\pgfusepath{clip}%
\pgfsetbuttcap%
\pgfsetroundjoin%
\pgfsetlinewidth{0.501875pt}%
\definecolor{currentstroke}{rgb}{0.501961,0.501961,0.501961}%
\pgfsetstrokecolor{currentstroke}%
\pgfsetdash{{1.000000pt}{3.000000pt}}{0.000000pt}%
\pgfpathmoveto{\pgfqpoint{0.624509in}{2.113795in}}%
\pgfpathlineto{\pgfqpoint{5.987009in}{2.113795in}}%
\pgfusepath{stroke}%
\end{pgfscope}%
\begin{pgfscope}%
\pgfsetbuttcap%
\pgfsetroundjoin%
\definecolor{currentfill}{rgb}{0.000000,0.000000,0.000000}%
\pgfsetfillcolor{currentfill}%
\pgfsetlinewidth{0.501875pt}%
\definecolor{currentstroke}{rgb}{0.000000,0.000000,0.000000}%
\pgfsetstrokecolor{currentstroke}%
\pgfsetdash{}{0pt}%
\pgfsys@defobject{currentmarker}{\pgfqpoint{0.000000in}{0.000000in}}{\pgfqpoint{0.055556in}{0.000000in}}{%
\pgfpathmoveto{\pgfqpoint{0.000000in}{0.000000in}}%
\pgfpathlineto{\pgfqpoint{0.055556in}{0.000000in}}%
\pgfusepath{stroke,fill}%
}%
\begin{pgfscope}%
\pgfsys@transformshift{0.624509in}{2.113795in}%
\pgfsys@useobject{currentmarker}{}%
\end{pgfscope}%
\end{pgfscope}%
\begin{pgfscope}%
\pgfsetbuttcap%
\pgfsetroundjoin%
\definecolor{currentfill}{rgb}{0.000000,0.000000,0.000000}%
\pgfsetfillcolor{currentfill}%
\pgfsetlinewidth{0.501875pt}%
\definecolor{currentstroke}{rgb}{0.000000,0.000000,0.000000}%
\pgfsetstrokecolor{currentstroke}%
\pgfsetdash{}{0pt}%
\pgfsys@defobject{currentmarker}{\pgfqpoint{-0.055556in}{0.000000in}}{\pgfqpoint{0.000000in}{0.000000in}}{%
\pgfpathmoveto{\pgfqpoint{0.000000in}{0.000000in}}%
\pgfpathlineto{\pgfqpoint{-0.055556in}{0.000000in}}%
\pgfusepath{stroke,fill}%
}%
\begin{pgfscope}%
\pgfsys@transformshift{5.987009in}{2.113795in}%
\pgfsys@useobject{currentmarker}{}%
\end{pgfscope}%
\end{pgfscope}%
\begin{pgfscope}%
\pgftext[x=0.568953in,y=2.113795in,right,]{\sffamily\fontsize{10.000000}{12.000000}\selectfont 250}%
\end{pgfscope}%
\begin{pgfscope}%
\pgfpathrectangle{\pgfqpoint{0.624509in}{0.759629in}}{\pgfqpoint{5.362500in}{1.625000in}} %
\pgfusepath{clip}%
\pgfsetbuttcap%
\pgfsetroundjoin%
\pgfsetlinewidth{0.501875pt}%
\definecolor{currentstroke}{rgb}{0.501961,0.501961,0.501961}%
\pgfsetstrokecolor{currentstroke}%
\pgfsetdash{{1.000000pt}{3.000000pt}}{0.000000pt}%
\pgfpathmoveto{\pgfqpoint{0.624509in}{2.384629in}}%
\pgfpathlineto{\pgfqpoint{5.987009in}{2.384629in}}%
\pgfusepath{stroke}%
\end{pgfscope}%
\begin{pgfscope}%
\pgfsetbuttcap%
\pgfsetroundjoin%
\definecolor{currentfill}{rgb}{0.000000,0.000000,0.000000}%
\pgfsetfillcolor{currentfill}%
\pgfsetlinewidth{0.501875pt}%
\definecolor{currentstroke}{rgb}{0.000000,0.000000,0.000000}%
\pgfsetstrokecolor{currentstroke}%
\pgfsetdash{}{0pt}%
\pgfsys@defobject{currentmarker}{\pgfqpoint{0.000000in}{0.000000in}}{\pgfqpoint{0.055556in}{0.000000in}}{%
\pgfpathmoveto{\pgfqpoint{0.000000in}{0.000000in}}%
\pgfpathlineto{\pgfqpoint{0.055556in}{0.000000in}}%
\pgfusepath{stroke,fill}%
}%
\begin{pgfscope}%
\pgfsys@transformshift{0.624509in}{2.384629in}%
\pgfsys@useobject{currentmarker}{}%
\end{pgfscope}%
\end{pgfscope}%
\begin{pgfscope}%
\pgfsetbuttcap%
\pgfsetroundjoin%
\definecolor{currentfill}{rgb}{0.000000,0.000000,0.000000}%
\pgfsetfillcolor{currentfill}%
\pgfsetlinewidth{0.501875pt}%
\definecolor{currentstroke}{rgb}{0.000000,0.000000,0.000000}%
\pgfsetstrokecolor{currentstroke}%
\pgfsetdash{}{0pt}%
\pgfsys@defobject{currentmarker}{\pgfqpoint{-0.055556in}{0.000000in}}{\pgfqpoint{0.000000in}{0.000000in}}{%
\pgfpathmoveto{\pgfqpoint{0.000000in}{0.000000in}}%
\pgfpathlineto{\pgfqpoint{-0.055556in}{0.000000in}}%
\pgfusepath{stroke,fill}%
}%
\begin{pgfscope}%
\pgfsys@transformshift{5.987009in}{2.384629in}%
\pgfsys@useobject{currentmarker}{}%
\end{pgfscope}%
\end{pgfscope}%
\begin{pgfscope}%
\pgftext[x=0.568953in,y=2.384629in,right,]{\sffamily\fontsize{10.000000}{12.000000}\selectfont 300}%
\end{pgfscope}%
\begin{pgfscope}%
\pgftext[x=0.234413in,y=1.572129in,,bottom,rotate=90.000000]{\sffamily\fontsize{10.000000}{12.000000}\selectfont Count}%
\end{pgfscope}%
\begin{pgfscope}%
\pgfsetrectcap%
\pgfsetroundjoin%
\pgfsetlinewidth{2.007500pt}%
\definecolor{currentstroke}{rgb}{0.000000,0.000000,0.000000}%
\pgfsetstrokecolor{currentstroke}%
\pgfsetdash{}{0pt}%
\pgfpathmoveto{\pgfqpoint{2.066558in}{0.234668in}}%
\pgfpathlineto{\pgfqpoint{2.386558in}{0.234668in}}%
\pgfusepath{stroke}%
\end{pgfscope}%
\begin{pgfscope}%
\pgftext[x=2.533225in,y=0.188001in,left,base]{\sffamily\fontsize{12.000000}{14.400000}\selectfont Entering}%
\end{pgfscope}%
\begin{pgfscope}%
\pgfsetbuttcap%
\pgfsetroundjoin%
\pgfsetlinewidth{1.003750pt}%
\definecolor{currentstroke}{rgb}{0.000000,0.000000,0.000000}%
\pgfsetstrokecolor{currentstroke}%
\pgfsetdash{{6.000000pt}{6.000000pt}}{0.000000pt}%
\pgfpathmoveto{\pgfqpoint{3.544969in}{0.234668in}}%
\pgfpathlineto{\pgfqpoint{3.864969in}{0.234668in}}%
\pgfusepath{stroke}%
\end{pgfscope}%
\begin{pgfscope}%
\pgftext[x=4.011636in,y=0.188001in,left,base]{\sffamily\fontsize{12.000000}{14.400000}\selectfont Exiting}%
\end{pgfscope}%
\end{pgfpicture}%
\makeatother%
\endgroup%

%% file: figures/summary_stats.tex
\begin{tabular}{lrrrrr}
  \toprule
 & All & Cannabis & MDMA & Heroin & Cocaine \\ 
  \midrule
Number of vendors & 1,482 & 824 & 508 & 192 & 473 \\ 
  Number of reviews per vendor & 386.1 & 338.7 & 236.2 & 288.7 & 249 \\ 
  Average life span per vendor (days) & 349.1 & 344.3 & 337.4 & 307.8 & 319.2 \\ 
  Share of vendors active when market closed & 0.38 & 0.35 & 0.33 & 0.43 & 0.33 \\ 
  Average price per gram & 49.83 & 12.11 & 37.45 & 162.32 & 98.86 \\ 
  Average rating & 4.93 & 4.94 & 4.97 & 4.80 & 4.92 \\ 
  \end{tabular}

%% file: figures/regression.tex
\begin{tabular}{lllll}
\hline\\[-0.05in]
                                         &   (1)    &   (2)    &   (3)    &   (4)      \\[-0.05in]\\\hline \\\emph{Dependent variable} & $\log(p)$ & $\log(p)$ & $\log(p)$ & $\log(p)$ \\ \\ 
\hline
\hline\\
Rating                                   & 0.10***  &          & -0.03*** &           \\
                                         & (0.00)   &          & (0.00)   &           \\
Rating (large sellers)                                   &          & 1.80***  &          & -0.25***  \\
                                         &          & (0.03)   &          & (0.01)    \\
Rating (small sellers)                                   &          & 0.07***  &          & -0.01***  \\
                                         &          & (0.00)   &          & (0.00)    \\
Total reviews (1,000s)                                    & 0.03***  & 0.03***  & 0.00***  & 0.00***   \\
                                         & (0.00)   & (0.00)   & (0.00)   & (0.00)    \\
Quantity (g)                                 & -0.15*** & -0.15*** & -0.11*** & -0.11***  \\
                                         & (0.00)   & (0.00)   & (0.00)   & (0.00)    \\
Year                                     & -0.02*** & -0.03*** & -0.02*** & -0.01***  \\
                                         & (0.00)   & (0.00)   & (0.00)   & (0.00)    \\
Age (years)                                      & 0.02***  & 0.01*    & 0.00***  & 0.00***   \\
                                         & (0.00)   & (0.00)   & (0.00)   & (0.00)    \\
Total sales (1,000s)                                & -0.05*** & -0.04*** & 0.01***  & 0.01***   \\
                                         & (0.00)   & (0.00)   & (0.00)   & (0.00)    \\
Intercept                                & 0.03***  & 0.03***  & 0.01***  & 0.01***   \\
                                         & (0.00)   & (0.00)   & (0.00)   & (0.00)    \\
\emph{Fixed effects}\textcolor{white}{b} & No       & No       & Yes      & Yes       \\[0.01in]\hline\\[-0.1in]
$N$                                      & 280366   & 280366   & 280366   & 280366    \\
$R^2$                                    & 0.09     & 0.10     & 0.08     & 0.08      \\
\hline
\end{tabular}

%% file: section/model.tex

\section{Model\label{sec:model}}

\begin{dictionary}
    \begin{itemize}
        \item The publicly observable state of a seller is $\omega \in \Omega$.
        \item The transition matrix for a seller of type $\theta$ is denoted $\Pi_\theta$.
        \item Indices are small roman letters $t$ for time, $h$ for age, $i$ for general indices.
        \item Masses, fractions, measures are small greek letters: $\mu, \eta, \zeta$.
        \item Indices are subscripts, quality is a subscript.
        \item $F, G, H$ are CDFs.
        \item Sellers are he, buyers are she.
        \item Tildes denote random variables.
    \end{itemize}
\end{dictionary}

In this section, we develop a model of a market platform that features ratings, prices, and exit.
Sellers are long lived and differ in their privately observed quality.
The market platform provides publicly observable information about each seller.
This publicly observable information may include the seller's average rating and the seller's total number of sales made.
Instead of modeling buyers and their behavior directly, we will make two simplifying assumptions.

First, our model takes as a primitive a \emph{rating system} \citep{Ekmekci11}.
The rating system specifies how a seller's publicly available information evolves over time conditional on the true quality of the seller.
Second, our model takes as a primitive a demand process.\footnote{
  We take demand as exogenous because, in practice, endogenizing demand proves to be intractable.
  If a seller is allowed to choose a price, then a seller's choice of price should be informative about his quality.
  In practice this results in multiple equilibria.
  ~\cite{Saeedi14} takes an alternate approach and assumes sellers choose a price, but buyers do not form inferences about the quality of the seller from that price.
}
We assume that the price that a seller charges and the quantity sold is a function of market beliefs of the seller's quality.
Market beliefs are determined in equilibrium by the rating system and seller's actions.
Each period, faced with a price, a per-period cost shock, and some number of expected sales, each seller makes a decision whether or not to (irreversibly) exit.
Market beliefs over a seller's quality are formed, in equilibrium, conditional on the publicly observable information about the seller that is provided by the rating system, as well as seller's exit strategies.
In equilibrium, sellers optimally exit the market, taking market beliefs as given.

\subsection{Environment\label{sec:environment}}

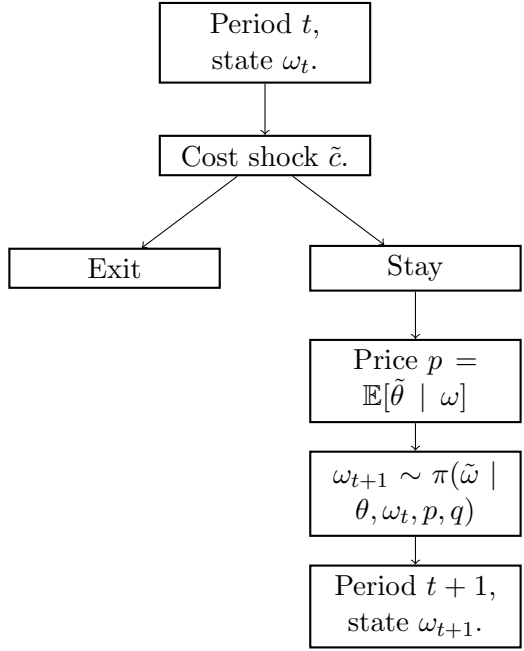
\begin{figure}
  \centerline{
  \begin{tikzpicture}[auto, block/.style ={rectangle, draw = black, thick, align=center, text width = 1in}]
    \node[block] (A) at (0, 0) {Period $t$, state $\omega_t$.};
    \node[block] (B) at (0, -1.5) {Cost shock $\tilde{c}$.};
    \node[block] (C) at (-2, -3) {Exit};
    \node[block] (D) at (2, -3) {Stay};
    \node[block] (E) at (2, -4.5) {Price $p = \E[\tilde{\theta} \mid \omega]$};
    \node[block] (F) at (2, -6) {$\omega_{t+1} \sim \pi(\tilde{\omega} \mid \theta, \omega_t, p, q)$};
    \node[block] (G) at (2, -7.5) {Period $t+1$, state $\omega_{t+1}$.};
    \draw[->] (A) edge (B) (B) edge (C) (B) edge (D) (D) edge (E) (E) edge (F) (F) edge (G);
  \end{tikzpicture}
  }
  \caption{Sequence of events for an individual seller within one period.\label{fig:modeltimeline}}
\end{figure}

Time is discrete and infinite, $t = 1, 2, \ldots$.
There is a continuum of sellers.
Sellers are long-lived, forward looking and maximize profits, and they discount the future with discount factor $\beta \in (0,1)$.
Each seller is characterized by a privately-known and \emph{permanent} seller quality, $\theta \in \Theta = \{\underline{\theta}, \overline{\theta}\} \subset \mathbb{R}$, satisfying $\underline{\theta} < \overline{\theta}$, and a publicly observable state $\omega \in \Omega$, which is \emph{time-varying} and which we interpret as the seller's rating and sales made.\footnote{
  In reality, buyers observe other characteristics of the seller, such as the five most recent reviews, and the position of the seller in the search ranking.
  We currently do not allow the buyers to condition on these characteristics, however, augmenting the model to allow for this is straightforward and a possible avenue for future work.
}\footnote{
  Allowing the seller to choose $\theta$, or allowing $\theta$ to be time-varying, is a potential way to introduce moral hazard into the model.
  Here, we focus on the problem of adverse selection, and so $\theta$ is fixed.
}
Seller's publicly observable state evolves according to a Markov process, conditional on $\theta$, the seller's type.
We denote the type-dependent transition probability matrix by $\Pi_\theta$, a square matrix of size $|\Omega| \times |\Omega|$, whose elements represent the transition probabilities between states $\omega$ and $\omega'$:
\begin{equation*}
    \Pi_\theta(\omega, \omega') = \P(\omega_{t+1} = \omega' \mid \omega_t = \omega, \theta).
\end{equation*}
Each period, sellers incur a fixed cost of operation $c$.
This cost is identically and independently distributed and drawn from a cumulative distribution function $F$.
Sellers can avoid paying the fixed cost $c$ by irreversibly exiting the market.
See Figure~\ref{fig:modeltimeline} for an illustration of the sequence of events within a period.

Each period, a continuum of young sellers enters the market
We denote the mass of entering sellers with private type $\theta$ and initial public state $\omega$ by $\eta_\theta(\omega)$.
Under the interpretation that $\omega$ represents the publicly available information about a seller, a reasonable assumption, which we impose when estimating the model, is that there is a initial state $\omega_0$, representing the information that a seller is new, such that
\begin{equation*}
  \eta_\theta(\omega_0) > 0, \eta_\theta(\omega) = 0 \,\forall \omega \neq \omega_0.
\end{equation*}
That is, all entering sellers are assigned state $\omega_0$.
For now, we allow $\eta_\theta(\omega)$ to be any measure on $\Omega$.

We model sellers as price takers and take demand as exogenously given.
Prices and demand depend only on the market beliefs about the average quality of a seller, conditional on the state.
Market beliefs about the average quality of a seller at time $t$ are denoted $\hat{\theta}_t(\omega) \in [\underline{\theta}, \overline{\theta}]$, conditional only on the publicly observable state variable.
Prices and demand are then denoted $p(\hat{\theta}_t(\omega))$ and $q(\hat{\theta}_t(\omega))$
Later, for purposes of estimation, we assume that prices and quantities are given by the specific functional form
\begin{align}
    p(\hat{\theta}_t(\omega)) & = \hat{\theta}_t(\omega), \label{eq:pricefunctional} \\
    q(\hat{\theta}_t(\omega)) & = \gamma_0 + \gamma_1 \hat{\theta}_t(\omega), \label{eq:quantityfunctional}
\end{align}
That is, we later assume that the price simply equals the market's beliefs about the quality of the seller, and that demand is a linear function of market beliefs, parameterized by $\beta_0, \beta_1$.
This is intended to capture, in a reduced-form way, some process through which prices are determined in which higher buyer beliefs about the quality of a seller results in higher prices to that seller.
Directly modeling such a process is complicated, and so, we abstract away from it.

The mass of sellers of quality $\theta$ in state $\omega$ at time $t$ is denoted $\mu_{t\theta}(\omega)$.
The age of a seller is denoted by $a \geq 0$, and a seller's history contains the history of seller states and cost shocks,
\begin{equation*}
  h^a = (\omega_1, c_1, \omega_2, c_2, \ldots, \omega_a, c_a).
\end{equation*}
The set of all seller histories is $\mathcal{H}$.
A seller's strategy, $\tau$, maps time, a seller's history, and his realized cost shocks into a probability of remaining in the market, denoted $(h^a) \overset{\tau_t}{\mapsto} [0, 1]$.

To represent transition probabilities in a compact way, we let $\vec{\tau}_{t\theta}, \vec{\mu}_{t\theta}, \vec{\eta}_{\theta}$ denote the $|\Omega|$-length vectors containing the exit probabilities, mass of sellers in each state, and mass of sellers entering into each state, that is
\begin{equation*}
    \vec{\tau}_{t\theta} = \langle \E_c[\tau_{t\theta}(\omega, c)] \rangle_{\omega \in \Omega}, \hspace{0.1in} \vec{\mu}_{t\theta} = \langle \mu_{t\theta}(\omega) \rangle_{\omega \in \Omega}, \hspace{0.1in} \text{ and } \vec{\eta}_{t\theta} = \langle \eta_{t\theta}(\omega) \rangle_{\omega \in \Omega}.
\end{equation*}

We may then write the distribution of sellers over states as satisfying the transition rules
\begin{equation}
    \vec{\mu}_{t+1\theta} = \Pi_\theta (\vec{\mu}_{t\theta} \circ \vec{\tau}_{t\theta}) + \vec{\eta}_{\theta},
\end{equation}
where $\circ$ denotes the pointwise product.\footnote{
  The \emph{pointwise product} of two vectors of length $n$, $\vec{x}^1, \vec{x}^2$, is the vector $\langle x_1^1 x_1^2, x_2^1 x_2^2, \ldots, x_n^1, x_n^2\rangle$.
}


If a seller exits, then he receives a payoff of zero for the remainder of the game.
If instead a seller remains in the game, then a seller in period $t$, in state $\omega_a$, with realized cost shock $c_a$ recieves a flow payoff of
\begin{equation*}
    u_t(h^a, \hat{\theta}_t) = p(\hat{\theta}_t(\omega_a)) q(\hat{\theta}_t(\omega_a)) - c_a.
\end{equation*}
Under the functional form assumptions~\eqref{eq:pricefunctional} and~\eqref{eq:quantityfunctional}, the flow payoff may be written
\begin{equation*}
  u_t(h^a, \hat{\theta}_t) = \hat{\theta}_t(\omega_a) (\gamma_0 + \gamma_1 \hat{\theta}_t(\omega_a)) - c_a.
\end{equation*}

A seller's continuation value is the discounted sum of his flow payoff, until exit.
The seller's entire payoff, given some exit strategy $\tau$, entrance at time $t$, starting state $\omega$, and market beliefs $\hat{\theta}$, may be written
\begin{equation*}
  U_t(\omega_0, \hat{\theta}, \tau) = \E\left[\sum_{a=0}^\infty \beta^a u_t(\tilde{h}^a, \hat{\theta}_{t+a}) \prod_{a'=0}^a \tau(\tilde{h}^{a'}) \mid \omega_0 \right],
\end{equation*}
where the expectation is taken with respect to the probability measure over all histories, beginning in state $\omega_0$, induced by $\Pi_\theta$ and the distribution of cost shocks; the random variable is denoted $\tilde{h}^a$.
The entire payoff of a seller is the discounted sum of his flow payoffs, $u_t(\tilde{h}^a, \hat{\theta}_{t+a})$, multiplied by the probability that the seller has remained in the market up to age $a$, $\prod_{a'=0}^a \tau(\tilde{h}^{a'})$.

In the remainder of the paper, we are mainly interested in stationary equilibria.
In a stationary equilibrium, market beliefs are independent of time, and seller strategies are independent of time and age.
To denote stationarity, we drop the dependence on time, that is, $\hat{\theta}_t(\omega) \equiv \hat{\theta}(\omega), \tau_t(h^a) \equiv \tau(\omega_a, c_a)\,\forall t, h^a$.
The corresponding stationary payoffs are then denoted $u(\omega, c, \hat{\theta})$, and $U(\omega, \hat{\theta})$, functions of the current state, $\omega$, the cost shock, $c$, and market beliefs, $\hat{\theta}$.
The transition rules under stationarity may be written simply as
\begin{equation}\label{eq:stationaritycondition}
    \vec{\mu}_{\theta}(\omega) = \Pi_\theta (\vec{\mu}_{\theta} \circ \vec{\tau}_\theta) + \vec{\eta}_{\theta}.
\end{equation}

\subsection{Equilibrium\label{sec:equilibrium}}

Our equilibrium notion requires that sellers be exiting optimally, and that the beliefs buyers form be consistent with behavior of sellers.
We focus on stationary equilibria, and so for compactness the general definition of equilibrium is omitted.
\begin{definition}
    An exit strategy and beliefs $\langle \tau, \hat{\theta} \rangle$ is a \emph{stationary equilibrium} iff $\tau$ solves the seller's problem,
    \begin{equation}\label{eq:equilibriumbr}
        \tau(\omega, c) \in \underset{\hat{\tau}}{\arg\max} \hspace{0.05in} U(\omega, \hat{\theta}, \hat{\tau}) \,\forall \omega \in \Omega
    \end{equation}
    and buyer beliefs are consistent with Bayes rule and equilibrium seller behavior, that is, $\mu$ satisfies the stationarity condition~\eqref{eq:stationaritycondition}, and
    \begin{equation}\label{eq:equilibriumbelief}
        \hat{\theta}(\omega) = \underline{\theta} + (\overline{\theta} - \underline{\theta})\frac{\mu_{\overline{\theta}}(\omega)}{\mu_{\underline{\theta}}(\omega) + \mu_{\overline{\theta}}(\omega)},
    \end{equation}
    where well-defined.\footnote{Assumptions~\ref{ass:connected} and~\ref{ass:costs} ensure that $\mu_{\underline{\theta}}(\omega) + \mu_{\overline{\theta}}(\omega) > 0$, so that~\eqref{eq:equilibriumbelief} is well-defined everywhere in equilibrium, so that the consistency requirement~\eqref{eq:equilibriumbelief} holds in all states.}
\end{definition}

We note finally that in a stationary equilibrium, in a standard result, the seller's payoff and strategy may be written recursively as a value function $V_\theta(\omega, c)$, and an exit decision, $\tau(\theta, c) \in [0, 1]$, functions of the current state, $\omega$, and the current realization of the cost shock, $c$, in the standard Bellman form,
\begin{equation}\label{eq:bellman}
    V_\theta(\omega, c) = \max_{\tau \in [0, 1]} u(\omega, c, \hat{\theta}) + \beta \E[V_\theta(\tilde{\omega}, \tilde{c}) \mid \omega],
\end{equation}
where $\tilde{\omega}, \tilde{c}$ denote the next-period draws of the state variable and the cost shock conditional on the current state.

\noindent We maintain the following assumptions for the remainder of the paper on the rating system, $\Pi$:

\begin{description}
  \item[A1\namedlabel{ass:connected}{A1}] (\emph{Irreducible and positive recurrent rating system}.) $\Pi_\theta$ is irreducible (that is, for every $\omega$, $\omega'$, there exists $n \in \mathbb{Z}$ so that $\Pi_\theta^{(n)}(\omega', \omega) > 0$), and $\Pi$ is positive recurrent, i.e.
  \begin{equation*}
    \sum_{n=1}^\infty n\cdot \Pi_\theta^{(n)}(\omega, \omega) < \infty, \forall \omega \in \Omega.
  \end{equation*}
  \item[A2\namedlabel{ass:costs}{A2}] (\emph{Unbounded costs}.) Seller costs satisfy
  \begin{equation*}
    0 < F(\underline{\theta}) < F(\overline{\theta}) < 1,
  \end{equation*}
  and the following technical assumption,
  \begin{equation}\label{eq:technicalcosts}
    \tfrac{d}{dc} \left(F(c)(c - \E(\tilde{c} \mid \tilde{c} \leq c))\right) > 0\,\forall c \in \mathbb{R}.
  \end{equation}
\end{description}

Together, assumptions~\ref{ass:connected} and~\ref{ass:costs} ensures that there is always a positive mass of sellers in every state (possibly very small) who never exit, and therefore that beliefs over seller types conditional on a state are well-defined for all states.



The value to the seller,~\eqref{eq:bellman}, of staying in the market is strictly decreasing in the cost shock, $c$, hence, in every equilibrium the exit decision is described by a cutoff strategy:

\begin{proposition}[Seller's optimal exit decision]
  Let $\langle \tau, \hat{\theta}\rangle$ be an equilibrium.
    Then $\tau$ is described (up to zero-probability events) by a cutoff strategy in $c$,
  \begin{equation*}
    \tau_\theta(\omega, c, \hat{\theta}) = \begin{cases} 0 & c < \underline{c}_\theta(\omega, \hat{\theta}) \\ 1 & c \geq \underline{c}_\theta(\omega, \hat{\theta})\end{cases},
  \end{equation*}
  where $\underline{c}_\theta(\omega)$ is the cutoff, satisfying
  \begin{equation}
    \underline{c}_\theta(\omega, \hat{\theta}) = \hat{\theta}(\omega) + \beta \E[V_\theta(\tilde{\omega}', \tilde{c}) \mid \omega].
    \label{eq:sellercutoff}
  \end{equation}
\end{proposition}
\begin{proof}
  Immediate from the Bellman equation~\ref{eq:bellman}.
    When $c = \underline{c}_\theta(\omega, \hat{\theta})$, the seller is indifferent between staying and exiting and the seller's optimization problem has multiple solutions; by assumption $F$ is continuous and so this is a zero probability event.
\end{proof}


In principle, we might compute the equilibrium value function by iterating on~\eqref{eq:bellman}.
However, it is useful to reduce the dimensionality of the problem and iterate on the expected value function instead, where we take expectation with respect to the cost shocks, $\tilde{c}$.
From~\eqref{eq:sellercutoff}, we can write the expected value function as
\begin{equation}
  \E[V_\theta(\omega, \tilde{c}, \hat{\theta})] = \underline{c}_\theta(\omega, \hat{\theta}) - \E[\tilde{c} \mid \tilde{c} \leq \underline{c}_\theta(\omega, \hat{\theta})].
\end{equation}
Together with~\eqref{eq:sellercutoff}, this yields
\begin{equation}\label{eq:expectedvalue}
  \underline{c}_\theta(\omega, \hat{\theta}) = \hat{\theta}(\omega) + \beta \E[\underline{c}_\theta(\tilde{\omega}, \hat{\theta}) - \E[\tilde{c} \mid \tilde{c} \leq \underline{c}_\theta(\tilde{\omega}, \hat{\theta})] \mid \theta, \omega].
\end{equation}
In the special case in which $\tilde{c} \sim U[0, 1]$ and $\Theta = \{0, 1\}$,~\eqref{eq:expectedvalue} may be written explicitly as
\begin{equation}\label{eq:expectedvaluesimple}
  \underline{c}_\theta(\omega) = \hat{\theta}(\omega) + \beta \sum_{\omega'} \frac{1}{2} \Pi_\theta(\omega, \omega') \underline{c}_\theta(\omega').
\end{equation}

A stationary equilibrium is therefore fully characterized by cutoff rules $\underline{c}$ and a distribution $\mu$, jointly satisfying the equilibrium conditions,~\eqref{eq:equilibriumbr} and~\eqref{eq:equilibriumbelief}.
In practice, we compute equilibria by iterating jointly on~\eqref{eq:expectedvalue} and~\eqref{eq:stationaritycondition} until convergence is reached.

Under Assumptions~\ref{ass:connected} and~\ref{ass:costs}, together with high discount rates, equilibria are unique, as summarized in the following proposition.\footnote{Equilibria are not, however, guaranteed to exist.}

\begin{restatable}{proposition}{propunique}\label{prop:unique}
  There exists $\overline{\beta} < 1$ such that, when $\beta \geq \overline{\beta}$, the model admits a unique equilibrium.
\end{restatable}
\begin{proof}
  {\sc tbc}.
\end{proof}

Proposition~\ref{prop:unique} establishes that, when the discount rate is sufficiently high, there is a unique equilibrium.
The proof technique, which applies a global inverse function theorem \citep{Gale65} to establish uniqueness, also yields a method for computing $\overline{\beta}$, the limit above which uniqueness is guaranteed.
Proposition~\ref{prop:unique} is a useful result empirically because generally, in models in which beliefs enter into payoffs as in~\eqref{eq:bellman}, there are multiple equilibria, complicating their empirical analysis.
In Appendix~\ref{app:illustration}, we illustrate computation of the model equilibrium with a simple, four-state example.


%% file: section/estimation.tex

\section{Estimation\label{sec:estimation}}

In this section, we discuss how we bring the model to the data.
When we estimate the model, we only consider the sale of Cannabis, the product category for which we have the most observations.
This leaves us with a total of $279,054$ reviews across $824$ vendors (for summary statistics see Table~\ref{tab:summary-stats}).
The restriction to one product category simplifies the analysis, albeit at the cost of ignoring reputation effects that span product categories.
A vendor who accumulates a reputation selling one product category, may benefit from this reputation when he sells a different product category.
However, for $730$ of the $824$ vendors in our data, Cannabis is the only product sold during the seller's lifetime.

As in Section~\ref{sec:data}, we use reviews left as a measure of sales made.
We aggregate our data into a weekly panel data set of vendors.
We normalize all prices by weight, which allows us to treat multiple listings of the same vendor as the same product.
We further normalize prices by removing a shipping location fixed effect.
This makes prices comparable across geographic regions, which we do not model explicitly.
For vendors with multiple sales per week, we use the end-of-week realizations of the number of sales and rating and the week's average realization of the price.

We estimate the model with maximum likelihood using a nested fixed point algorithm as in~\cite{Rust87}.
For the purpose of estimating the model, the cost shocks serve as the model's econometric error.
These shocks are observed by the vendor but unobserved by the econometrician.
A period in the model corresponds to a week.
For each vendor we observe how the vendor's ratings and sales evolve over time.
We also observe when a vendor decides to exit.
Importantly, we do not observe vendor types $\theta \in \{\underline{\theta}, \overline{\theta}\}$.
These are private information.
We model these types as finite mixtures as commonly done in single-agent dynamic discrete choice problems (e.g.\ in~\cite{Eckstein90, Eckstein99} or~\cite{Keane97}).

\subsection{Model Parameterization}\label{sec:parameterization}

To estimate the model, we impose some additional structure on parts of it. In particular, we parameterize the rating system and the demand process.
Each seller's publicly observable information, $\omega \equiv [r, s]$, consists of a rating, $r$, and the total number of sales made, $s$.
Ratings can take values in $\{0.00, 0.01, 0.02, \ldots, \ldots, 4.99, 5.00\}$, while sales can take values in $\{ 0-1, 1-5, 5-10, \ldots, 5000-\infty \}$.
We model the rating of seller $i$ at age $a$, $\tilde{r}_{ia}$, as evolving stochastically according to a Markov process dependent only on the total number of sales made, the seller's type, and the previous period's rating,
\begin{equation}
  \tilde{r}_{ia+1} = \xi r_{ia} + (1 - \xi) \rho_\theta + \varepsilon_{ia} \quad\text{with}\quad \varepsilon_{ia} \sim \N(0, \sigma_r),
  \label{eq:ratingprocess}
\end{equation}
which is governed by the parameters, $\xi$, $\langle \rho_\theta \rangle_{\theta \in \{\underline{\theta}, \overline{\theta}\}}$ and $\sigma_r$.
The parameter $\rho_\theta$ measures the `true' rating of a type to which a type $\theta$-seller's rating converges in the long-run.
The coefficient $\xi \in (0, 1)$ governs how slowly the rating adjusts.
When $\xi = 1$, the rating system is independent of a seller's type, and so is uninformative about type.
When $\xi = 0$, the rating of a seller is, on average, $\rho_\theta$.
When $\xi = 0$ and $\sigma_r$ is small, the rating almost entirely reveals the seller's type.
$\sigma_r > 0$ describes the standard deviation of the ratings process.
\footnote{
  We intend for this particular specification to capture the true underlying stochastic process, in which buyers purchase the product, sample it, and leave reviews, which are then averaged to form the rating.
  In an alternate specification, we model this stochastic process directly.
  However, this approach introduces computational difficulties, as well as some assumptions about the underlying process through which buyers leave reviews.
  To avoid this difficulty, we abstract away and use the specification~\eqref{eq:ratingprocess}.
  A more general specification of~\eqref{eq:ratingprocess} may allow the variance of the process to be time-varying, to capture the concept that as more reviews are left, each additional review will have less of an impact on the rating.
  On the other hand, as more reviews are left, sellers tend to receive more reviews, increasing the variance of the ratings process, and so it is not clear \emph{ex-ante} whether the variance of the ratings process should increase or decrease as more sales are made.
  In an alternate specification of the model, we allow the variance of the rating process to be $\sigma_r \lambda^{s_{ia}}$, where $\lambda > 0$ is a parameter capturing a time-varying variance of the ratings process, but in practice, we estimate $\lambda$ very close to 1, and so omit it here for simplicity.
}
The process is then discretized using the technique of~\cite{Tauchen86}.

Sales are assumed to evolve exogenously and independently of a seller's type and rating.
Specifically, with probability $\gamma$, sellers graduate from one sales level to the next.
\footnote{
  More generally, we might wish to allow the sales made to depend on market beliefs about the seller.
  Our specification of demand fits the data well.
  The model, when estimated with the belief-dependent demand curve, generally fails to find evidence of any relationship between buyer beliefs and total sales.
  Since price is assumed to already capture the expected value of a seller to a buyer, it may be that any variations in sales due to buyer beliefs are of second-order importance.
  Regardless, even if more favorable beliefs were to lead to higher sales, the effect on reputation is then of third-order importance in the model, and this is the effect we are interested in.
}
We denote the resulting transition probability matrices that characterize the joint evolution of publicly observable information $\omega$ by $\pi_{\overline \theta}(\omega, \omega')$ and $\pi_{\underline \theta}(\omega, \omega')$.
We parameterize the initial distribution of publicly observable information for new sellers by $\eta(\omega)$, which we assume does not depend on the seller's quality $\theta$.
Costs shocks are assumed to be normally distributed with mean $\mu_c$ and standard deviation $\sigma_c$.
The distribution of cost shocks is assumed to be independent of the seller's type, $\theta$.

In the model, all vendors with the same publicly observable state, $\omega$, receive the same price.
In the data, this is clearly not true.
To ensure that the model is statistically non-degenerate, we add identically and independently distributed measurement error to the price.
We assume that the measurement error follows a log-normal distribution with the location parameter fixed at zero and scale $\sigma_p$.
The log-normal specification both excludes non-positive prices, and as well, fits the true distribution of prices (conditional on state) well.
This measurement error is independent of the seller's quality $\theta$.
We denote the density of the measurement error by $\phi$.

\subsection{Likelihood}\label{sec:likelihood}

We now construct the likelihood function.
We observe each vendor $i$ for $A$ periods.
$A$ is not the same for all vendors, because different vendors exit the market at different ages.
For each vendor we observe a sequence of the publicly observable state variable, $\omega_{i1}, \ldots, \omega_{iA}$, a sequence of prices, $p_{i1}, \ldots, p_{iA}$, and whether the vendor exits in the last period $d_{iA}$.
Because of the sudden closure of the market place, we we do not observe the exit decision for all vendors.

First, consider a vendor that we see exiting the market at time $A$, i.e.\ $d_{iA}=1$.
The likelihood contribution for this vendor conditional on $\theta$ is given by
\begin{align*}
p(\omega_{i1}, \ldots,& \omega_{iA}, d_{iA} = 1 | \theta) = \quad\quad\quad\quad\quad\quad\quad\quad\quad\quad\quad & \\
& \eta(\omega_{i1}) \P(\tilde{c} < \underline c_\theta(\omega_{i1})) \phi(p_{i1} - \mathbb E_\mu[\theta | \omega_{i1}])  & \text{\emph{initial period}} \\
&\prod_{a = 2}^{A-1} \left[ \pi_\theta(\omega_{ia-1}, \omega_{ia}) \P(\tilde{c} < \underline c_\theta(\omega_{ia})) \right]\phi(p_{ia} - \mathbb E_\mu[\theta | \omega_{ia}])  & \text{\emph{intervening periods}} \\
&\pi_\theta(\omega_{iA-1}, \omega_{iA}) \P(\tilde{c} \geq \underline c_{\theta}(\omega_{iA}))  & \text{\emph{exit period}}
\end{align*}
Next, consider a vendor that we observe up until time $T$ without exiting the market ($d_{iA}=0$).
The likelihood contribution for such a vendor equals
\begin{align*}
p(\omega_{i1}, \ldots,& \omega_{iA}, d_{iA} = 0 | \theta) = \quad\quad\quad\quad\quad\quad\quad\quad\quad\quad\quad & \\
& \eta(\omega_{i1}) \P(\tilde{c} < \underline c_\theta(\omega_{i1})) \phi(p_{i1} - \mathbb E_\mu[\theta | \omega_{i1}])  & \text{\emph{initial period}} \\
&\prod_{a = 2}^{A} \left[ \pi_\theta(\omega_{ia-1}, \omega_{ia}) \P(\tilde{c} < \underline c_\theta(\omega_{ia})) \right]\phi(p_{ia} - \mathbb E_\mu[\theta | \omega_{ia}])  & \text{\emph{all other periods}}
\end{align*}
The entire log-likelihood contribution for a vendor is therefore given by
\begin{equation*}
\ell(\omega_{i1}, \ldots, \omega_{iA}, d_{iA}) = \log\left[\alpha p(\omega_{i1}, \ldots, \omega_{iA}, d_{iA} | \overline \theta)
+ (1-\alpha) p(\omega_{i1}, \ldots, \omega_{iA}, d_{iA} | \underline \theta)\right],
\end{equation*}
where we integrate out the vendor's type $\theta \in \{ \overline \theta, \underline \theta \}$ using the weights $\alpha$ and $1-\alpha$.
The log-likelihood function is then given by
\begin{equation*}
\sum_{i=1}^N \ell(\omega_{i1}, \ldots, \omega_{iA}, d_{iA}).
\end{equation*}
We do not face the traditional initial conditions problem in the dynamic discrete choice literature, because we assumed that the initial distribution of the publicly observable information $\omega_{i1}$ is independent of the private vendor type.
Also we only use vendors in the estimation for which we observe the initial sale in the data.
We thereby bypass issues resulting from left-censoring.

\subsection{Identification}\label{sec:identification}

In our model, a firm's only decision is when to leave the market platform.
Therefore, our model is essentially an optimal stopping problem with permanent unobserved heterogeneity.
Optimal stopping problems and their identification properties have been extensively studied in the literature (see~\cite{Abbring10} for a review).
Obtaining identification in this class of models is notoriously difficult when allowing for dynamic selection on unobservables (as we do in our case).
Our model differs from the traditional optimal stopping problem in the dynamic discrete choice literature in several important aspects.
We directly observe the evolution of the state variable, which we assume to be type-dependent.
Following~\cite{Kasahara09}, this allows us to identify the state-dependent type probabilities, as well as type-dependent transition probabilities between states.
We directly observe prices, $p(\hat{\theta}(\omega)) = \hat{\theta}(\omega)$, and the total mass of sellers in each state, $\mu_{\underline{\theta}}(\omega) + \mu_{\overline{\theta}}(\omega)$.

For another, we directly observe prices and a proxy for sales, which means that we observe vendors' flow payoffs.
Ideally, this data would identify the distribution of cost shocks, parameterized by $\mu_c$ and $\sigma_c$.
However, as in most dynamic discrete choice models, we cannot separately identify the location and scale of this distribution.
We therefore fix the scale $\sigma_c$ at 1, and only estimate the location $\mu_c$.

\subsection{Point Estimates\label{sec:estimates}}

We estimate ten model parameters.
The type-dependent parameters of the model are $\langle \theta, \rho_\theta \rangle_{\theta \in \{\underline{\theta}, \overline{\theta}\}}$.
The type-independent parameters are $\langle \alpha, \mu_c, \gamma, \xi, \sigma_r, \sigma_p \rangle$.
We do not attempt to estimate the discount factor $\beta$, which we fix to correspond to an annual interest rate of approximately $25\%$.
\footnote{
  One period in the model corresponds to a week.
  Little research has been done to ascertain the discount factor of drug dealers in online marketplaces.
  We assume that they are impatient for a variety of reasons.
  They presumably face impatient suppliers, have little access to formal credit, and furthermore face risks associated with law enforcement and the sudden shutdown of the marketplace.
}
We estimate the model using prices residual of seller observables, such as shipping location and the region a seller ships to.
Prices are then normalized so that they lie on the interval $[0, 1]$.

We estimate the model using a nested-fixed point algorithm, following a procedure as in~\cite{Rust87}.
We compute asymptotic standard errors using the outer-product of the gradients of the log likelihood to estimate the asymptotic covariance matrix.
The model estimates and standard errors are reported in Table~\ref{tab:estimationresults}.

The share of entering high types, $\alpha$, is estimated at $0.233$.
This estimate indicates that the market may indeed be prone to adverse selection, because there is a non-trivial share of high and low types.
The estimates of $\underline \theta$ and $\overline \theta$ indicate that the quality difference between low types and high types is substantial.
The low type value, $\underline \theta$, is estimated at $0.3$ and the value of the high type over that of the low type, $\overline \theta - \underline \theta$, is estimated at $0.226$.
Since we normalized prices to take on values on the unit interval, these estimates correspond to $10.5$ dollars per gram and $18.4$ dollars per gram, respectively.
(Compare to an average price of cannabis in the market is $12.11$ dollars per gram, as shown in Table~\ref{tab:summary-stats}.)
The probability of moving up in the sales ladder each week is estimated at $0.293$.
This corresponds to an average sales rate for young sellers (fewer than 80 sales) of approximately 23 sales per week, and for large, mature sellers (more than 10000 sales) of approximately 556 sales per week.

\begin{figure}[p]
  \centerline{
    \input{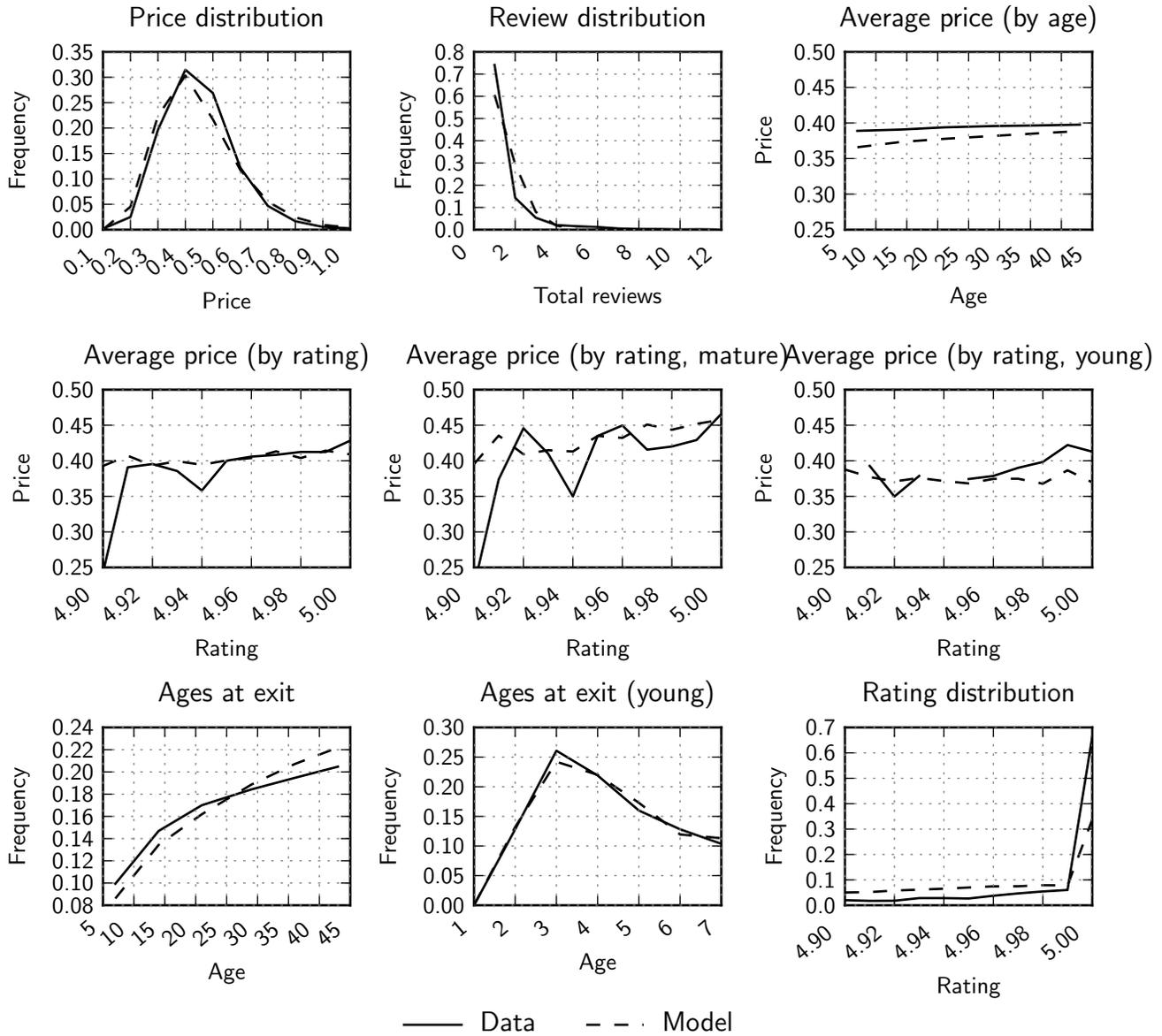}
  }
  \caption{Model fit.\label{fig:modelfit}}
  \floatfoot{\textbf{Note:} Model fit graphs produced by simulation. Mature sellers are defined as those for whom there are more than 1000 reviews. Young sellers are those for whom there are fewer than 300 reviews. For ages at exit among young sellers (bottom, center), the age at exit among sellers who had been in the market fewer than 50 days is plotted. Omitted in the plots of age at exit (bottom left, bottom center) are sellers who exited the market when it closed.}
\end{figure}

In Figure~\ref{fig:modelfit} we simulate data using the estimated model and compare the moments of the model to the moments of the data.
Overall, the model fits the data well.

\begin{table}[!tbhp]
  \centerline{\small
    \input{figures/estimates.tex}
  }
  \caption{Parameter estimates.\label{tab:estimationresults}}
\end{table}

\begin{figure}[p]
  \centerline{
    \resizebox{5.5in}{!}{\input{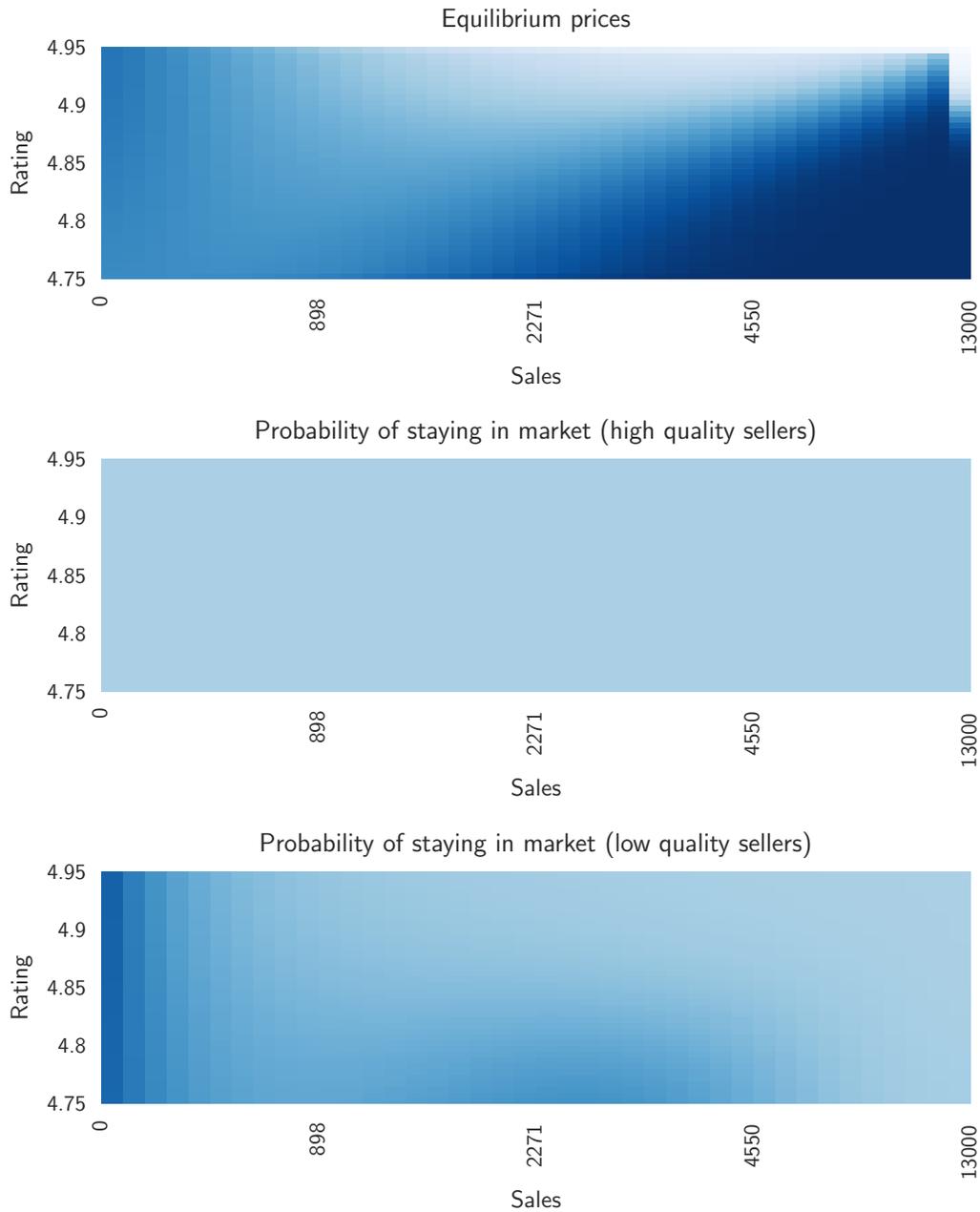}}
  }
  \caption{Prices and exit probabilities, in equilibrium, at estimated model parameters.\label{fig:etheta}}
  \floatfoot{
    \textbf{Note:}
    \textbf{Top}, price received in equilibrium, as a function of the rating (y-axis) and the sales (x-axis).
    Lighter is higher.
    \textbf{Middle}, the probability of remaining in the market, as a function of the rating and sales, for high quality sellers.
    \textbf{Bottom}, the probability of remaining in the market, as a function of the rating and sales, for low quality sellers.
    High quality sellers are more likely to stay in the market than low quality sellers.
    The gradient between price and rating is increasing in the number of sales made.
  }
\end{figure}

In Figure~\ref{fig:etheta} we plot the equilibrium price and exit probabilities, at the estimated parameters, as a function of the rating, the number of sales made, and the quality of a seller.
The top panel shows prices.
Prices are highest for sellers with high rating and large number of sales and lowest for sellers with low rating and a large number of sales.
For seller with few sales, prices are approximately independent of the seller's rating.
In the bottom two panels, we show the probability of staying in the market separately for high and low quality sellers. Overall, high quality sellers are much more likely to stay in the market than low quality sellers.
High quality sellers are most likely to leave when their rating is low and they have made a moderate number of sales (approximately between 1,000 and 2,000 sales).
A low quality seller's probability of leaving the market is increasing in sales made when the rating is low and decreasing in sales when the rating is high.

%% file: figures/estimates.tex
%
\begin{tabular}{llrr}
  \toprule
Parameter & Variable & Estimate & Std. error \\
  \midrule
Low type value & $\underline{\theta}$ & 0.300 & 0.001 \\
  High type value & $\overline{\theta}$ & 0.525 & 0.004 \\
  Fraction of entering high type & $\alpha$ & 0.233 & 0.006 \\
  Mean of cost shocks & $\mu_c$ & 0.386 & 0.001 \\
  Probability of making sale & $\gamma$ & 0.293 & 0.002 \\
  Average low type rating & $\rho_{\underline{\theta}}$ & 5.010 & 0.007 \\
  Average high type rating & $\rho_{\overline{\theta}}$ & 6.372 & 0.035 \\
  Rate of rating adjustment & $\xi$ & 0.060 & 0.001 \\
  Standard deviation of rating process & $\sigma_r$ & 0.037 & 0.001 \\
  Standard deviation of prices & $\sigma_p$ & 0.144 & 0.001 \\
\end{tabular}

%% file: section/results.tex

\section{Results\label{sec:results}}

We apply our estimated model to produce estimates of sales, profits, and life-span, conditional on quality.
As well, we simulate the data at different parameter values to perform counterfactual exercises.
Our main finding is that, in the absence of a rating system, the market would have collapsed due to adverse selection.
Dollar values in this section are stated for parameter estimates produced by estimating the model on cannabis sellers.

\subsection[Composition of Types]{Composition of Types in the Market Place}

In our estimated model, high and low type sellers differ substantially.
We find that an average entering high quality seller expected to make \$$50,011$ in total profits in the course of his lifetime.
Meanwhile, a low quality seller expected to make only \$$4292$.
The difference in profits is due to two features: First, high quality sellers remained in the market for a longer period of time, second, high quality sellers received a higher price, on average, due to the rating system.

\begin{figure}[p]
  \centerline{
    \input{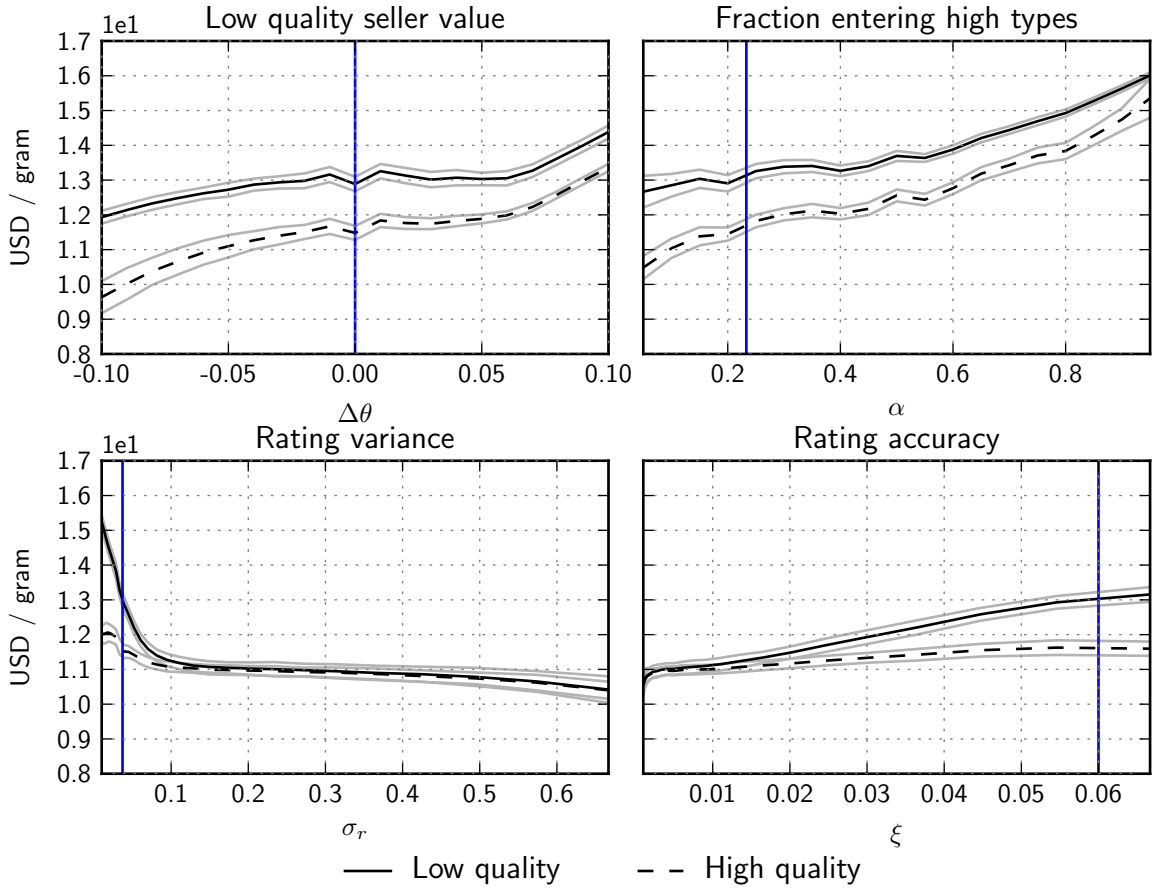}
  }
  \caption{Average price, for each type.\label{fig:pricebytype}}
  \floatfoot{
    \textbf{Note:}
    Vertical line represents estimated value.
    Values produced by simulation with $N = 500$.
    Gray lines $\pm$ two standard deviations of simulated values.
    \textbf{Top left}, the average price charged is increasing in the quality of the low type seller, since the average quality of seller in the market is higher.
    \textbf{Top right}, through a similar channel, the average price charged is increasing in the fraction of entering sellers who are high types.
    \textbf{Bottom left}, as the variance of the rating process increases, the average prices charged by high and low type decreases, since more low types remain in the market.
    Eventually, the two prices converge together, and they converge to the value of the low type seller, around \$10, since high type sellers prefer to exit at high values of $\sigma_r$ so that only low type sellers remain in the market.
    \textbf{Bottom right}, through a similar channel, the average price charged is increasing for both seller types, and diverging, as the accuracy of the rating process increases.
  }
\end{figure}

\begin{figure}[p]
  \centerline{
    \input{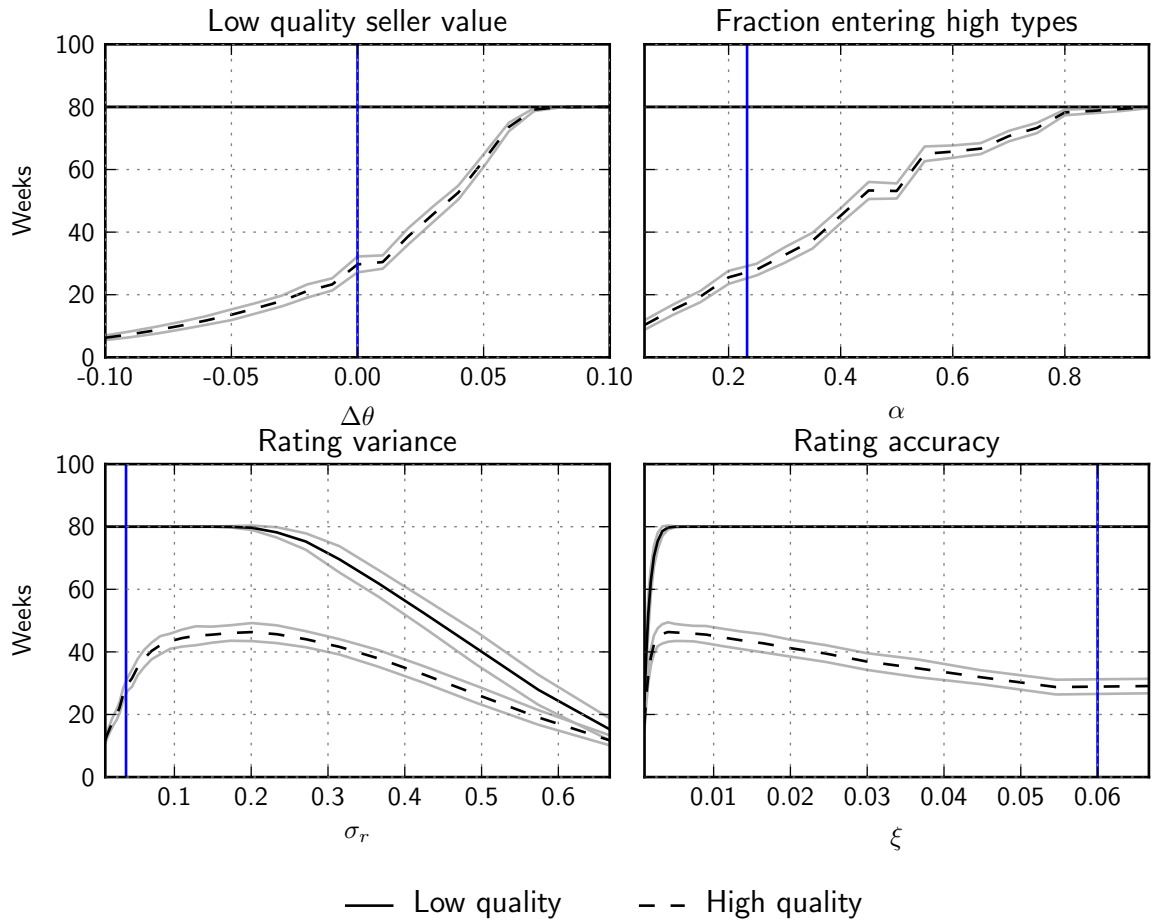}
  }
  \caption{Average age at time of death, for each type.\label{fig:ageofdeath}}
  \floatfoot{
    \textbf{Note:}
    Vertical line represents estimated value.
    Values produced by simulation with $N = 500$.
    Gray lines $\pm$ two standard deviations of simulated values.
    \textbf{Top left}, high quality sellers remain in the market until market closure, at 80 weeks.
    As the quality of the low types improves, they survive for longer periods of time.
    \textbf{Top right}, as more high types enter the market, low types survive for longer because the market is more lucrative.
    \textbf{Bottom left}, as the variance of the ratings process increases, corresponding to a decrease in informativeness of the rating, low types initially survive for longer, because they are better able to hide their quality.
    Past some point, however, high types begin to exit, since they are unable to demonstrate their high quality, which makes the market less lucrative, and both types exit sooner.
    \textbf{Bottom right}, as the accuracy of the ratings process increases, a similar channel results in first an increase then a decrease in the average age of low type sellers.
  }
\end{figure}

\begin{figure}[p]
  \centerline{
    \input{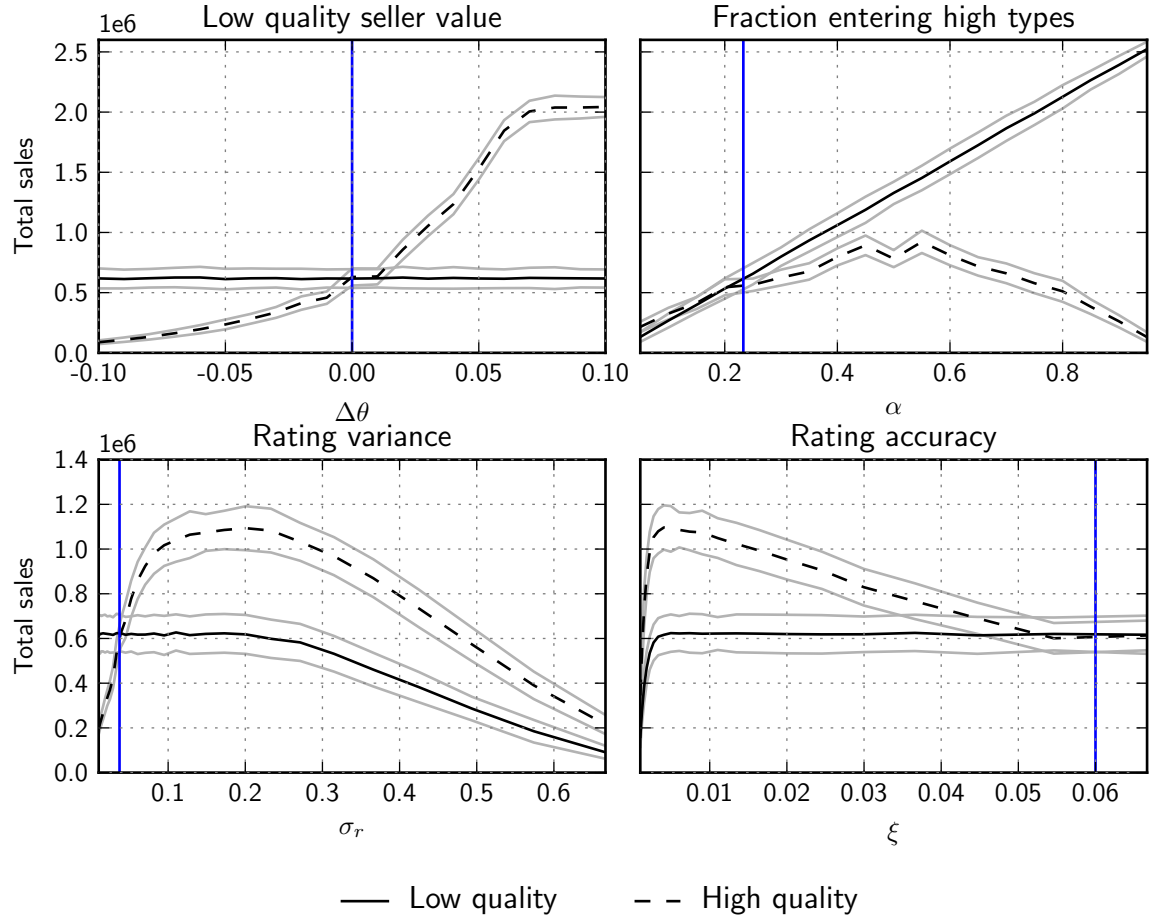}
  }
  \caption{Average sales in the market, for each type.\label{fig:salesbytype}}
  \floatfoot{
    \textbf{Note:}
    Vertical line represents estimated value.
    Values produced by simulation with $N = 500$.
    Gray lines $\pm$ two standard deviations of simulated values.
    \textbf{Top left}, as the quality of low type sellers increases, they make more sales since they remain in the market for longer.
    \textbf{Top right}, as the fraction of entering high type sellers increases, the number of low type sales first increases, since more low type sellers remain in the market, then decreases, since there are fewer low type sellers.
    \textbf{Bottom left}, as the variance of the rating process increases, the number of low type sales first increases, since low type sellers are less likely to exit, then decreases, as the market becomes less lucrative so that both seller types exit.
    \textbf{Bottom right}, through a similar channel, as the accuracy of the rating process increases, the number of low type sales first increases then decreases.
  }
\end{figure}

\begin{figure}[p]
  \centerline{
    \input{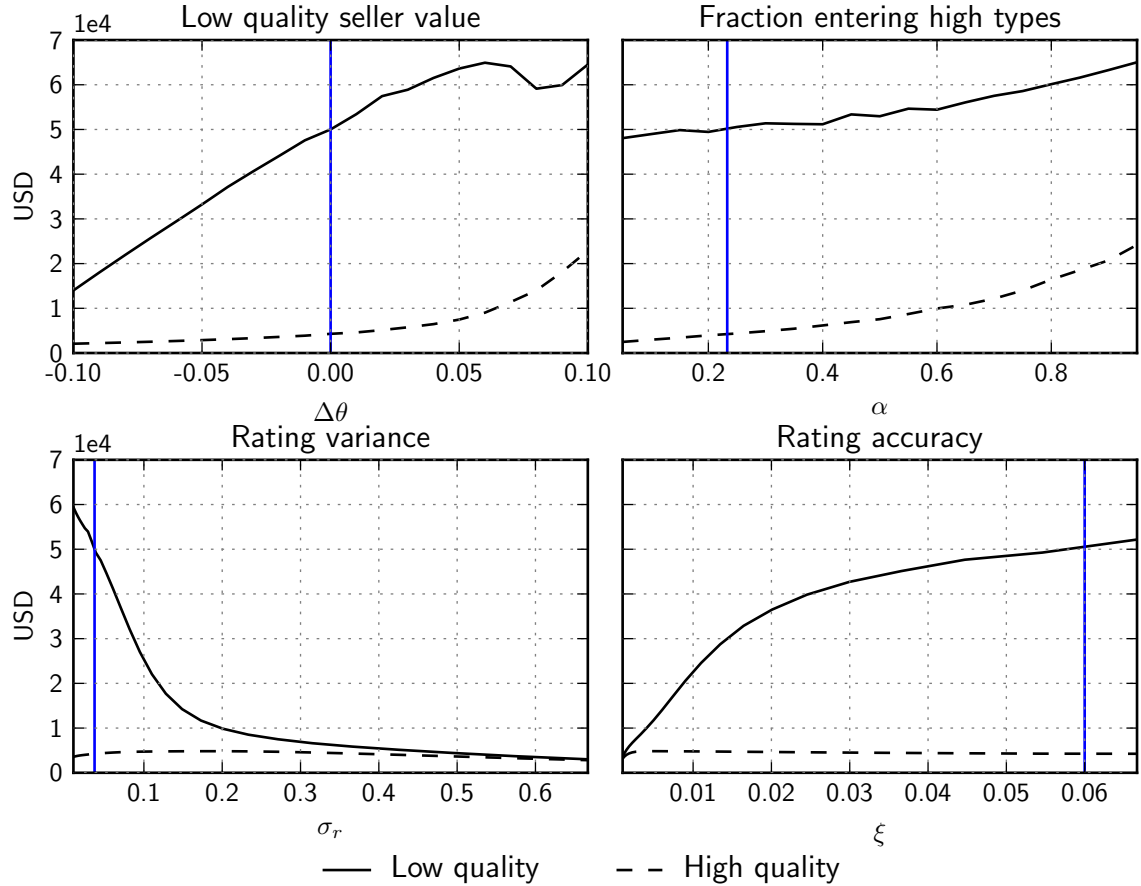}
  }
  \caption{Expected profits upon entry, for each type.\label{fig:expectedprofits}}
  \floatfoot{
    \textbf{Note:}
    Vertical line represents estimated value.
    Values produced by equilibrium computation.
    \textbf{Top left}, as the quality of the low type seller increases, both high and low type sellers see increases in profit.
    The marginal profit of the high type seller is higher, because they remain in the market for longer.
    \textbf{Top right}, as the fraction of high type sellers increases, both high and low type sellers see increases in profit, since the market is now more lucrative due to higher prices.
    \textbf{Bottom left}, as the variance of the rating process increases, the profits of high type sellers declines, while low type seller's increases then decreases.
    \textbf{Bottom right}, as the accuracy of the rating process increases, a similar dynamic results in first an increase, than a decrease of low type seller profits.
  }
\end{figure}


\begin{figure}[p]
  \centerline{
    \input{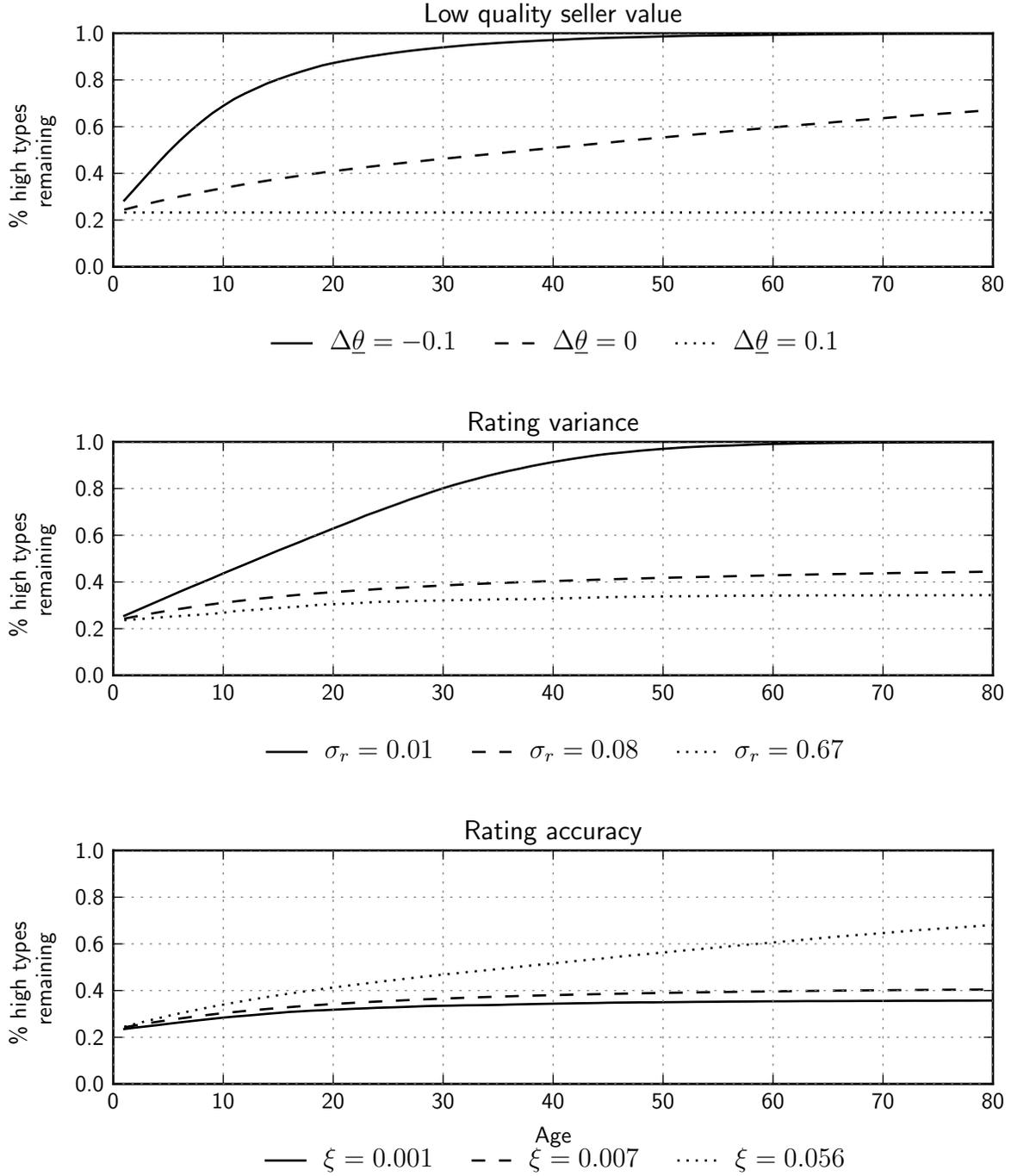}
  }
  \caption{Percentage high types remaining in the market, by age of seller.\label{fig:hightypes}}
  \floatfoot{
    \textbf{Note:}
    Vertical line represents estimated value.
    Values produced by simulation with $N = 500$.
    \textbf{Top}, when low type sellers are very low quality (solid line), mature sellers are very likely to be high quality sellers.
    When low type sellers are high quality (dotted line), the fraction of high quality sellers remains unchanged with time.
    \textbf{Middle}, when the variance of the rating process is low (solid line), mature sellers are very likely to be high quality sellers.
    When the variance of the ratings process is high (dotted line), the fraction of mature sellers does not asymptote to \%100.
    \textbf{Bottom}, when the accuracy of the rating process is high (dotted line), mature sellers are more likely to be high quality sellers.
    When the accuracy of the rating process is low (solid line), the long-run quality of the sellers in the market is lower.
  }
\end{figure}

\subsection[Returns to Reputation]{Returns to Reputation Implied by the Model}

Using the estimated model, it is straightforward to compute the returns to reputation that it implies.
Following previous literature, we ask how much of a loss in profits a seller whose rating falls suffers.
In our model, this effect is heterogeneous and depends on both the seller's private type and the seller's observable state.
We compute the expected loss in net present value of profits to a high type and low type seller, respectively, whose rating falls from $5.0$ to $4.99$, as a function of sales made.
First, consider a high type seller. A young seller making sales of 20 grams of cannabis per order, expects to lose \$150 from such a fall in ratings (or approximately 2\% in total expected discounted profits), while a mature seller with more than 10,000 sales is unaffected by such a change in rating.
In contrast, a young, low type seller only expects to lose \$350 in expected profits from such a fall, which corresponds to about 11\% of total expected discounted profits.

\subsection{Counterfactuals}

Two classes of counterfactuals are of interest in this environment.
First, we may ask what the effect of changing the rating system is.
We consider what happens if the ratings are made coarser or finer, and what happens if seller's ratings move more or less quickly to their long-run value.
(See the bottom left and right in Figures~\ref{fig:pricebytype},~\ref{fig:ageofdeath},~\ref{fig:salesbytype},~\ref{fig:expectedprofits}, and~\ref{fig:hightypes}.)
At the coarsest possible rating system, no information is communicated to buyers about the seller type, because there is only one possible rating.
We find that in this case, the market shuts down entirely, meaning that all sellers exit as soon as possible.
At estimated values, reducing the quality of the rating system can, perversely, increase the number of sales, since low-type sellers are less likely to exit.
(See Figure~\ref{fig:salesbytype}.)
This finding suggests that the market platform may have faced competitive pressures to provide more information to buyers.

Second, we may ask about the effect of changing the proportion of entering types who are low quality sellers, as well as the effect of reducing the quality of low-quality sellers.
(See the top left and right in Figures~\ref{fig:pricebytype},~\ref{fig:ageofdeath},~\ref{fig:salesbytype},~\ref{fig:expectedprofits}, and~\ref{fig:hightypes}.)
Such an exercise may be of interest to questions of how to undermine these market platforms by introducing `honeypot' sellers, false sellers whose only purpose is to identify buyers.

%% file: section/conclusion.tex

\section{Conclusion\label{sec:conclusion}}

This paper describes the collection of a large and novel dataset of price and reputation in a black market.
We replicate prior reduced form work on the relationship between price, exit, and reputation, with an emphasis on the dynamic incentives faced by firms in the market.
We show that the relationship between price and ratings is positive and increasing in the number of sales made, consistent with a model of adverse selection and consumers forming rational inferences about the quality of the product jointly from the rating and the number of sales made.
The results of our model suggest that adverse selection may be alleviated dynamically in only marketplaces, that a static analysis of regression will result in estimates of the return to reputation which are downward biased due to a composition bias, that these market platforms may face competitive pressures to provide information about the quality of sellers and that a monopolist market platform may optimally decide to reduce the quality of information about sellers.
We produce counterfactual estimates of the effects of changing the composition of sellers in the marketplace.
Our results suggest that the market platform could have increased total sales by reducing the amount of information provided about seller quality to buyers, implying that these markets face competitive pressures to inform buyers about seller quality.
We show that the market would not have functioned in the absence of a rating system, due to adverse selection.

%% file: section/simplemodel.tex
\subsection{Illustration of Model Equilibria\label{sec:illustration}}

To illustrate some dynamics of the model, we consider a special case in which the state space consists of two possible ratings, elements of $\{L, H\}$, and two possible sales levels, elements of $\{1, 2\}$, and there are two seller qualities, $\underline{\theta} = 0$ and $\overline{\theta} = 1$.
There are therefore four possible states, $\{L1, H1, L2, H2\}$.
A seller of quality $\theta$ with sales level $s$ transitions between ratings according to the transition matrices
\begin{equation}\label{eq:sptran}
  \Pi_\theta = \left(\begin{array}{cc} 1 - (1 - \tfrac{s}{2})(1 - \gamma_\theta) & (1 - \tfrac{s}{2})(1 - \gamma_\theta) \\ 1 - (1 - \tfrac{s}{2})(1 - \gamma_\theta) & (1 - \tfrac{s}{2})(1 - \gamma_\theta) \end{array}\right),
\end{equation}
where $\gamma_\theta \in (0, 1)$ parameterizes the quality of the rating system.
So that a high rating is associated with a high type, we assume $\gamma_0 > \tfrac{1}{2} > \gamma_1$.
When $s = 0$, the transition matrix is simply
\begin{equation*}
  \Pi_\theta = \left(\begin{array}{cc} \gamma_\theta & 1 - \gamma_\theta \\ \gamma_\theta & 1 - \gamma_\theta \end{array}\right),
\end{equation*}
when $s = 1$, it is
\begin{equation*}
  \Pi_\theta = \left(\begin{array}{cc} 1 - \frac{1 - \gamma_\theta}{2} & \frac{1 - \gamma_\theta}{2} \\ \frac{\gamma_\theta}{2} & 1 - \frac{1 - \gamma_\theta}{2} \end{array}\right).
\end{equation*}
With some exogenous probability $\rho \in (0, 1)$, sellers have a chance of transitioning from the low sales rate, $s = 0$, to the high sales rate, $s = 1$.
Once a seller reaches a high sales rate, they cannot transition back to a low sales rate.
This parameterization of the model is one in which low quality sellers are more likely than the high quality sellers to transition to the low rating, but when $s = 1$, they are less likely to transition between ratings, capturing, in a reduced form way, the idea that a seller who has already made many sales needs more bad reviews to shift their average rating.

\begin{figure}[p]
    \centerline{
        \input{figures/simple_model_prices.pgf}
    }
    \caption{The price in each state, as function of the precision of the rating system, $\gamma$.\label{fig:simpleprices}}
    \floatfoot{\textbf{Note:} Left: The price, as a function of each of the possible four states, solid ($L1$, low rating, low sales), dashed ($H1$, high rating, low sales), dotted ($L2$, low rating, high sales), and dot-dashed ($H2$, high rating, high sales).
         As the rating system becomes more accurate, the prices received by low rated and highly-rated sellers diverges.
         Right: The relationship between rating and price (dotted) is increasing in the precision of the rating system, $\gamma$.
         Controlling for sales reveals evidence of a composition effect---sellers with higher sales see a larger relationship between price and rating.}

    \centerline{
        \input{figures/simple_model_masses.pgf}
    }
    \caption{Mass of sellers in each state, low rated, low sales (solid), high rated, low sales (dashed), low rated, high sales (dotted), and high rated, high sales (dash-dotted).\label{fig:simplemass}}
    \floatfoot{\textbf{Note:} The figures show how the mass of types varies (due to exit) as a function of the precision of the rating system, $\gamma$.
    The left figure shows the mass of high types, the right figure the mass of low types.}

\end{figure}

\begin{figure}[p]
    \centerline{
    \input{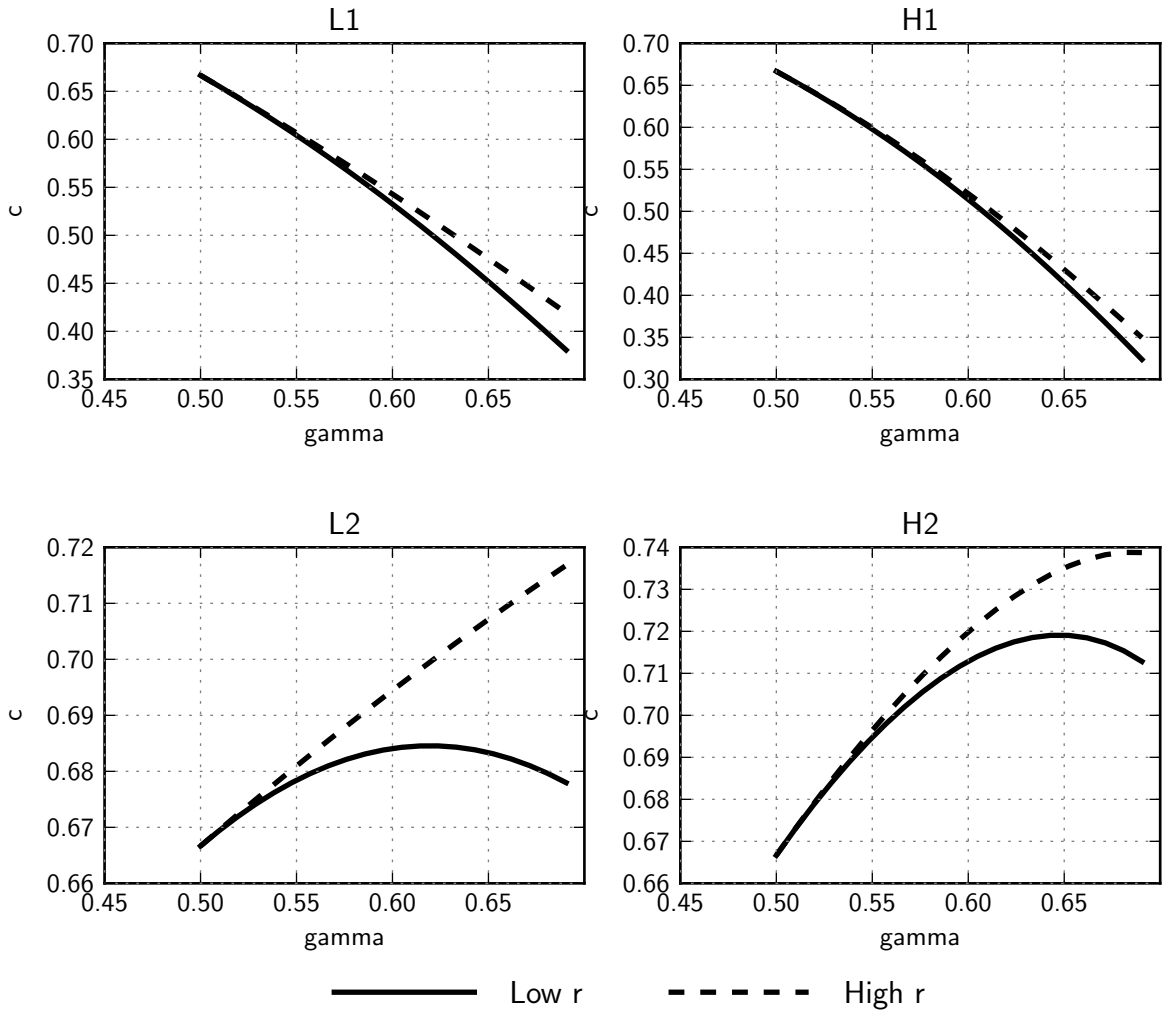}}

    \caption{The cost shock cutoff in each of the four possible states, for low quality sellers (solid), and high quality sellers (dashed), as a function of the quality of the rating system.\label{fig:simpleexit}}

    \floatfoot{\textbf{Note:} The figures show the cost shock cutoff $\underline{c}$ as a function of the precision of the rating system, $\gamma$.
    Solid lines refer to high quality sellers, dashed lines to low quality sellers.
    Each figure represents each of the four possible states, $L1$ (low rating, low sales), $H1$  (high rating, low sales), $L2$  (low rating, high sales), and $H2$  (high rating, high sales).}

\end{figure}

What is the equilibrium? Let $\underline{c}_{\omega_r\omega_s\theta}$ represent the exit cutoff for type $\theta$ in state $(\omega_r, \omega_s)$, and let $\mu_{\omega_r\omega_s\theta}$ denote the measure of type $\theta$ in state $(\omega_r, \omega_s)$.
The stationarity conditions may be written explicitly as
\begin{align}
  \overbrace{\mu_{H1\theta}\left((1-c_{H1\theta}) + c_{H1\theta}\left(\frac{1-\gamma_{1-\theta}}{2} + \rho\right)\right)}^{\text{Mass exiting}} &= \overbrace{\tfrac{1}{2} + \mu_{L1\theta}c_{L1\theta}\left(\frac{1-\gamma_\theta}{2}\right)}^{\text{Mass entering}} \tag{$H1\theta$} \\
  \mu_{L1\theta} \left((1-c_{L1\theta}) + c_{L1\theta}\left(\frac{1-\gamma_\theta}{2} + \rho\right)\right) &= \mu_{H1\theta} c_{H1\theta} \left(\frac{1 - \gamma_{1-\theta}}{2}\right) \tag{$L1\theta$}\\
  \mu_{H2\theta}\left((1-c_{H2\theta}) + c_{H2\theta} \left(\frac{1-\gamma_{1-\theta}}{4}\right)\right) &= \mu_{H1\theta}c_{H1\theta}\rho + \mu_{L2\theta}c_{L2\theta}\left(\frac{1-\gamma_\theta}{4}\right) \tag{$H2\theta$} \\
  \mu_{L2\theta}\left((1 - c_{L2\theta}) + c_{L2\theta}\left(\frac{1-\gamma_\theta}{4}\right)\right) &= \mu_{L1\theta}c_{L1\theta}\rho + \mu_{H2\theta}c_{H2\theta}\left(\frac{1-\gamma_{1-\theta}}{4}\right) \tag{$L2\theta$}
\end{align}
The corresponding optimality conditions, from~\eqref{eq:expectedvaluesimple}, are
\begin{align}
  \underline{c}_{H1\theta} & = \frac{\mu_{H11}}{\mu_{H11} + \mu_{H10}} + \frac{\beta}{2}\left(p \underline{c}_{H2\theta} + \left(\frac{1 - \gamma_{1-\theta}}{2}\right)\underline{c}_{L1\theta} + \left(1 - p - \frac{1 - \gamma_{1-\theta}}{2}\right)\underline{c}_{H1\theta}\right) \tag{$cH1\theta$} \\
  \underline{c}_{L1\theta} & = \frac{\mu_{L11}}{\mu_{L11} + \mu_{L10}} + \frac{\beta}{2}\left(p \underline{c}_{L2\theta} + \left(\frac{1 - \gamma_\theta}{2}\right)\underline{c}_{H1\theta} + \left(1 - p - \frac{1 - \gamma_\theta}{2}\right)\underline{c}_{L1\theta}\right) \tag{$cL1\theta$} \\
  \underline{c}_{H2\theta} & = \frac{\mu_{H21}}{\mu_{H21} + \mu_{H20}} + \frac{\beta}{2} \left(\left(\frac{1 - \gamma_{1-\theta}}{4}\right)\underline{c}_{L2\theta} + \left(1 - \frac{1 - \gamma_{1-\theta}}{4}\right)\underline{c}_{H2\theta}\right) \tag{$cH2\theta$} \\
  \underline{c}_{L2\theta} & = \frac{\mu_{L21}}{\mu_{L21} + \mu_{L20}} + \frac{\beta}{2}\left(\left(\frac{1 - \gamma_\theta}{4}\right)\underline{c}_{H2\theta} + \left(1 - \frac{1 - \gamma_\theta}{4}\right)\underline{c}_{L2\theta}\right).
    \tag{$cL2\theta$}
\end{align}
This is a system of eight equations in eight unknowns, and may be solved explicitly under the assumption that $\gamma_0 = 1 - \gamma_1 = \gamma$.
(The algebra is omitted.) Figures~\ref{fig:simpleprices} and~\ref{fig:simpleexit} display the prices, relationship between price and rating, and exit probabilities, respectively.
We highlight the following relevant features of the simple model:
\begin{enumerate}
  \item The relationship between the price and the rating is dependent on the number of sales made by the seller (Figure~\ref{fig:simpleprices}).
  \item The spread between the price in a high-rated state and a low-rated state is dependent on the quality of the rating system, $\gamma$ (Figure~\ref{fig:simpleprices}).
  \item High quality sellers are more likely to exit compared to low quality sellers, but the exit decision is non-monotonic in the quality of the rating system (Figure~\ref{fig:simpleexit}).
  \item The mass of low quality sellers who are low ranked is increasing in the quality of the rating system, and the mass of high quality sellers who are high ranked is decreasing in the quality of the rating system (Figure~\ref{fig:simplemass}).
\end{enumerate}